\documentclass[aps,prb,twocolumn,superscriptaddress,longbibliography,amsmath,amssymb,floatfix]{revtex4-2}

\usepackage[T1]{fontenc}
\usepackage[utf8]{inputenc}
\usepackage{lmodern}
\usepackage{graphicx}
\usepackage{bm}
\usepackage{braket}
\usepackage{mathtools}
\usepackage{microtype}
\usepackage{placeins}
\usepackage[hidelinks]{hyperref}
\usepackage{verbatim}
\usepackage{xcolor}

\allowdisplaybreaks[2]
\graphicspath{{./}{figures/}}

\newcommand{\ii}{\mathrm{i}}

\newcommand{\vk}{\mathbf{k}}

\newcommand{\bareps}{\bar{\epsilon}}
\newcommand{\mel}[3]{\left\langle #1\middle|#2\middle|#3\right\rangle}

\begin{document}
	
	\title{Generalized s-d model for Wannier-Mott excitons in layered magnetic semiconductors}
	
	\author{Sonu Verma}
	\email{sonu.vermaiitk@gmail.com}
	\affiliation{Department of Physics and Research Center OPTIMAS, Rheinland-Pf\"alzische Technische Universit\"at Kaiserslautern-Landau, 67663 Kaiserslautern, Germany}
	\affiliation{Institute for High Pressure, Department of Physics, Hanyang University, Seoul 04763, Republic of Korea }

	\author{Bashab Dey}
	\affiliation{Department of Physics and Research Center OPTIMAS, Rheinland-Pf\"alzische Technische Universit\"at Kaiserslautern-Landau, 67663 Kaiserslautern, Germany}
	
	\author{Akashdeep Kamra}
	\affiliation{Department of Physics and Research Center OPTIMAS, Rheinland-Pf\"alzische Technische Universit\"at Kaiserslautern-Landau, 67663 Kaiserslautern, Germany}
	
	\date{\today}

	\begin{abstract}
		The recent discovery of excitons coupled to the magnetic order, and the consequent strong magneto-optic responses, in some van der Waals magnetic semiconductors has triggered intense activity at the interface of magnetism and semiconductor optics. Here, we present an analytically tractable minimal model that describes magnetic order, electrons, holes, and excitons within a unified framework, thereby capturing a wide range of phenomena. It treats the magnetic order and itinerant carriers to be comprised by distinct electronic orbitals that are mutually coupled via orbital-dependent onsite exchange, similar to the treatment of metallic magnets using an s-d model. Investigating CrSBr bilayer as a case study, we benchmark our model and its predictions against recent experimental and ab-initio results finding good agreement as well as new insights enabled by the model's simplicity. Examining the optical selection rules, we find the conservation of a quantum number formed from a combination of spin and layer pseudospin to be a useful guiding principle, even in noncollinear magnetic configurations. Our analysis finds a series of bright and dark excitonic states in such layered A-type antiferromagnets. The presented framework should be valuable in achieving intuitive understanding of recently discovered excitonic phenomena and guiding the discovery of other excitonic states in layered magnetic semiconductors.
	\end{abstract}

	\maketitle
	
	\section{Introduction}\label{sec:intro}
	
	The recent discovery of layered van der Waals magnets has provided material platforms hosting a number of unique phenomena~\cite{Gibertini2019,Burch2018,Wang2022}, such as two-dimensional magnetism in mono- and bilayers, with profound scientific and technological implications. A key ingredient that underlies this opportunity is the coexistence of strong crystalline intralayer interactions, which characterize their robust properties, and weak interlayer van der Waals coupling that admits a convenient external control. A special place in this class is held by the magnetic semiconductors, a prominent example being CrSBr~\cite{Wilson2021,Ziebel2024,Brennan2024}, which are enabling unprecedented magneto-optic effects~\cite{Bae2022,Wilson2021,Dirnberger2023} due to their hosting robust bright excitons~\cite{Adak2026}, bound electron-hole pairs, that couple strongly to the magnetic order. Since their recent discovery, these excitonic states have been investigated intensively, experimentally and using ab-initio methods, uncovering a wealth of phenomena~\cite{Adak2026}. Among these, we note their strong tunability via the magnetic order control as well as their different spatial extents associated with their variable Frenkel-like and Wannier-like characters~\cite{Shao2025,Liebich2025,Smiertka2026}. While the early discoveries focused on optically bright excitons playing a role in the optical responses, recent research has gone beyond optics and also investigates excitonic transport~\cite{Dirnberger2025Transport,Iakovlev2026} as well as dark excitons~\cite{Krelle2025,Shree2021,Bork2026, Heissenbuettel2025,Acharya2026}.
	
	Complementary to the optics, these magnetic semiconductors also harbor magnonic and spintronic phenomena, such as magnetoresistance effect~\cite{Telford2020}, magnetic hysteresis~\cite{BoixConstant2023,BoixConstant2025,Mondal2026}, and magnon spin transport~\cite{deWal2024}. These effects can be understood in terms of the magnetic order parameter, described via the Landau-Lifshitz framework, combined with the independent particle description of electrons and holes developed in semiconducting spintronics~\cite{Jungwirth2006,Jungwirth2016,Fabian2007}. This class of effects is valuable for both probing the magnetic and crystalline order in the materials as well as for storing information magnetically~\cite{BoixConstant2023,BoixConstant2025,Mondal2026}. 
	
	In this article, we aim to unify these seemingly complementary phenomena and theoretical approaches into a simplified analytically tractable framework. We formulate a generalized s-d model to describe the magnetic order, electrons, holes, and excitons in layered A-type antiferromagnets with ferromagnetic intralayer exchange~\cite{Adak2026}. In developing this framework, we are inspired by the treatment of conventional metallic magnets via the s-d model that considers the magnetic order to be captured by the localized d electrons and electrical conduction to be contributed by the itinerant s electrons. Despite the actual metals often consisting of hybridized orbitals that do not allow this clear separation, the s-d model has been successful in analytically capturing and predicting a broad range of phenomena by incorporating an s-d exchange coupling treated as a phenomenological material parameter~\cite{Ralph2008,Manchon2019,Simensen2020}. 
	
	With the goal of accomplishing a minimal framework, we consider a tight-binding Hamiltonian with each site associated with a localized spin magnetic moment and two electronic orbitals that capture the conduction and valence bands. Their orbital-dependent onsite exchange coupling with the localized spin is found to be important for capturing the key material properties. Formulating our general model, we take up a detailed case study of CrSBr bilayers with the aim of benchmarking our approach against previous results~\cite{Heissenbuettel2025,Semina2024,TabatabaVakili2024,Klein2023,Wilson2021} and developing new insights enabled by our simple analytic model. We determine the phenomenological parameters, such as intersite hopping amplitudes and onsite exchange couplings, entering our model by comparing the previously established band structure with our evaluation. Our simple model enables formulation of optical selection rules thereby labeling bright and dark transitions in terms of a combined spin and layer-pseudospin quantum number. Investigating Wannier-Mott excitonic states, we discuss approximate treatments based on truncating the Hamiltonian in the spin-layer space and benchmark our results against a numerical solution of the Bethe-Salpeter equation finding good agreement. Our analysis uncovers a series of bright and dark excitons with their optical response depending on the combined spin-layer pseudospin conservation. We expect that the framework presented here will be useful in qualitatively understanding excitons in van der Waals magnets and guide the search for further excitons that still await discovery.

	The article is organized as follows. Section~\ref{sec:model} introduces the general
	s-d Hamiltonian,
	its reduction for systems with a magnetically ordered state, the CrSBr
	bilayer model, and the scope, limitations, and extensions of the model. Section~\ref{sec:singleparticle} derives symmetry properties, the exact \(\Gamma\)-point eigenstates and energies, and the analytical direct-gap expressions. Section~\ref{sec:selectionrules} derives the optical selection rules and band-edge matrix elements. Section~\ref{sec:exciton} formulates the layer-resolved Bethe-Salpeter
	equation and develops the analytical two-channel model for excitons,
	its layer decomposition, and the optical-response formalism. Section~\ref{sec:numerical_results} presents the numerical Bethe-Salpeter spectra, the field-dependent layer redistribution, and the dependence of the binding energy on screening, layer separation, and mass anisotropy. Section~\ref{sec:discussion} interprets the physical results and summarizes their experimental relevance and model scope.
	
	\section{Generalized s-d model employing tight-binding framework}
	\label{sec:model}
	
	We first introduce a general s-d Hamiltonian for itinerant electrons coupled to localized magnetic
	moments. We then obtain a single-particle Hamiltonian for a prescribed magnetically ordered state by replacing the localized-spin operators with
	their expectation values. We next specialize this Hamiltonian to the layer, orbital, and spin degrees of freedom relevant to the near-gap bands of
	a CrSBr bilayer. Finally, we outline the scope of the model, including more general
	multilayers and magnetic textures, and the extensions required to incorporate
	additional orbitals, interlayer tunneling, spin-orbit coupling, and dynamical-spin effects.
	
	\subsection{Generalized s-d Hamiltonian and ordered-state reduction}
	\label{sec:general_real_space_model}
	
	Let \(o=1,\ldots,N_{\rm orb}\) label the low-energy effective electronic orbitals retained
	in layer \(\ell=1,\ldots,N_L\), and let
	\(\hat c_{\mathbf R\ell o s}\) annihilate an electron with spin \(s\) in the
	in-plane unit cell \(\mathbf R\). The \(2N_{\rm orb}\)-component spinor
	\(\hat{\mathbf c}_{\mathbf R\ell}\) collects the orbital and electron-spin
	indices. The Hamiltonian of the generalized s-d model is
	\begin{align}
		\hat H
		&=\hat H_{\rm el}+\hat H_{\rm ex}+\hat H_{\rm spin},
		\label{eq:H_general}\\
		\hat H_{\rm el}
		&=\sum_{\mathbf R\mathbf R'}\sum_{\ell\ell'}
		\hat{\mathbf c}_{\mathbf R\ell}^{\dagger}
		\,\mathcal T_{\mathbf R\mathbf R'}^{\ell\ell'}\,
		\hat{\mathbf c}_{\mathbf R'\ell'},
		\nonumber\\
		\hat H_{\rm ex}
		&=-\sum_{\mathbf R\ell o}
		\sum_{\nu,\nu'=x,y,z}
		\hat s_{\mathbf R\ell o}^{\nu}
		\,\mathcal J_{\ell o}^{\nu\nu'}\,
		\hat S_{\mathbf R\ell}^{\nu'}.
		\nonumber\\
		\hat H_{\rm spin}
		&=
		\frac{1}{2}\sum_{\mathbf R\mathbf R'}\sum_{\ell\ell'}
		\hat{\mathbf S}_{\mathbf R\ell}^{\,T}
		\,\mathcal K_{\mathbf R\mathbf R'}^{\ell\ell'}\,
		\hat{\mathbf S}_{\mathbf R'\ell'}
		+\mu_{\rm B}\sum_{\mathbf R\ell}
		\bm{B}^{T}_{\rm ext}\mathbf g_{\ell}
		\hat{\mathbf S}_{\mathbf R\ell}.
		\nonumber
	\end{align}
	The matrix
	\(\mathcal T_{\mathbf R\mathbf R'}^{\ell\ell'}\) acts in the combined orbital
	and electron-spin space. It contains the on-site energies, intralayer hopping,
	interlayer tunneling, orbital mixing, and spin-dependent hopping.
	The term \(\hat H_{\rm ex}\) describes the generalized s-d
	exchange interaction between the electrons and localized spins. The electron-spin density in orbital \(o\) is
	\begin{equation}
		\hat{\mathbf s}_{\mathbf R\ell o}
		=\frac{1}{2}\sum_{ss'}
		\hat c_{\mathbf R\ell o s}^{\dagger}
		\boldsymbol\sigma_{ss'}
		\hat c_{\mathbf R\ell o s'},
		\label{eq:carrier_spin_density}
	\end{equation}
	and \(\hat{\mathbf S}_{\mathbf R\ell}\) is the dimensionless localized-spin
	operator, such that the corresponding angular-momentum operator is
	$\hbar\hat{\mathbf S}_{\mathbf R\ell}$. The exchange tensor
	\(\mathcal J_{\ell o}^{\nu\nu'}\) may depend on the orbital and layer and contain both isotropic and anisotropic exchange couplings.  The \(3\times3\) tensor
	\(\mathcal K_{\mathbf R\mathbf R'}^{\ell\ell'}\) describes the localized-spin
	interactions, including exchange and anisotropy. The second term in
	\(\hat H_{\rm spin}\) is the Zeeman coupling to the applied magnetic field
	\(\bm{B}_{\rm ext}\), where \(\mu_{\rm B}\) is the Bohr magneton and
	\(\mathbf g_\ell\) is the localized-spin \(g\) tensor.
	
	For a prescribed magnetically ordered state, we replace each localized-spin
	operator by its expectation value,
	\begin{equation}
		\hat{\mathbf S}_{\mathbf R\ell}
		\;\longrightarrow\;
		S_{\ell}\mathbf n_{\ell}(\mathbf R),
		\qquad |\mathbf n_{\ell}(\mathbf R)|=1.
		\label{eq:ordered_spin_expectation}
	\end{equation}
	Here, $S_{\ell}$ is the magnitude of the ordered spin expectation value
	and $\mathbf n_{\ell}(\mathbf R)$ specifies its direction in layer $\ell$ and
	in-plane unit cell $\mathbf R$. The exchange tensor then
	generates the local electronic exchange field
	\begin{equation}
		\mathbf h_{\ell o}(\mathbf R)
		=S_{\ell}\,\bm{\mathcal J}_{\ell o}\,
		\mathbf n_{\ell}(\mathbf R),
	\end{equation}
	where \((\bm{\mathcal J}_{\ell o})_{\nu\nu'}
	=\mathcal J_{\ell o}^{\nu\nu'}\). The electron-spin interaction becomes
	\begin{equation}
		\hat H_{\rm ex}^{\rm stat}
		=-\frac{1}{2}\sum_{\mathbf R\ell o}
		\hat c_{\mathbf R\ell o}^{\dagger}
		\left[\mathbf h_{\ell o}(\mathbf R)\cdot\boldsymbol\sigma\right]
		\hat c_{\mathbf R\ell o}.
		\label{eq:ordered_exchange_reduction}
	\end{equation}
	For isotropic exchange,
	\(\mathcal J_{\ell o}^{\nu\nu'}
	=\mathcal J_{\ell}^{o}\delta_{\nu\nu'}\),
	we define the exchange-energy scale
	\(J_{\ell}^{o}=\mathcal J_{\ell}^{o}S_{\ell}/2\).
	Equation~\eqref{eq:ordered_exchange_reduction} then reduces to
	\(-\sum_{\mathbf R\ell o}J_{\ell}^{o}
	\hat c_{\mathbf R\ell o}^{\dagger}
	[\mathbf n_{\ell}(\mathbf R)\cdot\boldsymbol\sigma]
	\hat c_{\mathbf R\ell o}\).
	
	This construction applies to multilayer systems with collinear, canted, or noncollinear magnetic
	order. Crystal momentum is well defined when the electronic coefficients and
	the magnetic order share a common translation group. For an \(N_L\)-layer system with nearest-neighbor hopping along the stacking direction,
	the corresponding Bloch Hamiltonian takes the block form
	\begingroup
	\small
	\setlength{\arraycolsep}{3pt}
	\begin{equation}
		H_{N_L}(\mathbf k)=
		\begin{pmatrix}
			h_1(\mathbf k) & \mathsf T_{12}(\mathbf k) & 0 & \cdots\\
			\mathsf T_{12}^{\dagger}(\mathbf k) & h_2(\mathbf k) &
			\mathsf T_{23}(\mathbf k) & \ddots\\
			0 & \mathsf T_{23}^{\dagger}(\mathbf k) & \ddots & \ddots\\
			\vdots & \ddots & \ddots & h_{N_L}(\mathbf k)
		\end{pmatrix},
		\label{eq:general_multilayer_bloch_matrix}
	\end{equation}
	\endgroup
	where each layer block has dimension \(2N_{\rm orb}\).
	Here, \(h_\ell(\mathbf k)\) contains the intralayer electronic terms
	and the exchange field in layer \(\ell\), while
	\(\mathsf T_{\ell,\ell+1}(\mathbf k)\) describes hopping between adjacent
	layers.
	
	\subsection{Reduction to a CrSBr bilayer}
	\label{sec:crsbr_bilayer}
	
\begin{figure}[t]
		\centering
		\includegraphics[width=\columnwidth]{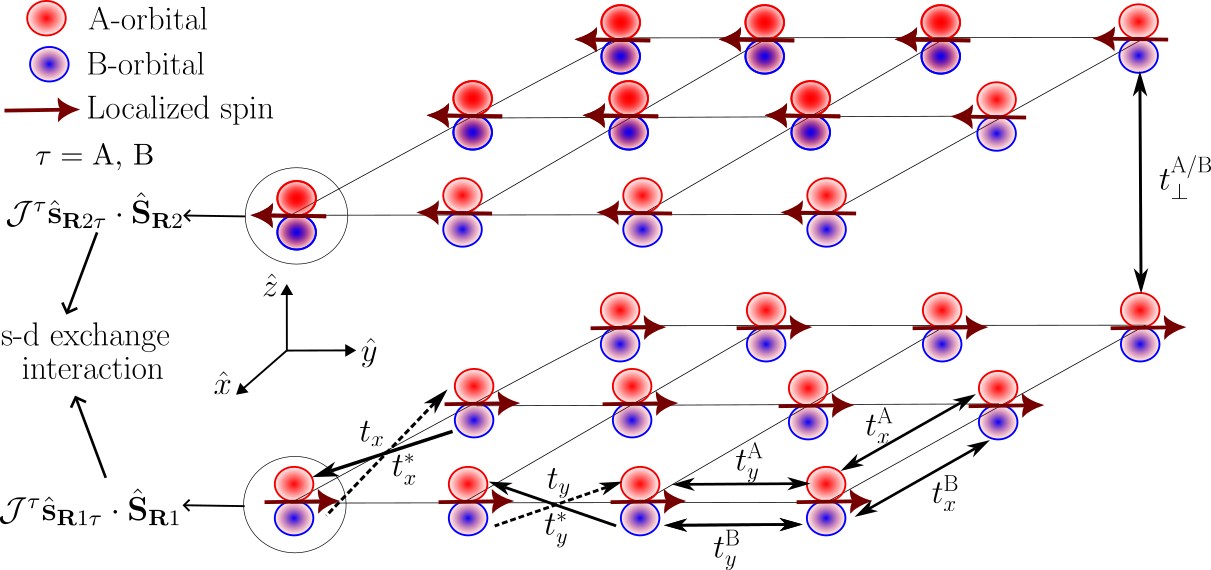}
		\caption{Schematic of the reduced low-energy model for a magnetically ordered CrSBr bilayer.
			Each layer \(\ell=1,2\) contains two effective orbital sectors,
			\(\tau=\textrm{A},\textrm{B}\), representing the conduction- and
			valence-band-edge manifolds, respectively. The localized Cr spin
			\(\hat{\mathbf S}_{\mathbf R\ell}\) couples to the electron spin
			\(\hat{\mathbf s}_{\mathbf R\ell\tau}\) through the orbital-dependent
			s-d exchange interaction
			\(-\mathcal J^\tau
			\hat{\mathbf s}_{\mathbf R\ell\tau}\cdot
			\hat{\mathbf S}_{\mathbf R\ell}\).
			The parameters \(t_x^\tau\) and
			\(t_y^\tau\) describe intralayer hopping along the crystallographic
			\(a\) and \(b\) directions, respectively, while \(t_\perp^\tau\)
			denotes spin-conserving interlayer hopping between equivalent orbital
			sectors. The interorbital hoppings are denoted by \(t_x\) and
			\(t_y\), with \(t_x=0\) and \(t_y=i g_y/2\). Red and blue circles represent the effective
			\(\textrm{A}\)- and \(\textrm{B}\)-orbital sectors, respectively, and
			the arrows indicate the localized Cr spins.}
		\label{fig:lattice}
	\end{figure}

	For CrSBr, we specialize the generic orbital space to two effective orbital sectors, A and B, that span
	the conduction- and valence-edge manifolds after the remote bands have been
	integrated out. We choose \(x\parallel a\), \(y\parallel b\), and
	\(z\parallel c\). The \(b\) axis is the magnetic easy axis, whereas \(c\) is
	the hard out-of-plane axis. Figure~\ref{fig:lattice} shows the degrees of
	freedom retained in the reduced bilayer model.
	
	Within the reduced model, interlayer hopping preserves both the effective
	orbital sector and the electron spin. We retain nearest-neighbor intralayer hopping along the crystallographic
	\(a\) and \(b\) directions. For the interorbital sector, we retain the
	leading coupling along \(b\), consistent with the near-\(\Gamma\)
	description of the CrSBr band edges and their strongly \(b\)-polarized
	optical response~\cite{Klein2023,Qian2023,Watson2024}.
	
	Let \(\mathbf m_1\) and
	\(\mathbf m_2\) be unit vectors along the uniform ordered moments in the two
	layers, with \(|\mathbf m_\ell|=1\).
	The two layers are taken to be electronically equivalent, while their
	magnetic configurations are specified by \(\mathbf m_\ell\). In the
	isotropic ordered-state limit of
	Eq.~\eqref{eq:ordered_exchange_reduction}, the exchange-energy scale is
	orbital dependent, \(J_\ell^{o}\rightarrow J^\tau\). We denote the orbital
	opposite to \(\tau\) by \(\bar\tau\), with
	\(\bar{\rm A}=\rm B\) and \(\bar{\rm B}=\rm A\). The ordered-state electronic Hamiltonian is
	\(\hat H_{\rm CrSBr}
	=\hat H_{\rm el}^{\rm CrSBr}+\hat H_{\rm ex}^{\rm CrSBr}\), where
	\begingroup
	\setlength{\jot}{2pt}
	\begin{equation}
		\begin{aligned}
			\hat H_{\rm el}^{\rm CrSBr}
			&=
			\sum_{\mathbf R\ell\tau s}
			\epsilon^\tau
			\hat c_{\mathbf R\ell\tau s}^{\dagger}
			\hat c_{\mathbf R\ell\tau s}
			\\
			&-
			\sum_{\mathbf R\ell\tau s}
			\sum_{\nu=x,y}\sum_{\delta=\pm1}
			t_\nu^\tau
			\hat c_{\mathbf R\ell\tau s}^{\dagger}
			\hat c_{\mathbf R+\delta\mathbf a_\nu,\ell\tau s}
			\\
			&-
			\sum_{\mathbf R\tau s}
			t_\perp^\tau
			\left(
			\hat c_{\mathbf R1\tau s}^{\dagger}\hat c_{\mathbf R2\tau s}
			+
			\hat c_{\mathbf R2\tau s}^{\dagger}\hat c_{\mathbf R1\tau s}
			\right)
			\\
			&+
			\frac{i g_y}{2}
			\sum_{\mathbf R\ell\tau s}
			\hat c_{\mathbf R\ell\tau s}^{\dagger}
			\left(
			\hat c_{\mathbf R+\mathbf a_y,\ell\bar\tau s}
			-
			\hat c_{\mathbf R-\mathbf a_y,\ell\bar\tau s}
			\right).
		\end{aligned}
		\label{eq:CrSBr_real_space_el}
	\end{equation}
	\endgroup
	while the ordered-state s-d exchange is
	\begin{equation}
		\hat H_{\rm ex}^{\rm CrSBr}
		=
		-\sum_{\mathbf R\ell\tau}\sum_{ss'}
		J^\tau
		\hat c_{\mathbf R\ell\tau s}^{\dagger}
		\left(
		\mathbf m_\ell\cdot\boldsymbol{\sigma}
		\right)_{ss'}
		\hat c_{\mathbf R\ell\tau s'} .
		\label{eq:CrSBr_real_space_ex}
	\end{equation}
	Here, \(\mathbf a_\nu=a_\nu\hat{\boldsymbol\nu}\) denotes the
	nearest-neighbor displacement along \(\nu=x,y\). The parameters
	\(\epsilon^\tau\), \(t_\nu^\tau\), and \(t_\perp^\tau\) denote the onsite,
	intralayer, and interlayer matrix elements of orbital sector \(\tau\),
	respectively, and specify the corresponding entries of
	\(\mathcal T_{\mathbf R\mathbf R'}^{\ell\ell'}\) in
	Eq.~\eqref{eq:H_general}. The sum over \(\delta=\pm1\) runs over the two
	nearest neighbors at \(\mathbf R\pm\mathbf a_\nu\).
	The \(g_y\) term instead describes interorbital hopping along the
	crystallographic \(b\) direction with opposite signs for displacements
	\(+\mathbf a_y\) and \(-\mathbf a_y\), consistent with the strongly
	\(b\)-polarized interband coupling of CrSBr
	\cite{Klein2023,Qian2023}.
	Equation~\eqref{eq:CrSBr_real_space_ex} is the CrSBr form of the
	ordered-state s-d exchange in
	Eq.~\eqref{eq:ordered_exchange_reduction}.
	
	For the momentum-space representation, we denote Pauli matrices in
	layer, orbital, and electron-spin space by
	\(\eta_i\), \(\tau_i\), and \(\sigma_i\), respectively. The layer projectors
	are \(\Pi_{1,2}=(\eta_0\pm\eta_z)/2\), and the orbital projectors are
	\(\tau_{\rm A,B}=(\tau_0\pm\tau_z)/2\). We define the orbital-dependent exchange
	and interlayer hopping matrices as
	\begin{equation}
		J_{\rm orb}=J^{\rm A}\tau_{\rm A}+J^{\rm B}\tau_{\rm B},
		\qquad
		\mathcal T_\perp=t_\perp^{\rm A}\tau_{\rm A}+t_\perp^{\rm B}\tau_{\rm B}.
	\end{equation}
	
	With
	\(\hat c_{\mathbf R\ell\tau s}
	=N^{-1/2}\sum_{\mathbf k}
	e^{i\mathbf k\cdot\mathbf R}
	\hat c_{\mathbf k\ell\tau s}\),
	the corresponding single-particle Bloch Hamiltonian is
	\begin{align}
		H_{\rm CrSBr}^{\rm orb}(\mathbf k;\mathbf m_1,\mathbf m_2)
		&=\eta_0\otimes h_{\rm orb}(\mathbf k)\otimes\sigma_0
		\nonumber\\
		&\quad-\sum_{\ell=1,2}\Pi_\ell\otimes J_{\rm orb}
		\otimes(\mathbf m_\ell\cdot\boldsymbol\sigma)
		\nonumber\\
		&\quad-\eta_x\otimes\mathcal T_\perp\otimes\sigma_0,
		\label{eq:Hcrsbr_general_ordered}
	\end{align}
	with
	\begin{align}
		h_{\rm orb}(\mathbf k)
		&=\epsilon_{\rm A}(\mathbf k)\tau_{\rm A}+\epsilon_{\rm B}(\mathbf k)\tau_{\rm B}
		+d_y(\mathbf k)\tau_x,
		\nonumber\\
		\epsilon_\tau(\mathbf k)
		&=\bareps^\tau+2t_x^\tau[1-\cos(k_xa_x)]
		+2t_y^\tau[1-\cos(k_ya_y)],
		\nonumber\\
		d_y(\mathbf k)&=-g_y\sin(k_ya_y).
		\label{eq:crsbr_dispersion}
	\end{align}
	
	The real-space onsite energy is related to the band-edge reference
	energy by
	\(\bareps^\tau=\epsilon^\tau-2t_x^\tau-2t_y^\tau\).
	The opposite signs of the interorbital hopping for displacements
	\(\pm\mathbf a_y\) give
	\((ig_y/2)(e^{ik_ya_y}-e^{-ik_ya_y})
	=-g_y\sin(k_ya_y)\),
	which yields \(d_y(\mathbf k)\) in
	Eq.~\eqref{eq:crsbr_dispersion}. Near \(\Gamma\)-point, this term is linear in
	\(k_y\) and is the lattice counterpart of the leading \(k_y p_{cv}\)
	interband coupling in low-energy descriptions of CrSBr
	\cite{Klein2023,Semina2024,Smolenski2025}.

	The anisotropy between \(t_x^\tau\) and \(t_y^\tau\) reproduces the
	quasi-one-dimensional band curvature
	\cite{Klein2023,Semina2024,Smolenski2025}. Because
	\(d_y(\Gamma)=0\), the two orbital sectors decouple at the \(\Gamma\) point.
	Its \(k_y\) derivative is nevertheless finite. The Bloch velocity operator is
	\(\hat V_\nu(\mathbf k)=\hbar^{-1}\partial_{k_\nu}
	H_{\rm CrSBr}^{\rm orb}(\mathbf k)\).
	In the layer-orbital-spin basis, its components have a matrix representation. At
	\(\Gamma\)-point, \(\hat V_x=0\), whereas
	\(\hat V_y=-(a_yg_y/\hbar)\eta_0\otimes\tau_x\otimes\sigma_0\).
	Hence, the \(b\)-directed velocity operator connects the A and B
	orbital sectors at the band edge, and its interband matrix elements determine
	the \(b\)-polarized optical coupling.
	This is the origin of the strong linear dichroism of CrSBr
	\cite{Klein2023,Qian2023}.
	
	Equation~\eqref{eq:Hcrsbr_general_ordered} does not select a direction for the
	applied magnetic field. The field acts on the localized moments through
	\(\hat H_{\rm spin}\), and minimization of the magnetic free energy determines
	\(\mathbf m_1\) and \(\mathbf m_2\). 
	
	For the analytical results developed below, we consider a magnetic field
	along the crystallographic \(c\) axis. Minimization of the two-sublattice
	magnetic free energy gives the symmetric canting trajectory
	\begin{align}
		\mathbf m_1
		&=(0,\cos\alpha,\sin\alpha),
		\nonumber\\
		\mathbf m_2
		&=(0,-\cos\alpha,\sin\alpha),
		\nonumber\\
		\sin\alpha
		&=\frac{B_{\rm ext}}{B_{\rm sat}},
		\qquad
		0\leq B_{\rm ext}\leq B_{\rm sat}.
		\label{eq:hard_axis_texture}
	\end{align}
	Here, \(\alpha\) is the canting angle, \(B_{\rm ext}\) is the applied field,
	and \(B_{\rm sat}\) is the saturation field. Appendix~\ref{app:canting}
	derives this trajectory from the magnetic free energy.
	
\begin{figure}[t]
		\centering
		\includegraphics[width=0.48\textwidth]{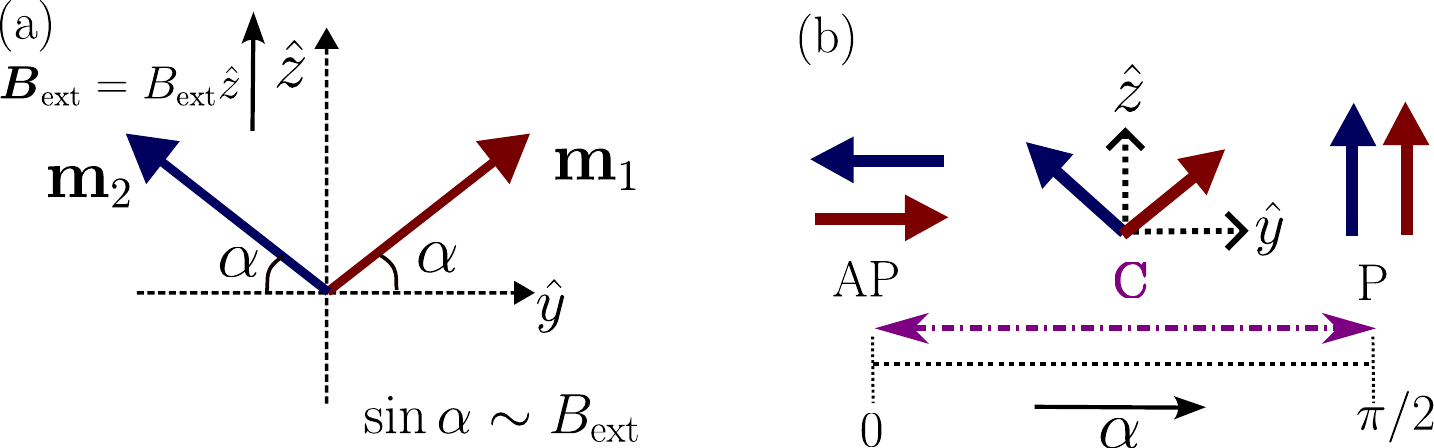}
		\caption{Field-induced canting of the ordered moments in a CrSBr bilayer
			for an applied magnetic field \(B_{\rm ext}\parallel c\).
			(a) Definition of the symmetric canting configuration. The ordered
			moment directions \(\mathbf m_1\) and \(\mathbf m_2\)
			have antiparallel components along the magnetic easy axis \(b\) and equal
			components along the applied-field direction \(c\). The canting angle
			\(\alpha\) is measured from the \(b\) axis.
			(b) Evolution of the magnetic configuration
			from the AFM state at \(\alpha=0\), through the canted regime, to the
			field-polarized state at \(\alpha=\pi/2\). AP and P denote antiparallel and parallel layer-moment
			configurations, respectively.}
		\label{fig:M_vs_B_hard}
	\end{figure}

	At zero field (\(\alpha=0\)), the ordered layer moments are antiparallel (AP) along the easy
	\(b\) axis, corresponding to interlayer antiferromagnetic (AFM) order. For
	\(0<B_{\rm ext}<B_{\rm sat}\) (\(0<\alpha<\pi/2\)), the equilibrium state is a canted AFM
	configuration with the \(b\)-axis components remaining antiparallel, whereas the \(c\)-axis components are parallel to the applied field. At
	\(B_{\rm ext}=B_{\rm sat}\) (\(\alpha=\pi/2\)), the \(b\)-axis components vanish and the two
	moments become parallel (P) along \(c\), producing the field-polarized
	interlayer ferromagnetic (FM) configuration. Above saturation, the moments
	remain aligned along \(c\). Figure~\ref{fig:M_vs_B_hard} summarizes this
	field-driven evolution of the magnetic order.
	
	Substituting the equilibrium moment directions in
	Eq.~\eqref{eq:hard_axis_texture} into
	Eq.~\eqref{eq:Hcrsbr_general_ordered} gives
	\begin{align}
		H_{\rm CrSBr}(\mathbf k)
		&=\eta_0\otimes h_{\rm orb}(\mathbf k)\otimes\sigma_0
		\nonumber\\
		&\quad-\eta_0\otimes J_{\rm orb}\otimes
		\sin\alpha\,\sigma_z
		\nonumber\\
		&\quad-\eta_z\otimes J_{\rm orb}\otimes
		\cos\alpha\,\sigma_y
		\nonumber\\
		&\quad-\eta_x\otimes\mathcal T_\perp\otimes\sigma_0.
		\label{eq:Hcrsbr}
	\end{align}
	The first line is the spin-independent intralayer band Hamiltonian. The
	second line is the uniform \(c\)-axis component of the exchange field, which
	has the same sign in both layers. The third line is the staggered \(b\)-axis
	component, which has opposite signs in the two layers. The last line is the
	interlayer tunneling. At \(\alpha=0\), the uniform component
	vanishes and the exchange field is purely staggered. At \(\alpha=\pi/2\), the
	staggered component vanishes and the exchange field is uniform.
	
	The hopping matrix is field independent in the present model. The magnetic
	field enters the electronic Hamiltonian through the equilibrium canting angle
	and changes the layer and spin composition of the Bloch states, thereby
	modifying their effective interlayer hybridization. The minimal electronic
	Hamiltonian does not include a separate bare Zeeman term for the itinerant
	electrons. The orbital dependence of \(J^\tau\) and \(t_\perp^\tau\) accounts
	for the different exchange splittings and interlayer hybridization of the
	conduction and valence edges
	\cite{Wilson2021,Watson2024,Heissenbuettel2025,Smiertka2026}.
	Appendix~\ref{app:parameterization} relates the reduced lattice parameters to
	the band-edge masses and optical matrix elements.
	
	\subsection{Generality, scope, and extensions}
	\label{sec:model_generality}
	
	Equation~\eqref{eq:H_general} provides a general
	s-d framework for
	layered magnetic systems with arbitrary orbital content, layer number, hopping
	structure, and magnetic configuration. For compactness, we introduce one
	effective localized moment per in-plane unit cell and layer. Systems with
	multiple magnetic sublattices can be incorporated by adding a corresponding site
	index to \(\hat{\mathbf S}\) and \(\mathcal J\). Different materials and
	stacking geometries can be described by the appropriate choices of
	\(N_{\rm orb}\), \(N_L\), \(\mathcal J\), and \(\mathcal T\).
	
	Equation~\eqref{eq:Hcrsbr_general_ordered} is the low-energy realization of
	this framework for a CrSBr bilayer. It retains two effective
	near direct gap orbital sectors, equivalent layers, a uniform ordered moment in each
	layer, orbital-dependent isotropic electron-spin exchange, and
	orbital-preserving spin-conserving interlayer hopping. Remote electronic
	bands are incorporated through the effective parameters of the reduced
	Hamiltonian. Equation~\eqref{eq:Hcrsbr} further specializes this model to the
	symmetric \(c\)-axis canting trajectory defined in
	Eq.~\eqref{eq:hard_axis_texture}.
	
	The general Hamiltonian can also describe noncollinear and spatially varying
	magnetic textures. A Bloch representation applies when the electronic
	structure and magnetic order share a common translation group. Commensurate
	magnetic or moir\'e modulations can then be treated using an enlarged
	supercell or a continuum moir\'e basis, whereas incommensurate structures
	require a real-space formulation. Inequivalent layers, additional orbital
	sectors, twist-dependent tunneling, spin-dependent hopping, and stronger
	spin-orbit interactions can be included by enlarging the basis and extending
	the corresponding hopping and exchange matrices.
	
	Replacing the localized-spin operators by their ordered expectation values
	treats the magnetic texture as static on the electronic time scale. Quantum
	and dynamical spin fluctuations are therefore not included explicitly in the
	present single-particle Hamiltonian, but they can be incorporated through a
	dynamical treatment of the localized moments and their coupling to the
	electrons. 
	
	\section{Spin-resolved layer parity symmetry and band structure}
	\label{sec:singleparticle}
	We now focus on the CrSBr bilayer described by
	Eq.~\eqref{eq:Hcrsbr}. Magnetic canting changes the layer and spin composition of the band-edge states and thereby modifies their interlayer hybridization. Along the symmetric \(c\)-axis canting trajectory, however, the Hamiltonian retains a composite layer-spin symmetry throughout the evolution from the antiferromagnetic state to the field-polarized state. This symmetry provides a natural basis for diagonalizing Eq.~\eqref{eq:Hcrsbr} and labeling its
	eigenstates.
	\begin{figure*}[t]
		\centering
		\includegraphics[width=0.78\linewidth]
		{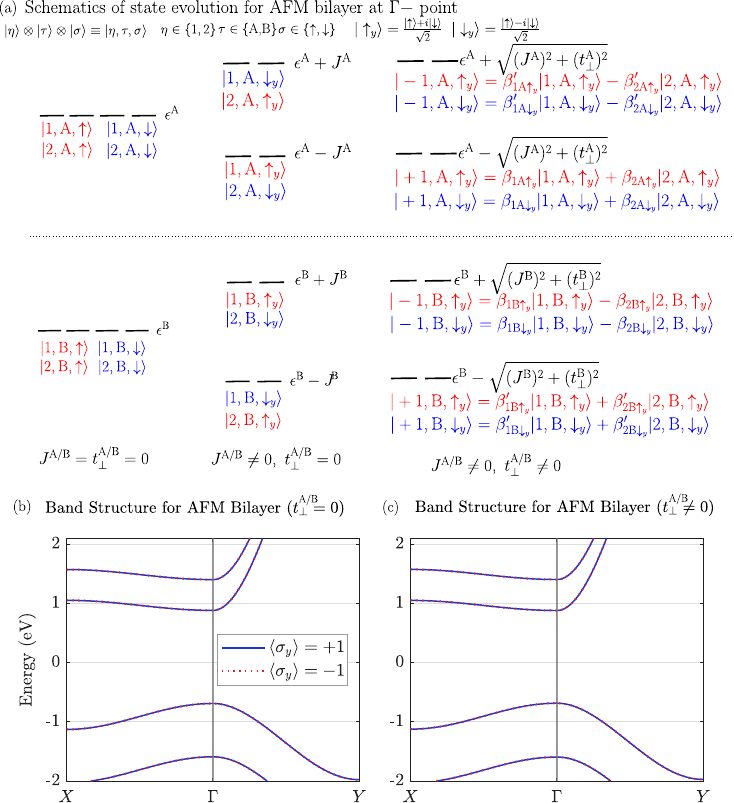}
		\caption{Single-particle states and band structure of the interlayer
			antiferromagnetic bilayer at \(\alpha=0\).
			(a) Evolution of the \(\Gamma\)-point states as the orbital-dependent
			exchange and interlayer hopping are introduced. The left column shows the
			layer-degenerate states for \(J^\tau=t_\perp^\tau=0\); the middle column
			shows the layer-staggered exchange splitting for \(J^\tau\neq0\) and
			\(t_\perp^\tau=0\); and the right column shows the hybridized states for
			finite \(J^\tau\) and \(t_\perp^\tau\), with
			\(\tau=\textrm{A},\textrm{B}\). In this case, the exchange term
			\(-\eta_z\otimes J_{\rm orb}\otimes\sigma_y\) produces opposite
			Zeeman-like shifts in the two layers. For a fixed \(\sigma_y\) eigenvalue,
			interlayer hopping therefore couples layer states separated in energy by
			\(2|J^\tau|\). The wave functions mix to first order in
			\(t_\perp^\tau/J^\tau\), whereas the leading energy shift is second order,
			with scale \((t_\perp^\tau)^2/|J^\tau|\).
			(b) Band structure along \(X\!-\!\Gamma\!-\!Y\) for
			\(t_\perp^\tau=0\).
			(c) Corresponding band structure for finite \(t_\perp^\tau\).
			Solid and dotted curves denote states with
			\(\langle\sigma_y\rangle=+1\) and \(-1\), respectively. The labels
			\(1,2\), \(\textrm{A},\textrm{B}\),
			\(\uparrow,\downarrow\), and
			\(\uparrow_y,\downarrow_y\) denote layer states, orbital sectors,
			\(\sigma_z\) basis states, and \(\sigma_y\) eigenstates, respectively. Energies are shown in eV, and the parameters are
			listed in Appendix~\ref{app:crsbr_parameter_registry}.}
		\label{fig:band_structure_AFM}
	\end{figure*}
	\subsection{Conserved spin-resolved layer-parity symmetry}
	\label{sec:srlp_symmetry}
	
	Along the symmetric canting trajectory in
	Eq.~\eqref{eq:hard_axis_texture}, no fixed Cartesian spin component is
	conserved for \(0<\alpha<\pi/2\). Nevertheless, the composite operator
	\begin{equation}
		\zeta
		=
		\eta_x\otimes\tau_0\otimes\sigma_z
		\label{eq:srlp_operator}
	\end{equation}
	commutes with the Bloch Hamiltonian,
	\([\zeta,H_{\rm CrSBr}(\mathbf k)]=0\) for \(0\leq\alpha\leq\pi/2\).
	The symmetry follows from the combined action of the layer and spin factors
	in \(\zeta\) [see the proof in Appendix~\ref{app:srlp_proof}]. The layer-exchange operator \(\eta_x\) changes the sign of
	\(\eta_z\), while \(\sigma_z\) changes the sign of \(\sigma_y\). These two
	sign changes cancel in the staggered exchange term
	\(\eta_z\otimes J_{\rm orb}\otimes\sigma_y\). The operator \(\zeta\) also
	commutes separately with the orbital, uniform-exchange, and interlayer-hopping
	terms in Eq.~\eqref{eq:Hcrsbr}.
	
	We refer to \(\zeta\) as the spin-resolved layer-parity (SRLP) operator.
	Since \(\zeta^2=\mathbb{I}_8\), its eigenvalues are \(\lambda=\pm1\), and the
	eight-dimensional single-particle Hilbert space separates into two
	four-dimensional sectors \(\mathcal H=\mathcal H_{+}\oplus\mathcal H_{-}\).
	Each Bloch eigenstate \(\ket{\xi_\lambda(\mathbf k)}\) can be chosen as a
	simultaneous eigenstate of \(H_{\rm CrSBr}(\mathbf k)\) and \(\zeta\), with
	\(\zeta\ket{\xi_\lambda(\mathbf k)}=\lambda\ket{\xi_\lambda(\mathbf k)}\).
	The SRLP eigenvalue remains a good quantum number in the canted regime, even
	though spin and layer parity are not separately conserved.
	
	To make the composite character of \(\lambda\) explicit, we define the
	layer-parity states
	\begin{equation}
		\ket{p_{\pm}}
		=
		\frac{\ket{1}\pm\ket{2}}{\sqrt{2}},
		\qquad
		\eta_x\ket{p_{\pm}}
		=
		\pm\ket{p_{\pm}}.
	\end{equation}
	For each orbital basis state \(\ket{\tau}\), a convenient basis for the two
	SRLP sectors is
	\begin{equation}
		\begin{aligned}
			\mathcal H_{+}:&\quad
			\ket{p_+,\tau,\uparrow_z},
			\quad
			\ket{p_-,\tau,\downarrow_z},
			\\
			\mathcal H_{-}:&\quad
			\ket{p_-,\tau,\uparrow_z},
			\quad
			\ket{p_+,\tau,\downarrow_z},
			\qquad
			\tau=\textrm{A,~B}.
		\end{aligned}
		\label{eq:srlp_basis}
	\end{equation}
	Thus, the SRLP eigenvalue is the product of the layer-parity eigenvalue and
	the \(\sigma_z\) eigenvalue of each basis state.
	
	The arrows in Eq.~\eqref{eq:srlp_basis} label eigenstates of \(\sigma_z\).
	For a general canting angle, they are basis labels rather than separately
	conserved quantum numbers. At the field-polarized state,
	\(\sigma_z\) and \(\eta_x\) are separately conserved. At the
	antiferromagnetic state, the conserved spin component is instead
	\(\sigma_y\).
	
	A perturbation preserves the SRLP decomposition only if it commutes with
	\(\zeta\). Spin-dependent hopping, spin-flip tunneling, and
	lower-symmetry spin-orbit terms can in general mix the two SRLP sectors. In
	that case, \(\lambda\) is no longer an exact quantum number.
	
	\subsection{Band-labeling convention}
	\label{sec:band_labels}
	
	Within each SRLP sector, we label the four bands according to their energies
	at the \(\Gamma\) point:
	\begin{equation}
		E_{v'_\lambda}(\Gamma)
		<
		E_{v_\lambda}(\Gamma)
		<
		E_{c_\lambda}(\Gamma)
		<
		E_{c'_\lambda}(\Gamma).
		\label{eq:band_ordering}
	\end{equation}
	Here, \(v_\lambda\) and \(v'_\lambda\) denote the upper and lower valence
	bands, respectively, while \(c_\lambda\) and \(c'_\lambda\) denote the lower
	and upper conduction bands. The SRLP eigenvalue \(\lambda=\pm1\) is the
	symmetry label of the state. The symbols \(c\), \(c'\), \(v\), and \(v'\)
	specify the energy ordering within each SRLP sector and do not denote a fixed
	spin character.
	
	This convention allows the bands to be followed continuously as the magnetic
	configuration changes. It also avoids ambiguities when the exchange couplings
	and interlayer hopping amplitudes differ between the conduction-like A
	sector and the valence-like B sector.
	
	\subsection{Antiferromagnetic and ferromagnetic limits}
	\label{sec:afm_fm_limits}
	
	At the two collinear states, the exchange interaction acts as a
	Zeeman-like field along a fixed spin axis. In the AFM state, this exchange
	field has opposite signs in the two layers. In the field-polarized FM state,
	it has the same sign in both layers. This difference determines whether
	interlayer hopping produces a second-order or first-order change in the
	band-edge energies.
	\begin{figure*}[t]
		\centering
		\includegraphics[width=0.72\linewidth]
		{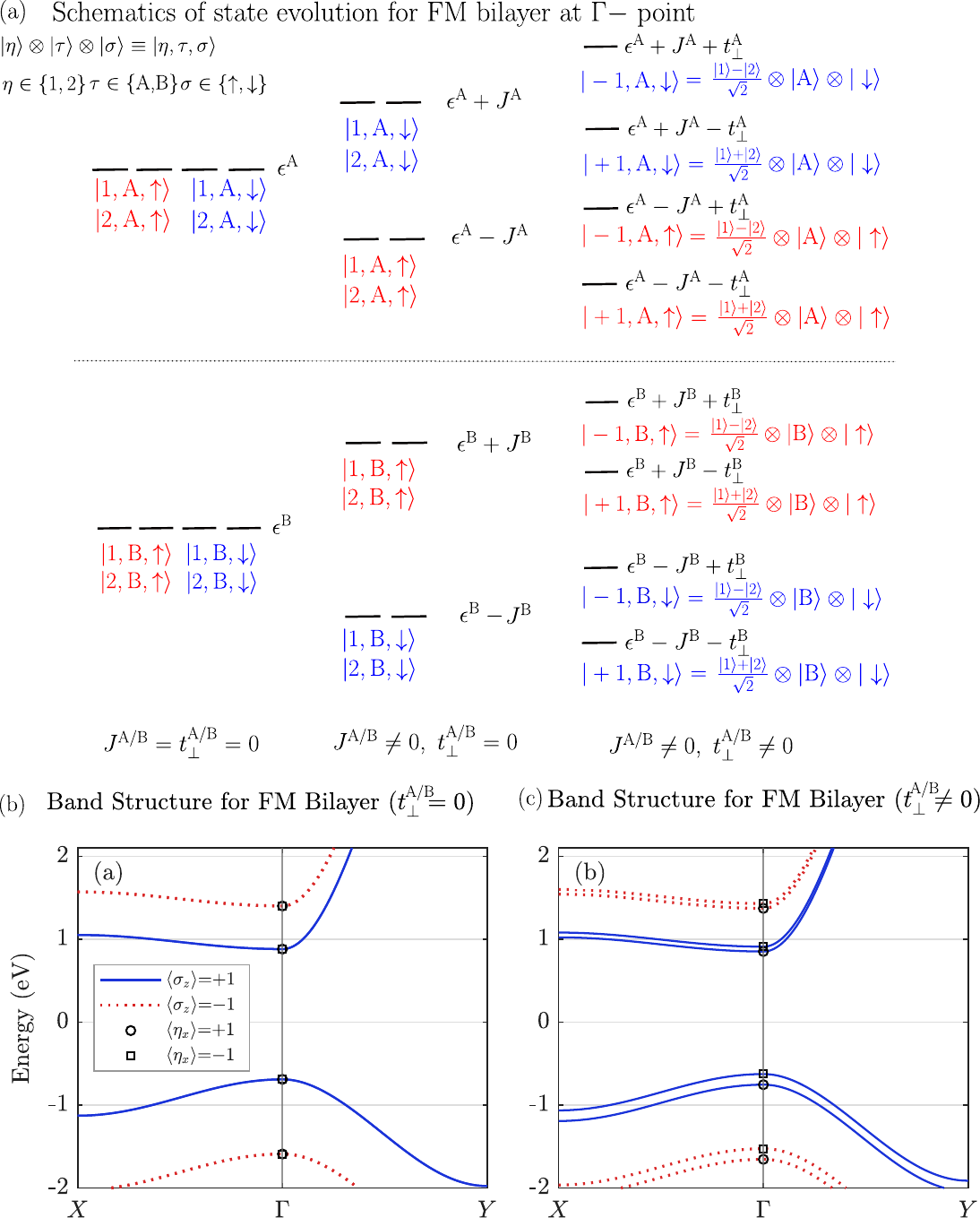}
		\caption{ Single-particle states and band structure of the field-polarized
			interlayer-ferromagnetic bilayer at \(\alpha=\pi/2\).
			(a) Evolution of the \(\Gamma\)-point states as the orbital-dependent
			exchange and interlayer hopping are introduced. The left column shows the
			layer-degenerate states for \(J^\tau=t_\perp^\tau=0\); the middle column
			shows the exchange-split states for \(J^\tau\neq0\) and
			\(t_\perp^\tau=0\); and the right column shows the bonding and
			antibonding states for finite \(J^\tau\) and \(t_\perp^\tau\), with
			\(\tau=\textrm{A},\textrm{B}\). In this case, the
			exchange term
			\(-\eta_0\otimes J_{\rm orb}\otimes\sigma_z\) produces the same
			Zeeman-like shift in both layers. Consequently, for a fixed
			\(\sigma_z\) eigenvalue, the two layer states are degenerate before
			interlayer hopping is introduced. Spin-conserving hopping couples these
			states resonantly and produces eigenstates with layer parity
			\(p=\pm1\), first-order energy shifts
			\(-p\,t_\perp^\tau\), and a splitting \(2|t_\perp^\tau|\).
			(b) Band structure along \(X\!-\!\Gamma\!-\!Y\) for
			\(t_\perp^\tau=0\).
			(c) Corresponding band structure for finite \(t_\perp^\tau\).
			Solid and dotted curves denote states with
			\(\sigma_z=+1\) and \(-1\), respectively. The labels
			\(1,2\), \(\textrm{A},\textrm{B}\),
			\(\uparrow,\downarrow\), and \(\pm1\) denote eigenstates or
			eigenvalues of \(\eta_z\), \(\tau_z\), \(\sigma_z\), and
			\(\eta_x\), respectively. Energies are shown in eV, and the parameters
			are listed in Appendix~\ref{app:crsbr_parameter_registry}.}
		\label{fig:band_structure_FM}
	\end{figure*}
	\subsubsection{Antiferromagnetic limit}
	
	At \(\alpha=0\), the ordered layer moments are antiparallel along the
	\(b\) axis. The exchange term in Eq.~\eqref{eq:Hcrsbr} becomes
	\(-\eta_z\otimes J_{\rm orb}\otimes\sigma_y\). It therefore acts as a
	layer-staggered Zeeman-like exchange field along \(y\), with its
	orbital-dependent sign and magnitude contained in \(J^\tau\). The Hamiltonian
	commutes with \(\sigma_y\). At the \(\Gamma\) point,
	\(d_y(\Gamma)=0\), so the A and B orbital sectors also decouple. The
	exact band-edge energies are
	\begin{align}
		E^{\rm AFM}_{c'_\lambda}
		&=
		\bareps^{\rm A}
		+
		\sqrt{(J^{\rm A})^2+(t_\perp^{\rm A})^2},
		\nonumber\\
		E^{\rm AFM}_{c_\lambda}
		&=
		\bareps^{\rm A}
		-
		\sqrt{(J^{\rm A})^2+(t_\perp^{\rm A})^2},
		\nonumber\\
		E^{\rm AFM}_{v_\lambda}
		&=
		\bareps^{\rm B}
		+
		\sqrt{(J^{\rm B})^2+(t_\perp^{\rm B})^2},
		\nonumber\\
		E^{\rm AFM}_{v'_\lambda}
		&=
		\bareps^{\rm B}
		-
		\sqrt{(J^{\rm B})^2+(t_\perp^{\rm B})^2}.
		\label{eq:afm_energies}
	\end{align}
	These energies are independent of \(\lambda\). The resulting twofold
	degeneracy is exact within the minimal AFM Hamiltonian.
	
	The unitary operator
	\begin{equation}
		\mathcal U_{\rm AFM}
		=
		\eta_x\otimes\tau_0\otimes\sigma_x
		\label{eq:afm_partner_operator}
	\end{equation}
	enforces this degeneracy because
	\begin{equation}
		\left[
		\mathcal U_{\rm AFM},
		H_{\rm CrSBr}(\mathbf k,\alpha=0)
		\right]
		=
		0,
		\qquad
		\left\{
		\mathcal U_{\rm AFM},\zeta
		\right\}
		=
		0.
		\label{eq:afm_partner_symmetry}
	\end{equation}
	If \(\ket{\psi_\lambda}\) is an eigenstate with energy \(E\) and SRLP
	eigenvalue \(\lambda\), then
	\(\mathcal U_{\rm AFM}\ket{\psi_\lambda}\) is an orthogonal eigenstate with
	the same energy and SRLP eigenvalue \(-\lambda\).
	
	For a fixed \(\sigma_y\) eigenvalue, the opposite Zeeman-like exchange fields
	shift the corresponding states in the two layers in opposite directions.
	Before interlayer hopping is introduced, these same-spin states are separated
	in energy by \(2|J^\tau|\). Interlayer hopping preserves spin and therefore
	couples two nondegenerate layer states. Its diagonal matrix element vanishes,
	so there is no first-order energy correction.
	
	For \(|t_\perp^\tau|\ll|J^\tau|\), the energies in orbital sector
	\(\tau\) have the expansion
	\begin{equation}
		E_{\tau,\beta}^{\rm AFM}
		=
		\bareps^\tau
		+
		\beta\left[
		|J^\tau|
		+
		\frac{(t_\perp^\tau)^2}{2|J^\tau|}
		\right]
		+
		O\!\left[
		\frac{(t_\perp^\tau)^4}{|J^\tau|^3}
		\right],
		~
		\beta=\pm1.
		\label{eq:afm_expansion}
	\end{equation}
	The leading energy shift is therefore second order in the interlayer hopping,
	with scale
	\((t_\perp^\tau)^2/|J^\tau|\). The wave functions acquire an
	opposite-layer component already at first order in
	\(t_\perp^\tau/J^\tau\), even though the energy changes only at second
	order. Figure~\ref{fig:band_structure_AFM}(a) illustrates this off-resonant
	hybridization, while Figs.~\ref{fig:band_structure_AFM}(b) and
	\ref{fig:band_structure_AFM}(c) compare the band structures without and with
	interlayer hopping.
	
	\begin{figure*}[t]
		\centering
		\includegraphics[width=\textwidth]
		{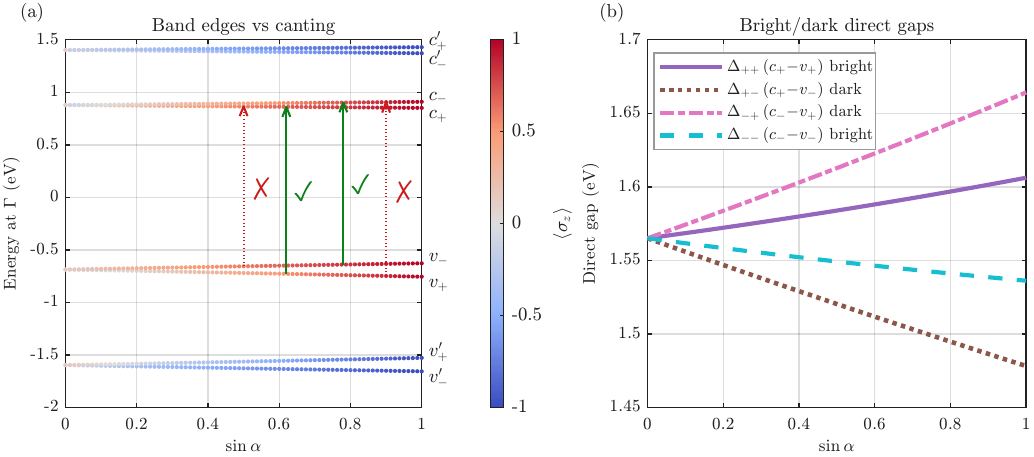}
		\caption{Evolution of the \(\Gamma\)-point band edges and direct gaps
			along the symmetric \(c\)-axis canting trajectory. The horizontal
			coordinate is
			\(\mathcal B=\sin\alpha=B_{\rm ext}/B_{\rm sat}\) below saturation.
			(a) Band-edge energies calculated from
			Eq.~\eqref{eq:gamma_energies}. Each band is labeled by its
			spin-resolved layer-parity (SRLP) eigenvalue
			\(\lambda=\pm1\), and the color scale represents
			\(\langle\sigma_z\rangle\). At \(\mathcal B=0\), the exchange field is
			layer staggered and directed along \(y\), and the bands form exactly
			degenerate SRLP partner pairs within the reduced Hamiltonian. At
			\(\mathcal B=1\), the exchange field is layer uniform and directed along \(z\),
			and the states have definite \(\sigma_z\). At intermediate canting,
			spin and layer parity are separately mixed, while SRLP remains conserved.
			Green solid arrows connect conduction and valence states with equal SRLP,
			whereas red dotted arrows connect states with opposite SRLP. Check marks
			and crosses denote symmetry-allowed and symmetry-forbidden transitions,
			respectively, according to the selection rule derived in
			Sec.~\ref{sec:selectionrules}.
			(b) The four direct gaps
			\(\Delta_{\lambda\lambda'}=
			E_{c_\lambda}(\Gamma)-E_{v_{\lambda'}}(\Gamma)\), calculated from
			Eq.~\eqref{eq:direct_gaps}. Equal-SRLP gaps
			\(\Delta_{++}\) and \(\Delta_{--}\) correspond to the two lowest
			symmetry-allowed channels, whereas
			\(\Delta_{+-}\) and \(\Delta_{-+}\) correspond to opposite-SRLP
			channels. The numerical parameters are listed in
			Appendix~\ref{app:crsbr_parameter_registry}.}
		\label{fig:Band_band_gap_Gamma_vs_alpha}
	\end{figure*}
	
	\subsubsection{Ferromagnetic limit}
	
	At \(\alpha=\pi/2\), the ordered layer moments are parallel along the
	\(c\) axis. The exchange term becomes
	\(-\eta_0\otimes J_{\rm orb}\otimes\sigma_z\). It therefore acts as a
	layer-uniform Zeeman-like exchange field along \(z\), again with the
	orbital-dependent sign and magnitude contained in \(J^\tau\). The operators
	\(\sigma_z\) and \(\eta_x\) are separately conserved, and the SRLP eigenvalue
	is their product,
	\begin{equation}
		\lambda=sp,
		\label{eq:fm_lambda}
	\end{equation}
	where \(s=\pm1\) and \(p=\pm1\) are the eigenvalues of \(\sigma_z\) and
	\(\eta_x\), respectively.
	
	With the interlayer hopping convention of Eq.~\eqref{eq:Hcrsbr}, the exact
	\(\Gamma\)-point energies are
	\begin{align}
		E^{\rm FM}_{\tau,s,p}
		&=
		\bareps^\tau
		-
		sJ^\tau
		-
		p t_\perp^\tau,
		\nonumber\\
		&=
		\bareps^\tau
		-
		sJ^\tau
		-
		\lambda s\,t_\perp^\tau.
		\label{eq:fm_energies}
	\end{align}
	The spin
	eigenvalue \(s\), layer-parity eigenvalue \(p\), and SRLP eigenvalue
	\(\lambda=sp\) are kept distinct because their relation to the ordered band
	labels depends on the signs and relative magnitudes of \(J^\tau\) and
	\(t_\perp^\tau\).
	For the CrSBr parameterization used here, the
	conduction-like and valence-like sectors have exchange couplings of opposite
	sign, \(J^{\rm A}>0\) and \(J^{\rm B}<0\), consistent with the opposite exchange response
	of the near-gap conduction and valence states
	\cite{Wilson2021,Watson2024,Heissenbuettel2025,Smiertka2026}.
	Consequently, for a fixed \(\sigma_z\) eigenvalue, the conduction- and
	valence-band edges are shifted in opposite energy directions. Within a given
	orbital sector, however, the uniform exchange field produces the same
	Zeeman-like shift in both layers. The two layer states are therefore
	degenerate before interlayer hopping is introduced. Interlayer hopping must
	then be diagonalized within this degenerate layer subspace, producing bonding
	and antibonding states with first-order energy shifts
	\(-p\,t_\perp^\tau\) and a splitting \(2|t_\perp^\tau|\).
	
	Figure~\ref{fig:band_structure_FM}(a) illustrates this resonant
	hybridization. Figures~\ref{fig:band_structure_FM}(b) and
	\ref{fig:band_structure_FM}(c) show that switching on interlayer hopping
	splits each pair of same-spin layer states already in first order.

	The two limits show how magnetic order controls interlayer hybridization. In
	the AFM state, the layer-staggered Zeeman-like exchange field detunes
	same-spin states in opposite layers, so interlayer hopping produces a
	second-order energy shift. In the field-polarized FM state, the
	layer-uniform exchange field leaves the same-spin layer states degenerate, so
	the same hopping produces a first-order bonding and antibonding splitting.
	The comparison between Figs.~\ref{fig:band_structure_AFM} and
	\ref{fig:band_structure_FM} directly demonstrates this change from
	off-resonant to resonant interlayer hybridization.
	
	\subsection{Exact \texorpdfstring{\(\Gamma\)}{Gamma}-point energies and direct gaps at arbitrary canting}
	\label{sec:gamma_arbitrary_canting}
	
	We now consider the continuous evolution between the AFM and field-polarized
	FM limits. At an arbitrary canting angle, the exchange field contains a
	layer-uniform component along \(z\) and a layer-staggered component along
	\(y\). At the \(\Gamma\) point, the odd interorbital coupling vanishes,
	\(d_y(\Gamma)=0\), so the A and B orbital sectors decouple. Resolving each orbital
	sector by SRLP reduces the eight-dimensional Hamiltonian to four independent
	\(2\times2\) blocks. Appendix~\ref{app:gamma_solution} gives these blocks and
	their normalized eigenstates.
	
	For orbital sector \(\tau=\) A,~B and SRLP eigenvalue \(\lambda=\pm1\), define
	\begin{align}
		R_{\tau\lambda}(\alpha)
		&=
		\sqrt{
			\left[t_\perp^\tau+\lambda J^\tau\sin\alpha\right]^2
			+
			\left[J^\tau\cos\alpha\right]^2
		}
		\nonumber\\
		&=
		\sqrt{
			(J^\tau)^2
			+
			(t_\perp^\tau)^2
			+
			2\lambda J^\tau t_\perp^\tau\sin\alpha
		}.
		\label{eq:R_tau_lambda}
	\end{align}
	The first term under the square root combines interlayer hopping with the
	layer-uniform \(z\)-directed exchange field within a fixed SRLP sector. The
	second term is the contribution of the layer-staggered \(y\)-directed
	exchange field. The quantity \(2R_{\tau\lambda}\) is the exact splitting
	between the two states in the corresponding orbital and SRLP block.
	
	The four \(\Gamma\)-point band energies in each SRLP sector are
	\begin{align}
		E_{c'_\lambda}(\Gamma)
		&=
		\bareps^{\rm A}+R_{\rm A\lambda}(\alpha),
		&
		E_{c_\lambda}(\Gamma)
		&=
		\bareps^{\rm A}-R_{\rm A\lambda}(\alpha),
		\nonumber\\
		E_{v_\lambda}(\Gamma)
		&=
		\bareps^{\rm B}+R_{\rm B\lambda}(\alpha),
		&
		E_{v'_\lambda}(\Gamma)
		&=
		\bareps^{\rm B}-R_{\rm B\lambda}(\alpha).
		\label{eq:gamma_energies}
	\end{align}
	These exact energies interpolate continuously between the two collinear
	limits. At \(\alpha=0\),
	\(R_{\tau\lambda}=\sqrt{(J^\tau)^2+(t_\perp^\tau)^2}\) is independent of
	\(\lambda\), reproducing the exact AFM energies in
	Eq.~\eqref{eq:afm_energies}. At \(\alpha=\pi/2\),
	\(R_{\tau\lambda}=|J^\tau+\lambda t_\perp^\tau|\), which reproduces the FM
	energies in Eq.~\eqref{eq:fm_energies} after applying the band-ordering
	convention in Eq.~\eqref{eq:band_ordering}.
	
	The four direct transitions formed from \(v_{\lambda'}\) and
	\(c_\lambda\) have the single-particle gaps
	\begin{align}
		\Delta_{\lambda\lambda'}(\alpha)
		&\equiv
		E_{c_\lambda}(\Gamma)
		-
		E_{v_{\lambda'}}(\Gamma)
		\nonumber\\
		&=
		\Delta_\Gamma^0
		-
		R_{A\lambda}(\alpha)
		-
		R_{B\lambda'}(\alpha),
		\label{eq:direct_gaps}
	\end{align}
	where
	\begin{equation}
		\Delta_\Gamma^0
		=
		\bareps^{\rm A}-\bareps^{\rm B}
		\label{eq:bare_gamma_gap}
	\end{equation}
	is the separation between the bare A- and B-orbital energies at
	\(\Gamma\). The pair of SRLP indices \((\lambda,\lambda')\) labels four
	distinct direct transitions. At the AFM state,
	\(R_{\tau,+}=R_{\tau,-}\), so all four gaps are degenerate within the
	minimal model. Canting lifts this degeneracy while each single-particle state
	retains a definite SRLP eigenvalue. The resulting evolution of the band-edge
	energies and direct gaps is shown in
	Fig.~\ref{fig:Band_band_gap_Gamma_vs_alpha}.

	To identify the leading canting dependence of the direct gaps, we write
	\(\mathcal B=\sin\alpha\) and define
	\begin{align}
		R_\tau
		&=
		\sqrt{(J^\tau)^2+(t_\perp^\tau)^2},
		~
		r_\tau
		=
		\frac{2J^\tau t_\perp^\tau}{R_\tau^2},
		~
		\tau=\textrm{A, B}.
		\label{eq:mixing_parameter}
	\end{align}
	Expanding the exact result in the dimensionless mixing parameters \(r_\tau\), with $\mathcal B=\sin\alpha$, gives
	\begin{equation}
		\Delta_{\lambda\lambda'}(\mathcal B)
		=
		\Delta_\Gamma'
		+
		\chi_{\lambda\lambda'}\mathcal B
		+
		\xi_{\rm sp}\mathcal B^2
		+
		O\!\left(
		\max_\tau R_\tau |r_\tau \mathcal B|^3
		\right),
		\label{eq:gap_weakfield}
	\end{equation}
	where
	\begin{align}
		\Delta_\Gamma'
		&=
		\Delta_\Gamma^0-R_{\rm A}-R_{\rm B},
		\nonumber\\
		\chi_{\lambda\lambda'}
		&=
		-\lambda\frac{J^{\rm A} t_\perp^{\rm A}}{R_{\rm A}}
		-\lambda'\frac{J^{\rm B} t_\perp^{\rm B}}{R_{\rm B}},
		\nonumber\\
		\xi_{\rm sp}
		&=
		\frac{(J^{\rm A}t_\perp^{\rm A})^2}{2R_{\rm A}^3}
		+
		\frac{(J^{\rm B}t_\perp^{\rm B})^2}{2R_{\rm B}^3}.
		\label{eq:gap_coefficients}
	\end{align}
	The coefficient \(\chi_{\lambda\lambda'}\) gives the
	channel-dependent linear variation with \(\sin\alpha\), whereas the positive
	coefficient \(\xi_{\rm sp}\) gives the leading quadratic single-particle
	contribution common to all four channels.
	
	Equation~\eqref{eq:gap_weakfield} provides a controlled approximation to the exact gaps throughout the
	canting interval when
	\(\max_{\tau={\rm A,B}}|r_\tau|\ll1\), as occurs in the exchange-dominated regime
	\(|t_\perp^\tau|\ll|J^\tau|\).
	The exact energies are given by Eq.~\eqref{eq:gamma_energies}, while
	Eq.~\eqref{eq:gap_weakfield} separates the leading linear and quadratic
	canting dependences.
	Its derivation is given in Appendix~\ref{app:gamma_solution}.
	
	Equations~\eqref{eq:gamma_energies} through
	\eqref{eq:gap_coefficients} provide the exact band-edge energies and direct
	gaps along the symmetric canting trajectory. They separate the
	channel-dependent linear detuning from the common quadratic
	single-particle contribution and allow each SRLP-resolved transition to be
	followed continuously between the AFM and field-polarized FM limits.

	\section{Single-particle optical excitation selection rules}
	\label{sec:selectionrules}
	
	The band-edge optical response is determined by the orbital structure of the
	velocity operator and by the SRLP symmetry of the Bloch states. For valence
	and conduction eigenstates of Eq.~\eqref{eq:Hcrsbr}, we define
	\begin{equation}
		M_{cv}^{\nu}(\mathbf k)
		=
		\mel{c(\mathbf k)}
		{\partial_{k_\nu}H_{\rm CrSBr}(\mathbf k)}
		{v(\mathbf k)}
		=
		\hbar V_{cv}^{\nu}(\mathbf k).
		\label{eq:M_velocity_def}
	\end{equation}
	Here, \(M_{cv}^{\nu}\) is the interband matrix element of the momentum derivative of
	the Bloch Hamiltonian~\cite{Aversa1995,Onida2002,Rohlfing2000} and its magnitude largely governs the oscillator strength of the corresponding optical transition as discussed further below. 
	
	At the \(\Gamma\) point, the derivatives of the diagonal orbital dispersions vanish.
	The momentum-odd interorbital term
	\(d_y(\mathbf k)=-g_y\sin(k_ya_y)\) therefore gives
	\begin{equation}
		\hat V_x(\Gamma)=0,\qquad
		\hat V_y(\Gamma)=
		-\frac{a_yg_y}{\hbar}\,
		\eta_0\otimes\tau_x\otimes\sigma_0.
		\label{eq:gamma_velocity_operators}
	\end{equation}
	Thus, within the reduced model, the band-edge optical matrix element is
	finite for polarization along \(y\parallel b\), whereas it vanishes for
	polarization along \(x\parallel a\). The operator \(\tau_x\) connects the
	A and B orbital sectors, while acting as the identity in layer and
	spin space. This orbital structure produces the strong linear dichroism
	characteristic of CrSBr
	\cite{Klein2023,Qian2023,Heissenbuettel2025,Smiertka2026}.
	
	The SRLP symmetry supplies an additional exact selection rule. Because
	\(\zeta\) is momentum independent and commutes with
	\(H_{\rm CrSBr}(\mathbf k)\) throughout the symmetric canting trajectory,
	\begin{equation}
		\left[
		\zeta,\partial_{k_\nu}H_{\rm CrSBr}(\mathbf k)
		\right]
		=
		0.
	\end{equation}
	Consequently,
	\begin{align}
		[\zeta,\hat V_\nu(\mathbf k)]
		&=0,
		\nonumber\\
		\lambda_c\neq\lambda_v
		&\Longrightarrow
		\mel{c_{\lambda_c}(\mathbf k)}
		{\hat V_\nu(\mathbf k)}
		{v_{\lambda_v}(\mathbf k)}
		=0.
		\label{eq:selection_rule}
	\end{align}
	An optical transition is therefore symmetry allowed only when the
	conduction and valence states have the same SRLP eigenvalue,
	\begin{equation}
		\lambda_c=\lambda_v.
	\end{equation}
	In particular, the two lowest direct transitions
	\(v_-\to c_-\) and \(v_+\to c_+\) are allowed, whereas
	\(v_-\to c_+\) and \(v_+\to c_-\) are forbidden in the minimal
	Hamiltonian. Appendix~\ref{app:selection_polarization} gives the formal
	proof and the exact band-edge matrix elements.
	
	Equal SRLP is necessary but not sufficient for a finite optical matrix
	element. Because \(\hat V_y(\Gamma)\) acts as the identity in the layer-spin
	subspace, the magnitude of an allowed matrix element is determined by the
	overlap between the layer-spin components of the corresponding A- and
	B-sector eigenstates. Destructive interference or orthogonality between
	these components can suppress an otherwise symmetry-allowed transition.
	
	For the CrSBr parameters used here, the lowest equal-SRLP transitions carry
	substantial oscillator strength throughout the canting trajectory. The
	equal-SRLP cross transitions involving a \(c'\) or \(v'\) band are weaker but
	generally finite in the AFM state and at intermediate canting. At the
	field-polarized FM state, these cross transitions vanish because spin and
	layer parity become separately conserved. This state suppression is
	therefore stronger than the SRLP selection rule alone.
	
	The two lowest symmetry-allowed direct gaps are
	\begin{align}
		\Delta_\lambda^{\rm opt}(\alpha)
		&=
		E_{c_\lambda}(\Gamma)
		-
		E_{v_\lambda}(\Gamma)
		\nonumber\\
		&=
		\Delta_\Gamma^0
		-
		R_{A\lambda}(\alpha)
		-
		R_{B\lambda}(\alpha)
		\nonumber\\
		&\simeq
		\Delta_\Gamma'
		+
		\chi_{\lambda\lambda}\mathcal B
		+
		\xi_{\rm sp}\mathcal B^2,
		\qquad
		\mathcal B=\sin\alpha.
		\label{eq:optical_direct_gaps}
	\end{align}
	The final line uses the expansion in
	Eq.~\eqref{eq:gap_weakfield}. Because
	\(\chi_{+,+}=-\chi_{-,-}\), the linear canting term shifts the two allowed
	gaps in opposite directions. The quadratic coefficient
	\(\xi_{\rm sp}\) is common to both channels and therefore shifts their
	average energy without changing their separation to this order. The exact
	expressions are used in the numerical results.
	
\begin{figure}[t]
		\centering
		\includegraphics[width=0.48\textwidth]{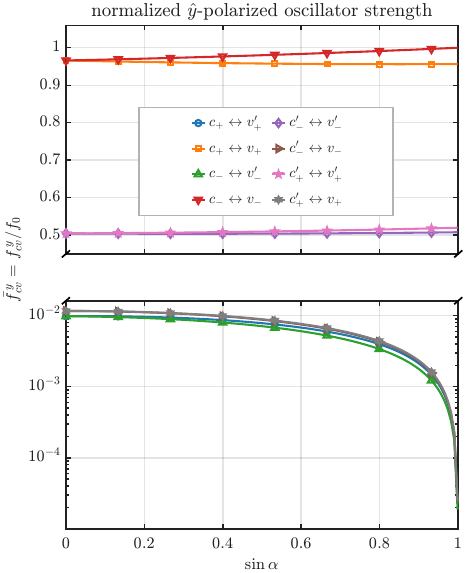}
		\caption{Normalized \(b\)-polarized single-particle oscillator strengths \(\bar f_{cv}^{y}=f_{cv}^{y}/f_0\)
			at the \(\Gamma\) point along the symmetric \(c\)-axis canting trajectory.
			The horizontal coordinate is
			\(\mathcal B=\sin\alpha=B_{\rm ext}/B_{\rm sat}\), with the interlayer-AFM state
			at \(\mathcal B=0\) and the field-polarized interlayer-FM state at \(\mathcal B=1\).
			The upper panel shows the strong equal-SRLP transitions on a linear scale,
			whereas the lower panel shows the weaker equal-SRLP cross transitions on a
			logarithmic scale. The two lowest transitions
			\(v_\lambda\rightarrow c_\lambda\) retain substantial oscillator strength
			throughout the canting trajectory. Equal-SRLP transitions involving an
			upper conduction band \(c'_\lambda\) or a lower valence band
			\(v'_\lambda\) are weaker and vanish at the FM state for the present
			parameterization, where spin and layer parity become separately conserved.
			Opposite-SRLP transitions are not shown because their oscillator strengths
			vanish exactly within the reduced Hamiltonian. The \(a\)-polarized band-edge oscillator
			strengths vanish because \(\hat V_x(\Gamma)=0\).}
		\label{fig:momentum_matrix_element_Gamma_point}
	\end{figure}

	To compare the strengths of the different optical transitions, we use
	the dimensionless single-particle oscillator strength for the
	transition from valence band \(v\) to conduction band \(c\), polarized
	along \(\nu\),
	\begin{equation}
		f_{cv}^{\nu}(\mathbf k)
		=
		\frac{2m_0}{\hbar^2}
		\frac{|M_{cv}^{\nu}(\mathbf k)|^2}
		{E_c(\mathbf k)-E_v(\mathbf k)},
		~
		\bar f_{cv}^{\nu}(\mathbf k)
		=
		\frac{f_{cv}^{\nu}(\mathbf k)}{f_0},
		\label{eq:single_particle_oscillator_strength}
	\end{equation}
	where \(m_0\) is the free-electron mass and
	\(f_0\) is the maximum
	\(b\)-polarized band-edge oscillator strength among the transitions shown
	in Fig.~\ref{fig:momentum_matrix_element_Gamma_point}.  Figure~
	\ref{fig:momentum_matrix_element_Gamma_point} shows the resulting
	\(y\)-polarized oscillator strengths at the \(\Gamma\) point.
	
	The exact absence of opposite-SRLP transitions and of the
	\(a\)-polarized band-edge response follows from the symmetries and orbital
	content of the minimal Hamiltonian. Spin-dependent tunneling,
	lower-symmetry spin-orbit terms, or additional orbital sectors can relax
	these restrictions and generate weak optical weight in channels that are
	dark in the reduced model.

	\section{Excitons}
	\label{sec:exciton}
	
	We now include the screened electron-hole interaction in the
	field-dependent single-particle basis developed above. Within the
	Tamm-Dancoff approximation, a zero-center-of-mass
	(\(\mathbf Q=0\)) exciton is a correlated electron-hole eigenstate
	expanded as a coherent superposition of vertical transitions from occupied
	valence bands to empty conduction bands
	\cite{Rohlfing2000,Onida2002,Rodl2008,Fuchs2008}. Its amplitudes describe
	both the relative electron-hole motion and the distribution of the state
	among the retained band-pair channels.
	
	Because the canted Bloch states are coherent mixtures of layer and spin
	components, a band-pair label does not by itself determine whether the
	electron and hole occupy the same layer or different layers. We therefore
	formulate the Bethe-Salpeter equation in the band basis while retaining the
	layer-resolved matrix elements of the screened interaction.
	
	\subsection{Layer-resolved Bethe-Salpeter equation}
	\label{sec:layer_resolved_bse}
	
	Within the Tamm-Dancoff approximation, a normalized
	\(\mathbf Q=0\) electron-hole state is written as
	\begin{equation}
		\lvert\Psi_n\rangle
		=
		\sum_{cv}
		\int\frac{d^2\mathbf k}{(2\pi)^2}
		\Psi_{n,cv}(\mathbf k)\,
		\hat c_{\mathbf k c}^{\dagger}
		\hat c_{\mathbf k v}
		\lvert GS\rangle,
		\label{eq:exciton_state}
	\end{equation}
	where \(\hat c_{\mathbf k c}^{\dagger}\) creates an electron in conduction
	band \(c\), while \(\hat c_{\mathbf k v}\) removes an electron from valence
	band \(v\). The exciton amplitudes satisfy
	\begin{equation}
		\sum_{cv}
		\int\frac{d^2\mathbf k}{(2\pi)^2}
		\left|\Psi_{n,cv}(\mathbf k)\right|^2
		=
		1.
		\label{eq:exciton_continuum_normalization}
	\end{equation}
	For compactness, we combine the conduction- and valence-band indices
	into an electron-hole channel index \(a=(c,v)\) and write
	\(\Psi_{n,a}(\mathbf k)\equiv\Psi_{n,cv}(\mathbf k)\). Retaining the statically screened direct electron-hole attraction, the
	Bethe-Salpeter equation is
	\begin{align}
		E_{n,X}\Psi_{n,a}(\mathbf k)
		&=
		\Delta_a(\mathbf k)\Psi_{n,a}(\mathbf k)\nonumber\\
		&-
		\sum_b
		\int\frac{d^2\mathbf k'}{(2\pi)^2}
		W_{\rm RK}(q)\,
		\mathcal F_{ab}(d;\mathbf k,\mathbf k')\,
		\Psi_{n,b}(\mathbf k'),
		\label{eq:BSE}\\
		\Delta_a(\mathbf k)
		&=
		E_c(\mathbf k)-E_v(\mathbf k),
		\nonumber
	\end{align}
	Here, \(b=(\bar c,\bar v)\) labels the intermediate electron-hole
	channel and \(q=\lvert\mathbf k-\mathbf k'\rvert\). The form factor
	\(\mathcal F_{ab}(d;\mathbf k,\mathbf k')\) contains the corresponding
	layer-resolved Bloch-state overlaps and is specified below. The function \(W_{\rm RK}(q)>0\) denotes the magnitude of
	the screened attraction.
	
	We model the screened interaction using the Rytova-Keldysh
	form~\cite{Rytova1967,Keldysh1979,Cudazzo2011},
	\begin{align}
		W_{\rm RK}(q)
		&=
		\frac{e^2}
		{2\epsilon_0q\left(\epsilon_{\rm env}+\rho_0q\right)}
		\nonumber\\
		&=
		\frac{e^2}
		{2\epsilon_0\epsilon_{\rm env}q\left(1+r_*q\right)},
		\qquad
		r_*=\frac{\rho_0}{\epsilon_{\rm env}}.
		\label{eq:scalar_separable_kernel}
	\end{align}
	Here, \(\rho_0\) is the intrinsic two-dimensional screening length and
	\(r_*\) is the effective screening radius in an environment with dielectric
	constant \(\epsilon_{\rm env}\). The conversion to the dimensionless
	reference units and the numerical screening parameters used below are given
	in Appendix~\ref{app:parameterization}.
	
	The form factor \(\mathcal F_{ab}(d;\mathbf k,\mathbf k')\) accounts for the
	layer composition of the initial electron-hole channel \(a\) and the
	intermediate channel \(b\). To resolve its same-layer and different-layer
	contributions, we use the one-particle layer projectors
	\begin{equation}
		\mathbb P_{\ell}
		=
		\Pi_{\ell}\otimes\tau_0\otimes\sigma_0,
		\qquad
		\Pi_{1,2}=\frac{\eta_0\pm\eta_z}{2},
		\qquad
		\ell=1,2.
		\label{eq:full_layer_projectors}
	\end{equation}
	Here, \(\mathbb P_\ell\) projects a Bloch state onto layer \(\ell\)
	while leaving its orbital and spin components unchanged.
	Using these projectors, the layer-resolved form factor is separated into
	intralayer and interlayer contributions,
	\begin{equation}
		\mathcal F_{ab}(d;\mathbf k,\mathbf k')
		=
		\mathcal F_{ab}^{\rm intra}(\mathbf k,\mathbf k')
		+
		e^{-dq}\mathcal F_{ab}^{\rm inter}(\mathbf k,\mathbf k').
		\label{eq:form_factor}
	\end{equation}
	For \(a=(c,v)\) and \(b=(\bar c,\bar v)\), the corresponding matrix
	elements are
	\begin{align}
		\mathcal F^{\rm intra}_{c\bar c\bar v v}
		(\mathbf k,\mathbf k')
		&=
		\sum_{\ell=1,2}
		\langle c\mathbf k|
		\mathbb P_{\ell}
		|\bar c\mathbf k'\rangle
		\langle\bar v\mathbf k'|
		\mathbb P_{\ell}
		|v\mathbf k\rangle,
		\label{eq:F_intra}\\
		\mathcal F^{\rm inter}_{c\bar c\bar v v}
		(\mathbf k,\mathbf k')
		&=
		\sum_{\ell\neq m}
		\langle c\mathbf k|
		\mathbb P_{\ell}
		|\bar c\mathbf k'\rangle
		\langle\bar v\mathbf k'|
		\mathbb P_m
		|v\mathbf k\rangle.
		\label{eq:F_inter}
	\end{align}
	The intralayer contribution describes electron-hole configurations whose
	charge centers occupy the same layer, whereas the interlayer contribution
	describes configurations whose charge centers occupy different layers. The
	factor \(e^{-dq}\) accounts for the Fourier-space attenuation of the
	interaction between charge distributions separated by the effective
	distance \(d\)
	\cite{VanderDonck2018,Quintela2022}.
	
	When the intralayer screening is kept independent of \(d\),
	varying \(d\) changes only the interlayer attraction through the factor
	\(e^{-dq}\).
	The intralayer interaction and the single-particle Hamiltonian remain fixed. This dependence therefore describes the weakening of the interlayer
	attraction rather than the complete effect of changing the physical layer
	spacing. The case in which the second layer also screens the intralayer
	electron-hole interaction
	is discussed in Appendix~\ref{app:bilayer_kernel}.
	
	The in-plane screening of CrSBr can in general depend on the direction of
	the transferred momentum. Here, we retain the strongly anisotropic
	single-particle dispersion while approximating the interaction by the
	isotropic function \(W_{\rm RK}(q)\). An anisotropic interaction can
	be incorporated by replacing \(W_{\rm RK}(q)\) with \(W(\mathbf q)\)
	without changing the layer-resolved structure of
	Eqs.~\eqref{eq:F_intra} and \eqref{eq:F_inter}
	\cite{Semina2024}.
	
	The present BSE retains the screened direct interaction and neglects the
	electron-hole exchange kernel. The exact separation into
	equal- and opposite-SRLP sectors
	therefore refers to the direct-kernel minimal model. Long-range
	exchange can modify the fine structure and radiative properties without
	altering the single-particle SRLP classification \cite{Semina2024}. Details
	of the momentum-space discretization, eigenvector normalization, treatment
	of the \(q=0\) singularity, and numerical convergence are given in
	Appendix~\ref{app:numerics}.
	
	\begin{figure*}[t]
		\centering
		\includegraphics[width=0.94\textwidth]{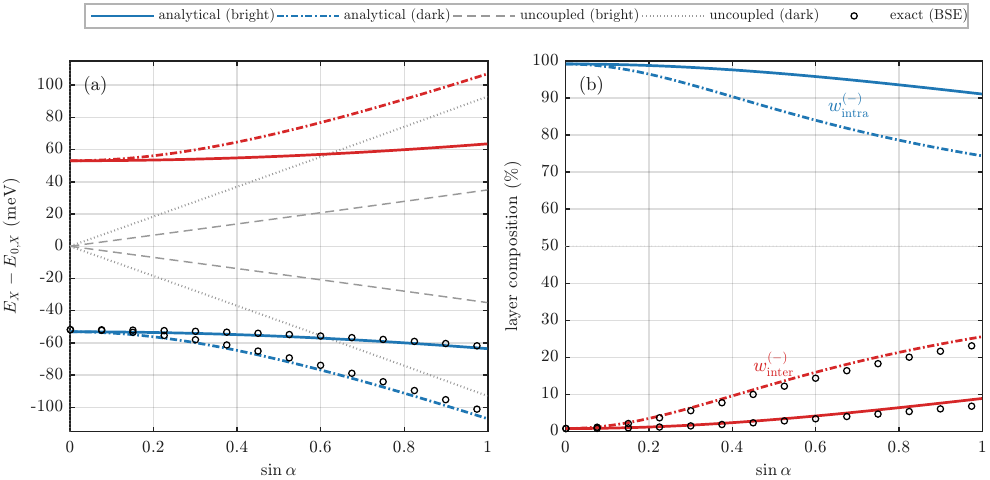}
		\caption{Two-channel exciton description and layer composition of the
			lowest excitons at $d=8.00~\text{\AA}$. Solid and dash-dotted curves denote
			the bright equal-SRLP
			and dark opposite-SRLP
			($\Lambda_{eh}=-1$)
			sectors. The horizontal coordinate is
			$\mathcal B=\sin\alpha=B_{\rm ext}/B_{\rm sat}$.
			(a) Exciton energies relative to
			$E_{0,X}$, the common energy at the AFM state when the coupling
			between the two excitons is omitted.
			The curves show the
			two-channel exciton models, the energies obtained by setting the
			coupling between the two excitons to zero in gray,
			and the full-BSE results as open circles.
			At the AFM state $\mathcal B=0$, the two excitons in each SRLP sector are
			degenerate when their mutual coupling is omitted. Including the coupling
			gives the equal-SRLP energies $E_{0,X}-\Omega_+$ and
			$E_{0,X}+\Omega_+$, separated by $2\Omega_+$. In the opposite-SRLP sector, the analogous coupling \(\Omega_-\)
			splits the two excitons by \(2\Omega_-\).
			(b) Intralayer and interlayer probabilities of the
			lower-energy exciton in each SRLP sector.
			Each retained electron-hole channel contains both intralayer and
			interlayer configurations.}
		\label{fig:two_channel_model_weights}
	\end{figure*}
	\subsection{Symmetry reduction and two-channel exciton model}
	\label{sec:bse_four_channel_projection}
	
	We restrict the layer-resolved Bethe-Salpeter equation to the two lowest
	conduction bands, \(c_+\) and \(c_-\), and the two highest valence bands,
	\(v_+\) and \(v_-\). These near-gap bands define four electron-hole
	channels,
	\begin{equation}
		(c_+v_-),\qquad
		(c_-v_-),\qquad
		(c_+v_+),\qquad
		(c_-v_+).
		\label{eq:four_bse_channels}
	\end{equation}
	The channel labels identify the conduction and valence bands forming the
	electron-hole pair. They do not specify its intralayer or interlayer
	character,
	because the conduction and valence states forming each electron-hole
	channel are themselves superpositions of the two layers.
	
	Each channel carries the two-particle SRLP eigenvalue
	\(\Lambda_{eh}=\lambda_c\lambda_v=\pm1\).
	We denote by \(W_{\ell m}(q)\) the screened direct interaction between
	an electron in layer \(\ell\) and a hole in layer \(m\). For equivalent
	layers, layer-exchange symmetry gives
	\(W_{11}(q)=W_{22}(q)\) and \(W_{12}(q)=W_{21}(q)\).
	
	Under this condition it
	conserves \(\Lambda_{eh}\), and channels belonging to different sectors do
	not mix. A layer-asymmetric dielectric environment or inequivalent electronic
	layers generally relax this exact decomposition. In the channel order of
	Eq.~\eqref{eq:four_bse_channels}, the BSE Hamiltonian has the block structure
	\begingroup
	\scriptsize
	\setlength{\arraycolsep}{0.8pt}
	\begin{equation}
		\mathcal H^{\rm BSE}_{\mathbf k\mathbf k'}
		=
		\begin{pmatrix}
			D_{+,-}&0&0&-K_{+,-;\,-,+}\\
			0&D_{-,-}&-K_{-,-;\,+,+}&0\\
			0&-K_{+,+;\,-,-}&D_{+,+}&0\\
			-K_{-,+;\,+,-}&0&0&D_{-,+}
		\end{pmatrix}_{\mathbf k\mathbf k'}.
		\label{eq:BSE_four_by_four_block}
	\end{equation}
	\endgroup
	These matrix elements are those of the BSE kernel in
	Eq.~\eqref{eq:BSE}. For channels \(a=(c_{\lambda_c},v_{\lambda_v})\) and
	\(b=(c_{\lambda_c'},v_{\lambda_v'})\), we define
	\begin{align}
		K_{ab}(\mathbf k,\mathbf k')
		&=
		W_{\rm RK}(q)\,
		\mathcal F_{ab}(d;\mathbf k,\mathbf k'),
		\nonumber\\
		D_a(\mathbf k,\mathbf k')
		&=
		(2\pi)^2\delta^{(2)}(\mathbf k-\mathbf k')\,
		\Delta_a(\mathbf k)
		-
		K_{aa}(\mathbf k,\mathbf k').
		\label{eq:BSE_DK_definition}
	\end{align}
	Thus, \(D_a\) is the intrachannel BSE kernel, whereas \(K_{ab}\) with
	\(a\neq b\) is the direct-interaction matrix element that couples two
	different electron-hole channels. The symmetry
	reduction is derived in Appendix~\ref{app:exciton_projection}.
	
	The channels \((c_-v_-)\) and \((c_+v_+)\) have
	\(\Lambda_{eh}=+1\) and form the
	equal-SRLP sector. The channels
	\((c_+v_-)\) and \((c_-v_+)\) have
	\(\Lambda_{eh}=-1\) and form an independent dark sector within the reduced
	Hamiltonian.
	The opposite-SRLP channels are dark by the SRLP electric-dipole
	selection rule. An individual exciton in the
	equal-SRLP sector can nevertheless
	have weak or vanishing oscillator strength because of destructive
	interference between its channel amplitudes.
	
	The binding energy is defined relative to the lowest independent
	electron-hole continuum in the same invariant sector,
	\begin{align}
		E_{b,n}^{(\Lambda_{eh})}(\alpha)
		&=
		E_{\rm th}^{(\Lambda_{eh})}(\alpha)
		-
		E_{n,X}^{(\Lambda_{eh})}(\alpha),
		\label{eq:binding_def}\\
		E_{\rm th}^{(\Lambda_{eh})}(\alpha)
		&=
		\min_{\mathbf k,\,(c,v)\in\mathcal C_{\Lambda_{eh}}}
		\left[
		E_c(\mathbf k,\alpha)-E_v(\mathbf k,\alpha)
		\right],
		\label{eq:binding_threshold}
	\end{align}
	where
	\begin{align}
		\mathcal C_{+}
		&=
		\left\{
		(c_-,v_-),(c_+,v_+)
		\right\},
		\nonumber\\
		\mathcal C_{-}
		&=
		\left\{
		(c_+,v_-),(c_-,v_+)
		\right\}.
	\end{align}
	A state is bound when
	\(E_{b,n}^{(\Lambda_{eh})}>0\). Because the
	equal- and opposite-SRLP sectors can have
	different continuum thresholds, their binding energies are evaluated
	separately.
	
	We first consider the two lowest excitons in the
	equal-SRLP sector, formed from the bright single-particle transitions
	\(v_-\rightarrow c_-\) and \(v_+\rightarrow c_+\).
	At the AFM state,
	if the part of the direct electron-hole interaction that couples these
	two transitions is neglected, the corresponding exciton states are exactly
	degenerate.
	Magnetic canting
	shifts the energies of these two excitons in opposite directions.
	
	To describe these two excitons analytically, we approximate their
	in-plane electron-hole wave functions as identical and retain their different
	band and layer-spin structure.
	Writing \(\mathcal B=\sin\alpha\), the
	effective Hamiltonian for the two equal-SRLP excitons is
	\begin{equation}
		H_{\rm eff}^{(+)}(\mathcal B)
		=
		\left(E_{0,X}+\xi_X\mathcal B^2\right)\mathbb I_2
		+
		\begin{pmatrix}
			-\chi_{+}\mathcal B & -\Omega_{+}\\
			-\Omega_{+} & +\chi_{+}\mathcal B
		\end{pmatrix},
		\label{eq:Heffbright}
	\end{equation}
	The two exciton branches are
	\begin{equation}
		E_{\pm,X}(\mathcal B)
		=
		E_{0,X}+\xi_X\mathcal B^2
		\pm
		\sqrt{\Omega_{+}^2+\chi_{+}^2\mathcal B^2}.
		\label{eq:brightbranch}
	\end{equation}
	where \(E_{0,X}\) is the
	common energy of the two excitons at the AFM state when the
	coupling between them is set to zero, and
	\(\xi_X\mathcal B^2\) is the quadratic energy shift common to both excitons.
	The parameters \(\chi_+\) and \(\Omega_+\) characterize, respectively, the
	canting-induced detuning and the coupling between the two equal-SRLP
	excitons. Their physical origins and explicit expressions are given below.
	
	The opposite energy shifts under canting arise from the interplay of
	the exchange field and interlayer hopping in the conduction and valence
	bands. The corresponding coefficient
	\begin{equation}
		\chi_{+}
		=
		-\frac{
			J^{\textrm{A}}t_\perp^{\textrm{A}}
		}{R_{\textrm{A}}}
		-\frac{
			J^{\textrm{B}}t_\perp^{\textrm{B}}
		}{R_{\textrm{B}}}
		\label{eq:chi_br}
	\end{equation}
	is the leading canting-induced detuning.
	The linear relative shift of the two exciton energies therefore
	vanishes when the interlayer hopping is absent.
	
	The coupling between the two equal-SRLP excitons has a different
	origin. Each of the two underlying SRLP transitions contains intralayer
	electron-hole configurations, in which the electron and hole occupy the same
	layer, and interlayer configurations, in which they occupy opposite layers.
	An interlayer electron-hole pair has a finite out-of-plane separation, and
	therefore a different Coulomb attraction from an intralayer pair. The two
	SRLP transitions combine these intralayer and interlayer configurations
	differently. Consequently, a difference between the intralayer and
	interlayer attraction energies couples the excitons formed from the two
	transitions and lifts their degeneracy at the AFM state.
	
	The strength of this coupling also depends on the layer composition of
	the conduction and valence band-edge states. At the AFM state, this
	dependence is quantified by
	\begin{equation}
		Q_0
		=
		\left|
		\langle c_+(\Gamma)|\eta_z|c_-(\Gamma)\rangle
		\langle v_+(\Gamma)|\eta_z|v_-(\Gamma)\rangle
		\right|_{\alpha=0}.
		\label{eq:main_Q_definition}
	\end{equation}
	Here, \(c_\pm(\Gamma)\) and \(v_\pm(\Gamma)\) denote the conduction-
	and valence-band-edge states with SRLP eigenvalue \(\lambda=\pm1\),
	respectively. The notation
	\(\left.\cdots\right|_{\alpha=0}\) specifies that these band-edge states are
	evaluated in the AFM configuration, and \(\eta_z\) distinguishes the two
	layers. Within the present model, \(Q_0=1\) in the absence of interlayer
	hopping and decreases as interlayer hopping mixes the layer character of the
	band-edge states.

	We denote by \(\mathcal V_{\rm intra}(0)\) and \(\mathcal V_{\rm inter}(0)\) the
	intralayer and interlayer electron-hole attraction energies, respectively,
	at the AFM state, obtained by averaging the corresponding Coulomb
	interaction over the common in-plane electron-hole wave function used above.
	Their average contributes equally to the energies of the two excitons,
	whereas their difference couples the two equal-SRLP excitons. The resulting
	coupling energy is
	\begin{equation}
		\Omega_{+}
		=
		Q_0
		\frac{
			\left|
			\mathcal V_{\rm intra}(0)-\mathcal V_{\rm inter}(0)
			\right|
		}{2}.
		\label{eq:Omega_br}
	\end{equation}

	The detailed expressions for \(Q_0\), \(\mathcal V_{\rm intra}\), and
	\(\mathcal V_{\rm inter}\) are derived in
	Appendix~\ref{app:exciton_projection}.
	The resulting exciton energies are compared with the full BSE
	calculation in Fig.~\ref{fig:two_channel_model_weights}(a).
	
	At the AFM state \(\mathcal B=0\), the two equal-SRLP excitons are degenerate
	when their mutual coupling is omitted. The coupling \(\Omega_+\) lifts this
	degeneracy and separates their energies by \(2\Omega_+\).
	
	The square-root term in Eq.~\eqref{eq:brightbranch} can be expanded
	in powers of $\mathcal B$ with expansion parameter
	$|\chi_+\mathcal B/\Omega_+|$. When $\Omega_+>|\chi_+|$, this parameter remains
	smaller than unity throughout the physical canting range
	$0\leq \mathcal B\leq1$. Retaining terms through second order gives
	\begin{equation}
		E_{-,X}(\mathcal B)
		=
		E_{0,X}-\Omega_{+}
		+
		\left(
		\xi_X-\frac{\chi_{+}^2}{2\Omega_{+}}
		\right)\mathcal B^2
		+
		O(\mathcal B^4).
		\label{eq:bright_quadratic}
	\end{equation}
	
	The quadratic coefficient therefore contains the common contribution
	\(\xi_X\) and the additional term
	\(-\chi_+^2/(2\Omega_+)\) generated by the coupling between the two
	equal-SRLP excitons.
	
	The opposite-SRLP sector
	is described by an analogous two-channel Hamiltonian with its own
	detuning and coupling. This Hamiltonian
	is derived in Appendix~\ref{app:dark_two_channel} and is compared with the
	full BSE results in
	Fig.~\ref{fig:two_channel_model_weights}(a).
	Its exciton energies, binding energies, and layer probabilities are included
	in the numerical results.
	
	\begin{figure*}[t]
		\centering
		\includegraphics[width=0.92\textwidth]{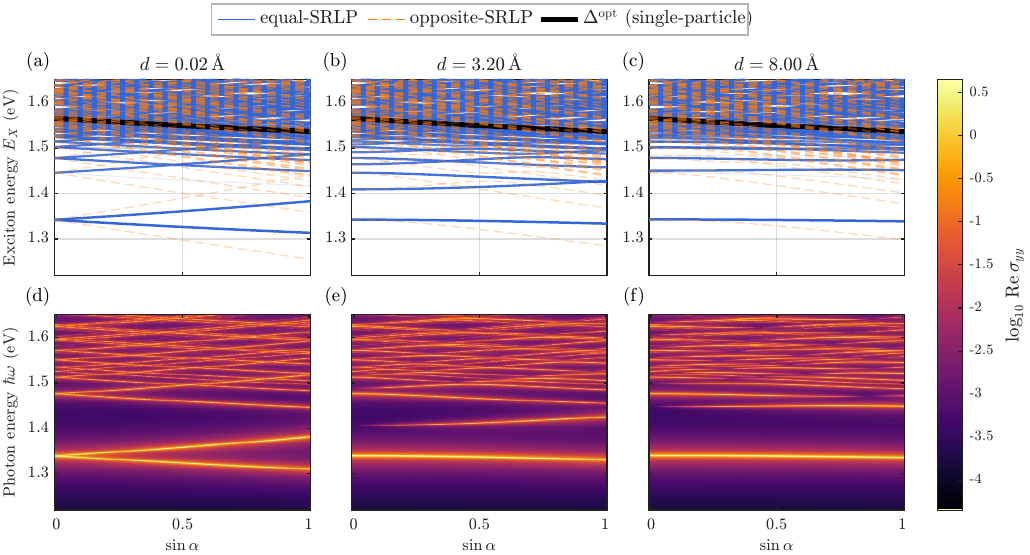}
		\caption{Exciton spectrum and normalized \(b\)-polarized optical response along
			the symmetric \(c\)-axis canting trajectory for
			three effective charge-center separations.
			The horizontal coordinate is
			\(\mathcal B=\sin\alpha=B_{\rm ext}/B_{\rm sat}\). Panels (a)-(c) show the exciton spectra, while panels
			(d)-(f) show the corresponding normalized optical response. The thick solid curves in panels (a)-(c) show the corresponding
			single-particle optical gaps \(\Delta_{\lambda}^{\rm opt}\).
			(a,d) Nearly equal intralayer and interlayer attractions,
			\(d=0.02~\text{\AA}\).
			(b,e) Intermediate separation, \(d=3.20~\text{\AA}\).
			(c,f) CrSBr calculation, \(d=8.00~\text{\AA}\).
			Increasing \(d\) attenuates only the interlayer attraction in this simulation and
			enlarges the avoided-crossing splitting.
			The quadratic dependence of the lower equal-SRLP exciton near the AFM
			state is resolved more directly in
			Fig.~\ref{fig:two_channel_model_weights}(a).
			The color scale in panels (d)-(f) is
			\(\log_{10}[\operatorname{Re}\widetilde{\sigma}_{yy}]\) in normalized model
			units. A Lorentzian broadening of \(\gamma=0.3~\mathrm{meV}\) is used as a
			numerical resolution parameter.} 
		\label{fig:exciton_energy_opt_con}
	\end{figure*}
	\subsection{Layer composition}
	\label{sec:layer_composition}
	
	The physical layer composition of an exciton is defined using the
	two-particle projectors
	\begin{align}
		\mathcal P_{\rm intra}
		&=
		\mathbb P_1^{(e)}\mathbb P_1^{(h)}
		+
		\mathbb P_2^{(e)}\mathbb P_2^{(h)},
		\nonumber\\
		\mathcal P_{\rm inter}
		&=
		\mathbb P_1^{(e)}\mathbb P_2^{(h)}
		+
		\mathbb P_2^{(e)}\mathbb P_1^{(h)},
		\label{eq:layer_pair_projectors}
	\end{align}
	where \(\mathbb P_\ell^{(e)}\) and
	\(\mathbb P_\ell^{(h)}\) act on the electron and hole coordinates,
	respectively. Their expectation values give
	\begin{align}
		w_{\rm intra}
		=
		\langle\Psi_n|
		\mathcal P_{\rm intra}
		|\Psi_n\rangle,
		~
		w_{\rm inter}
		&=
		\langle\Psi_n|
		\mathcal P_{\rm inter}
		|\Psi_n\rangle,
		\nonumber\\
		w_{\rm intra}+w_{\rm inter}
		&=1.
	\end{align}
	These quantities are physical
	intralayer and interlayer
	probabilities. They are not the squared coefficients of the
	\(\lvert c_-v_-\rangle\) and
	\(\lvert c_+v_+\rangle\) channel amplitudes, because each band-pair channel
	already contains both layer configurations.
	
	Within the
	equal-SRLP effective Hamiltonian in
	Eq.~\eqref{eq:Heffbright}, the interlayer probability of the
	lower exciton branch \(E_{-,X}\) is
	\begin{equation}
		w_{\rm inter}^{(-)}(\mathcal B)
		=
		\frac{1}{2}
		\left[
		1
		-
		Q_0
		\frac{\Omega_{+}}
		{\sqrt{\Omega_{+}^2+\chi_{+}^2\mathcal B^2}}
		\right],
		\label{eq:w_inter}
	\end{equation}
	with
	\(w_{\rm intra}^{(-)}(\mathcal B)=1-w_{\rm inter}^{(-)}(\mathcal B)\).
	The analytical layer probabilities are compared with the full BSE
	results in Fig.~\ref{fig:two_channel_model_weights}(b).
	The corresponding expression for the
	opposite-SRLP sector
	is given in
	Appendix~\ref{app:dark_two_channel}.
	
	The same coupling \(\Omega_+\) that opens the avoided crossing also
	changes the intralayer and interlayer composition of the equal-SRLP exciton
	states.
	Defining the mean energy and half-splitting,
	\begin{align}
		\overline E_X(\mathcal B)
		&=
		\frac{E_{+,X}(\mathcal B)+E_{-,X}(\mathcal B)}{2}
		=
		E_{0,X}+\xi_X\mathcal B^2,
		\nonumber\\
		S_X(\mathcal B)
		&=
		\frac{E_{+,X}(\mathcal B)-E_{-,X}(\mathcal B)}{2}
		=
		\sqrt{\Omega_{+}^2+\chi_{+}^2\mathcal B^2},
		\label{eq:experimental_two_channel_relations}
	\end{align}
	gives \(S_X(0)=\Omega_+\). Equation~\eqref{eq:w_inter} can therefore be
	written as
	\begin{equation}
		w_{\rm inter}^{(-)}(\mathcal B)
		=
		\frac{1}{2}
		\left[
		1-Q_0\frac{S_X(0)}{S_X(\mathcal B)}
		\right].
		\label{eq:w_inter_experimental}
	\end{equation}
	
	Hence, if both equal-SRLP exciton branches
	are spectroscopically resolved, their measured separation \(2S_X(\mathcal B)\),
	together with the single-particle factor \(Q_0\) obtained from the model,
	provides a model-based estimate of the intralayer and interlayer
	probabilities. This relation connects the field-dependent exciton spectrum
	to the evolution of its layer composition.

	\subsection{Optical response}
	\label{sec:optical_response}
	
	Magneto-optical spectroscopy probes how the energies and spectral weights
	of exciton resonances evolve with the magnetic configuration. In CrSBr,
	these resonances have been studied using polarization-resolved reflectance,
	differential-reflectance, absorption, and photoluminescence measurements
	\cite{Wilson2021,Heissenbuettel2025,Smiertka2026}. The present calculation
	describes the linear absorption response and therefore connects most
	directly to absorption and reflectance measurements. Photoluminescence
	intensities are not calculated because they additionally depend on exciton
	populations and relaxation processes.
	
	Using the interband matrix element
	\(M_{cv}^{\nu}(\mathbf k)\) defined in
	Eq.~\eqref{eq:M_velocity_def}, we define the single-particle dipole matrix
	element as
	\begin{equation}
		r_{cv}^{\nu}(\mathbf k)
		=
		-\ii\frac{M_{cv}^{\nu}(\mathbf k)}
		{E_c(\mathbf k)-E_v(\mathbf k)}
		=
		-\ii\hbar
		\frac{V_{cv}^{\nu}(\mathbf k)}
		{E_c(\mathbf k)-E_v(\mathbf k)}.
		\label{eq:dipole_matrix_element}
	\end{equation}
	
	For polarization along direction \(\nu\), the transition amplitude between
	the ground state and exciton \(n\) is
	\begin{align}
		X_{n0}^{\nu}
		&=
		\sum_{cv}
		\int\frac{d^2\mathbf k}{(2\pi)^2}
		\Psi_{n,cv}^*(\mathbf k)
		\,r_{cv}^{\nu}(\mathbf k),
		\nonumber\\
		X_{0n}^{\nu}
		&=
		\left(X_{n0}^{\nu}\right)^*,
		\label{eq:exciton_optical_amplitude}
	\end{align}
	where the sum runs over the retained conduction-valence band pairs
	\cite{Aversa1995,Rohlfing2000,Onida2002}.
	The sum is coherent, so contributions from different momenta and band-pair
	channels can interfere.
	Although the constituent equal-SRLP single-particle transitions are
	optically allowed, an exciton formed from them can have weak or vanishing
	optical weight because their contributions to
	\(X_{n0}^{\nu}\) can interfere destructively.
	In contrast, excitons in the opposite-SRLP sector have zero electric-dipole
	amplitude within the reduced Hamiltonian.
	
	The quantity \(\lvert X_{n0}^{\nu}\rvert^2\) is the squared transition-dipole
	amplitude. In the response convention used below, the resonant spectral
	weight is proportional to
	\(E_{n,X}\lvert X_{n0}^{\nu}\rvert^2\). The
	excitonic optical response is written as
	\cite{Aversa1995,Rohlfing2000,Onida2002}
	\begin{align}
		\widetilde{\sigma}_{\nu\nu'}(\omega)
		&=
		-\ii\sum_n
		\frac{
			E_{n,X}X_{0n}^{\nu}X_{n0}^{\nu'}
		}{
			E_{n,X}-\hbar\omega-\ii\gamma
		}
		\nonumber\\
		&\quad
		+\ii\sum_n
		\frac{
			E_{n,X}X_{n0}^{\nu}X_{0n}^{\nu'}
		}{
			E_{n,X}+\hbar\omega+\ii\gamma
		},
		\label{eq:sigma}
	\end{align}
	where \(\gamma\) is a phenomenological linewidth used to broaden the
	discrete BSE eigenvalues. The poles occur at the exciton energies
	\(E_{n,X}\), while their spectral weights are determined by the coherent
	optical amplitudes. The numerical spectra below show the \(b\)-polarized
	response \(\operatorname{Re}\widetilde{\sigma}_{yy}\).
	
	\section{Excitons: numerical results}
	\label{sec:numerical_results}
	
	\begin{figure*}[t]
		\centering
		\includegraphics[width=0.96\textwidth]{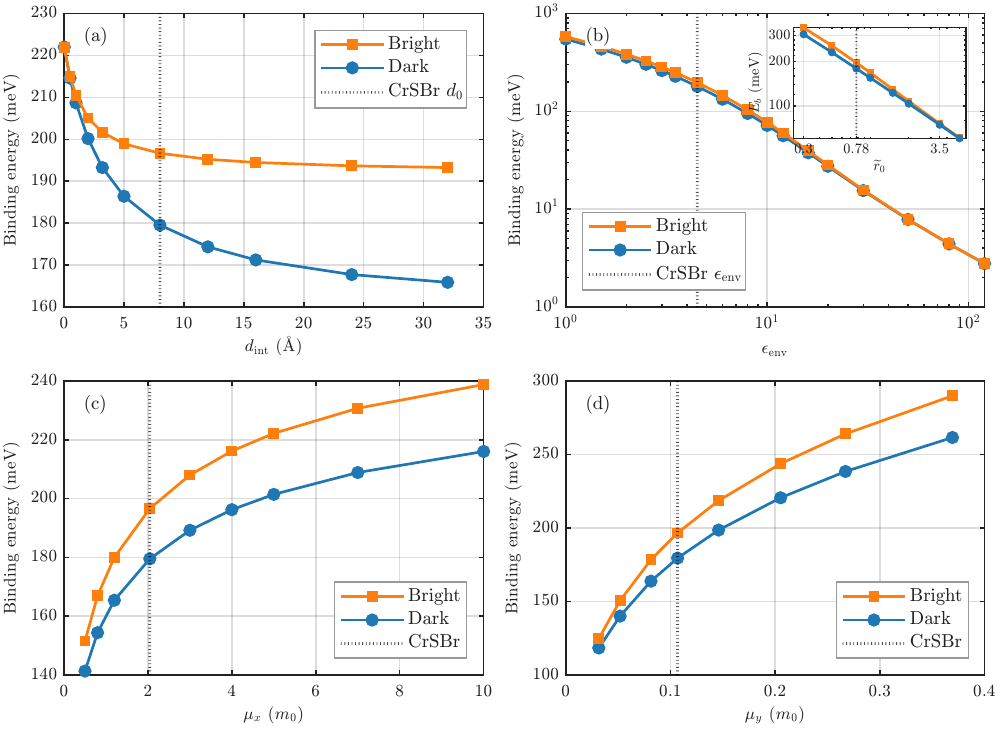}
		\caption{
			Binding energies at the field-polarized FM state of the lowest
			equal-SRLP
			($\Lambda_{eh}=+1$)
			and opposite-SRLP
			($\Lambda_{eh}=-1$)
			excitons.
			All panels are obtained by solving the momentum-dependent
			layer-resolved BSE for the four near-gap electron-hole channels in
			Eq.~\eqref{eq:four_bse_channels}, without the analytical two-channel
			exciton approximation.
			(a) Effective charge-center separation
			\(d\)
			when the modification of the intralayer screening by the other layer
			is neglected. Here
			\(W_{\rm intra}=W_{\rm RK}\) and
			\(W_{\rm inter}=e^{-\widetilde d\widetilde q}W_{\rm RK}\), so only the
			interlayer attraction changes.
			(b) Environmental dielectric constant
			\(\epsilon_{\rm env}\).
			The main panel varies \(\epsilon_{\rm env}\), which changes both the
			interaction strength and the effective Rytova-Keldysh screening radius.
			The inset instead varies the dimensionless effective screening radius
			\(\widetilde r_*\) at fixed interaction strength. The dotted lines mark the
			CrSBr values \(\epsilon_{\rm env}=4.5\) and
			\(\widetilde r_*=0.778\), respectively.
			(c) Reduced mass \(\mu_x\).
			Here \(\mu_x\) is varied while \(\mu_y\) and the interaction
			parameters are held fixed.
			(d) Reduced mass \(\mu_y\).
			Here \(\mu_y\) is varied while \(\mu_x\) and the interaction
			parameters are held fixed.
			All other parameters are fixed in each simulation. Dotted lines mark the reference
			parameter set.}
		\label{fig:binding_sensitivity}
	\end{figure*}
	
	\subsection{Exciton spectrum and layer evolution}
	\label{sec:numerical_spectrum}
	
	Figure~\ref{fig:two_channel_model_weights} compares the
	two-channel exciton model
	with the full BSE. At the
	AFM state, the
	two-channel model gives \(\Omega_+=53.1~\mathrm{meV}\), compared with
	\(51.3~\mathrm{meV}\) from the full BSE.
	The layer factor and attraction energies entering the two-channel model
	are defined in Appendix~\ref{app:exciton_projection}.
	
	The origin of the splitting is isolated by removing the off-diagonal BSE
	matrix elements between \(\lvert c_-v_-\rangle\) and
	\(\lvert c_+v_+\rangle\). The
	two equal-SRLP excitons
	are then degenerate at the
	AFM state within numerical precision.
	Restoring these BSE matrix elements
	opens the full \(102.6~\mathrm{meV}\) gap.
	The AFM splitting is therefore generated by
	level repulsion arising from the difference between the intralayer and
	interlayer electron-hole attractions
	rather than by a pre-existing detuning of the
	two excitons.

	The two-channel exciton model also reproduces the field dependence and physical layer
	composition of the lower branch. The rms differences from the full BSE are
	less than \(0.1\%\) for the exciton energy and \(1\%\)
	points for the interlayer probability.
	
	For the reference parameter set, the interlayer probability of the lowest
	equal-SRLP
	exciton increases from \(0.8\%\) in the AFM state to \(7.1\%\) near
	saturation. The state remains predominantly intralayer, but its interlayer
	component grows substantially relative to its zero-field value.
	
	Applying the same projection to the opposite-SRLP sector gives a
	two-channel Hamiltonian
	with different detuning and coupling parameters. It also reproduces the
	full-BSE
	opposite-SRLP branches
	and their layer probabilities.
	Thus, the level-repulsion mechanism occurs in both SRLP sectors, although
	only excitons in the equal-SRLP sector can carry electric-dipole weight
	in the direct-kernel minimal model.

	Figure~\ref{fig:exciton_energy_opt_con} shows the calculated
	exciton spectrum and \(b\)-polarized optical response along the
	symmetric canting trajectory for
	three effective charge-center separations
	of the interaction model.
	Increasing the effective
	charge-center separation \(d\) suppresses the interlayer attraction through
	\(e^{-dq}\) while leaving the intralayer attraction unchanged. The resulting
	difference between the intralayer and interlayer attractions strengthens
	the coupling between the two equal-SRLP excitons
	and enlarges their avoided-crossing gap. The nearly coincident limit \(d=0.02~\text{\AA}\) represents almost equal
	intralayer and interlayer attractions,
	while \(d=3.20~\text{\AA}\) gives a moderate reduction of the
	interlayer attraction,
	whereas \(d=8.00~\text{\AA}\) is used for the CrSBr
	calculation.
	In this comparison, \(d\) changes only the interlayer part of the
	electron-hole interaction and does not represent a physical change of the
	crystallographic layer spacing.
	
	The bare orbital separation \(\Delta_\Gamma^0\) is chosen such that
	the lowest equal-SRLP exciton at the
	AFM state
	lies near \(1.344~\mathrm{eV}\), within the energy range of the low-lying optical excitations reported
	for CrSBr
	\cite{Klein2023,TabatabaVakili2024}.
	The same separation enters the
	relation between the band-edge masses, the interorbital matrix element,
	and the lattice hoppings
	in Appendix~\ref{app:parameterization}. Changing \(\Delta_\Gamma^0\)
	while keeping the physical masses and \(p_{cv}\)
	fixed also changes the effective hoppings and can modify
	the exciton binding energy and wave function.
	A quantitative assignment to a particular experimental exciton
	therefore requires the band and electron-hole interaction parameters to be
	determined consistently for the same sample.

	At the AFM state, the splitting between the two lowest
	equal-SRLP
	excitons increases from below \(1~\mathrm{meV}\) in the nearly coincident
	limit to about \(103~\mathrm{meV}\) at the reference separation
	[Figs.~\ref{fig:exciton_energy_opt_con}(a,c) and
	\ref{fig:two_channel_model_weights}(a)].
	For the same parameter set, the lower branch redshifts by approximately
	\(4.3~\mathrm{meV}\) between the AFM and field-polarized FM states.
	The nearly quadratic field dependence, shown more directly by the lower equal-SRLP branch in
	Fig.~\ref{fig:two_channel_model_weights}(a),
	and the simultaneous increase in interlayer character agree
	qualitatively with magneto-optical and GW-BSE results for CrSBr
	\cite{TabatabaVakili2024,Heissenbuettel2025}.
	
	The optical maps
	in Fig.~\ref{fig:exciton_energy_opt_con}
	show both the resonance shifts and the transfer of spectral
	weight produced by coherent mixing of the two
	equal-SRLP excitons.
	They can be compared with relative peak positions and intensities in
	\(b\)-polarized absorption or magneto-reflectance.

	\subsection{Dependence of the binding energy on screening, separation, and mass anisotropy}
	\label{sec:binding_sensitivity}
	
	We now examine how the interaction and anisotropic dispersion control the
	binding energy. To separate these effects from the magnetic evolution, we
	evaluate the lowest excitons at the field-polarized FM state and vary one
	parameter at a time. The binding energy is defined relative to the lowest
	independent electron-hole continuum in the same \(\Lambda_{eh}\) sector, as
	in Eq.~\eqref{eq:binding_def}. Numerical details are given in
	Appendix~\ref{app:numerics}.

	Within the
	calculation that neglects the modification of intralayer screening by
	the other layer, increasing \(d\) weakens only the attraction between electron
	and hole configurations in different layers
	[Fig.~\ref{fig:binding_sensitivity}(a)].
	The binding therefore decreases but remains finite because the intralayer
	interaction is unchanged and the exciton becomes predominantly intralayer.
	A
	calculation that includes the modification of intralayer screening by
	the other layer
	also changes the screening of the intralayer interaction and
	gives the opposite trend. The binding increases with \(d\) over most
	of the range considered because increasing the layer separation reduces the
	screening of the intralayer interaction by the other layer.
	This comparison is discussed in Appendix~\ref{app:bilayer_kernel} and
	Fig.~\ref{fig:kernel_compare}.

	Increasing \(\epsilon_{\rm env}\) screens both intralayer and interlayer
	attractions and reduces the binding of the exciton. The finite value
	at the largest permittivity in Fig.~\ref{fig:binding_sensitivity}(b) reflects
	the finite simulation range; the binding approaches zero for asymptotically strong
	environmental screening.
	
	The directional reduced masses satisfy
	\begin{equation}
		\frac{1}{\mu_\nu}=\frac{1}{m_\nu^e}+\frac{1}{m_\nu^h},
		\qquad \nu=x,y.
	\end{equation}
	Increasing either reduced mass lowers the kinetic-energy cost of localizing
	the relative motion and increases the binding
	[Figs.~\ref{fig:binding_sensitivity}(c) and
	\ref{fig:binding_sensitivity}(d)].
	At the reference point, the FM
	equal-SRLP exciton
	logarithmic slopes are
	\begin{equation}
		\frac{d\ln E_b}{d\ln\mu_y}\approx0.35,
		\qquad
		\frac{d\ln E_b}{d\ln\mu_x}\approx0.16.
	\end{equation}
	The stronger sensitivity to \(\mu_y\) follows from the small light-axis mass,
	which makes localization along \(b\) the larger kinetic-energy cost. Since
	the interaction is isotropic in this calculation, the directional response
	comes from the band dispersion rather than anisotropic screening. This trend
	is consistent with effective-mass treatments of CrSBr excitons
	\cite{Semina2024,Liebich2025}.
	
	\section{Discussion and conclusions}
	\label{sec:discussion}
	
	Magnetic order controls whether interlayer hopping is off-resonant or
	resonant. In the AFM state, the layer-staggered exchange field detunes
	same-spin states in opposite layers, and hopping shifts their energies only
	in second order. In the field-polarized FM state, the exchange field is layer
	uniform, so the corresponding states are resonant and acquire a first-order
	bonding-antibonding splitting. Canting continuously connects these limits.
	The wave functions change together with the energies. In particular,
	the conduction and valence states acquire a different layer composition as
	the moments cant, and these field-dependent electronic states form the basis
	of the exciton problem.
	
	In the
	equal-SRLP sector, the
	excitons formed from the \(c_-v_-\) and \(c_+v_+\) transitions are
	degenerate at the AFM state when the BSE matrix elements coupling these
	two transitions are omitted. Canting shifts their energies in opposite
	directions.
	Because each
	transition contains both intralayer and interlayer electron-hole
	configurations, the
	difference between the intralayer and interlayer electron-hole
	attraction energies couples the two excitons.
	The resulting avoided crossing produces the quadratic curvature of the lower
	exciton and changes its physical layer composition.

	Previous experiments and GW-BSE calculations established magnetic
	control of interlayer hybridization and exciton energies in CrSBr
	\cite{Wilson2021,Heissenbuettel2025}.
	The present model identifies the corresponding analytical structure.
	SRLP remains conserved along the symmetric canting trajectory, the
	\(\Gamma\)-point eigenstates and direct gaps can be obtained at arbitrary
	canting, and for equivalent layers the direct BSE separates into two
	\(\Lambda_{eh}\) sectors. 
	This construction complements observations of magnetically confined excitons
	\cite{Shao2025}, magnetic-order control of Coulomb correlations
	\cite{Liebich2025}, distinct Frenkel- and Wannier-like excitonic responses
	\cite{Smiertka2026}, and bulk and surface excitons
	\cite{Choi2026}.
	
	If both
	equal-SRLP exciton branches
	are resolved, their AFM separation
	is \(2\Omega_+\) within the two-channel exciton model, while their
	field-dependent separation tests the predicted detuning. The transfer of
	spectral weight provides a second signature of coherent
	mixing between the two excitons.
	Inferring layer probabilities additionally requires the calculated overlap
	factor \(Q_0\)
	and is therefore not a model-independent extraction.
	Likewise, an optical line shift cannot be equated directly with a
	binding-energy change because the independent electron-hole continuum also
	moves with magnetic order.
	Resolving both branches would therefore provide information that is
	not contained in the shift of a single optical line.
	
	The
	absolute binding energy is more sensitive to the screened
	electron-hole interaction than the avoided-crossing splitting.
	Replacing the
	interaction that keeps the intralayer screening independent of layer
	separation by one that includes the screening of one layer by the other
	changes the FM equal-SRLP binding energy by about \(10\%\) for the
	reference parameter set, whereas the AFM exciton splitting changes by about
	\(3.5\%\).
	The dependence on layer separation is more sensitive. If the
	intralayer interaction is kept fixed, increasing the separation weakens the
	interlayer attraction and reduces the binding. When the screening of the
	intralayer interaction by the other layer is also included, increasing the
	separation reduces this additional screening and the binding increases over
	most of the range considered.
	Both interactions satisfy
	\(W_{11}=W_{22}\) and \(W_{12}=W_{21}\), so the direct BSE retains its two
	\(\Lambda_{eh}\) sectors. The avoided crossing, the quadratic variation near
	the AFM state, and the increase of interlayer character toward the
	field-polarized state also remain, although their numerical magnitudes
	depend on the screened interaction.
	
	The exact single-particle SRLP result relies on the symmetric canting
	trajectory and the minimal hopping structure. Spin-dependent tunneling,
	additional spin-orbit terms, inequivalent layers, or lower-symmetry magnetic
	configurations can mix its sectors.
	At the exciton level, separation into the two
	\(\Lambda_{eh}\) sectors additionally requires equivalent layers and a
	layer-symmetric direct interaction. Layer-asymmetric screening can therefore
	mix sectors that are independent here, while electron-hole exchange and
	additional electronic bands can modify the exciton fine structure and
	optical weights.

	The generalized s-d model is not limited to the exciton
	problem. For a specified magnetic configuration, the exchange and hopping
	terms determine the electronic energies, interlayer hybridization, and spin
	and layer composition of the Bloch states. These states determine the
	optical matrix elements considered here and also enter electronic transport.
	Changes in magnetic order can therefore modify spin-dependent interlayer
	transport and magnetoresistance through the same electronic Hamiltonian,
	although an explicit transport calculation additionally requires the
	relevant contacts and scattering processes \cite{Telford2020}. If the
	localized-spin operators are retained rather than replaced by their ordered
	expectation values, fluctuations of the magnetic moments couple directly to
	the electron spin density through \(\hat H_{\rm ex}\). Their expansion in
	magnon modes gives electron-magnon coupling, and after the electron-hole
	interaction is included the same exchange term can also couple excitons to
	magnons \cite{Iakovlev2026}.
	
	For another layered magnetic semiconductor, the orbital states,
	exchange couplings, interlayer hopping, and spin-orbit terms in
	Eq.~\eqref{eq:H_general} must be chosen for that material. Chromium
	trihalides such as CrBr$_3$ and CrI$_3$ provide examples in which these
	ingredients differ from those used for CrSBr
	\cite{Gibertini2019,Burch2018,Wang2022}. Additional layers can be included by
	adding the corresponding layer blocks, while several magnetic ions in the
	unit cell require additional magnetic-sublattice indices. Recent excitonic spectroscopy of four- and five-layer CrSBr has
	resolved layer-dependent magnetic configurations during magnetic switching
	\cite{krelle2026}, providing a direct setting in which the
	multilayer form of Eq.~\eqref{eq:general_multilayer_bloch_matrix} can be
	applied. Commensurate
	collinear, canted, and noncollinear magnetic orders can be represented by
	the corresponding ordered moment directions and magnetic unit cell. The
	resulting optical and excitonic selection rules must then be obtained from
	the symmetries of that Hamiltonian. In particular, the SRLP symmetry found
	here relies on the layer-spin structure of the reduced CrSBr Hamiltonian
	along the symmetric \(c\)-axis canting trajectory. In conclusion, we have developed an orbital-, layer-, and spin-resolved generalized
	s-d model to address the magnetically tunable electronic and excitonic phenomena in layered magnetic semiconductors  and applied it to the magnetic-order-dependent electronic
	structure, optical selection rules, and Wannier-Mott excitons of a CrSBr
	bilayer.

	\begin{acknowledgments}
		
		Financial support by the DFG (German Research Foundation) via Spin+X TRR 173-268565370 (project A13) is gratefully acknowledged. S.V. additionally acknowledges support from the Brain Pool Program funded by the Ministry of Science and ICT through the National Research Foundation of Korea (RS-2025-25446099).
	\end{acknowledgments}
	
	\section*{Data Availability}
	
	The numerical data underlying the figures in this article are openly
	available in Zenodo~\cite{vermazenodo2026}. The code used to generate
	and analyze the data is available from the authors upon reasonable request.

	\appendix
	
	\section{Hard-axis canting from a two-sublattice free energy}
	\label{app:canting}
	
	Let \(\mathbf m_1\) and \(\mathbf m_2\) be dimensionless unit vectors
	along the ordered layer moments, and define the physical magnetizations by
	\(\mathbf M_\ell=M_s\mathbf m_\ell\). We use a common free-energy-density
	normalization for every term below and set
	\(B_{\rm ext}=\mu_0H_{\rm ext}\). In the SI volumetric convention,
	\(M_s\) is the saturation magnetization in A/m and \(F\),
	\(\mathcal J_{\rm AF}\), and \(K_{h,e}\) are in J/m\(^3\); an areal
	convention is equivalent after integrating all terms through the same layer
	thickness. We write
	\cite{Yang2021,Scheie2022,Heissenbuettel2025}
	\begin{align}
		F
		&=
		\mathcal J_{\rm AF}\,\mathbf m_1\cdot\mathbf m_2
		+K_h(m_{1z}^{2}+m_{2z}^{2})
		\nonumber\\
		&\quad-K_e(m_{1y}^{2}+m_{2y}^{2})
		-M_sB_{\rm ext}(m_{1z}+m_{2z}).
		\label{eq:app_free_energy}
	\end{align}
	Here \(\mathcal J_{\rm AF}>0\), \(K_h>0\), and \(K_e>0\) have the
	same energy-density units as \(F\), while \(M_sB_{\rm ext}\) has those
	units in the chosen magnetic normalization. The notation
	\(\mathcal J_{\rm AF}\) distinguishes the magnetic interlayer exchange from
	the orbital-dependent electronic exchange couplings \(J^\tau\).
	
	The symmetric canting configuration is
	\begin{align}
		\mathbf m_1
		&=\cos\alpha\,\hat y+\sin\alpha\,\hat z,
		\nonumber\\
		\mathbf m_2
		&=-\cos\alpha\,\hat y+\sin\alpha\,\hat z.
		\label{eq:app_canting_ansatz}
	\end{align}
	Substitution gives
	\begin{align}
		F(\alpha)
		&=-\mathcal J_{\rm AF}\cos(2\alpha)
		+2(K_h+K_e)\sin^2\alpha
		\nonumber\\
		&\quad-2K_e-2M_sB_{\rm ext}\sin\alpha.
		\label{eq:app_canting_energy}
	\end{align}
	The minimization of free energy gives
	\begin{align}
		\frac{\partial F}{\partial\alpha}
		&=2\cos\alpha
		\left[
		2(\mathcal J_{\rm AF}+K_h+K_e)\sin\alpha
		-M_sB_{\rm ext}
		\right]
		=0.
		\label{eq:app_canting_stationarity}
	\end{align}

	Equation~\eqref{eq:app_canting_stationarity} has two solutions.
	One is \(\alpha=\pi/2\), for which the two layer moments are parallel to the
	applied field. For \(\cos\alpha\neq0\), the other solution is the canted
	configuration
	\begin{equation}
		\sin\alpha
		=\frac{M_sB_{\rm ext}}
		{2(\mathcal J_{\rm AF}+K_h+K_e)}.
		\label{eq:app_sin_alpha}
	\end{equation}
	
	For the canted solution,
	\[
	\frac{\partial^2F}{\partial\alpha^2}
	=
	4(\mathcal J_{\rm AF}+K_h+K_e)\cos^2\alpha>0,
	\]
	so this solution is a minimum as long as
	\(\sin\alpha<1\). For the parallel solution
	\(\alpha=\pi/2\),
	\[
	\left.
	\frac{\partial^2F}{\partial\alpha^2}
	\right|_{\alpha=\pi/2}
	=
	2\left[
	M_sB_{\rm ext}
	-2(\mathcal J_{\rm AF}+K_h+K_e)
	\right].
	\]
	The parallel state therefore becomes the stable solution when the applied
	field exceeds the value at which the canted solution reaches
	\(\sin\alpha=1\).
	
	This field is the saturation field,
	\begin{equation}
		B_{\rm sat}
		=\frac{2(\mathcal J_{\rm AF}+K_h+K_e)}{M_s}.
		\label{eq:app_saturation_field}
	\end{equation}
	
	Thus, for \(0\leq B_{\rm ext}<B_{\rm sat}\), the stable configuration
	is the canted state with
	\(\sin\alpha=B_{\rm ext}/B_{\rm sat}\). At
	\(B_{\rm ext}=B_{\rm sat}\), the moments become parallel to the applied
	field, and they remain parallel for \(B_{\rm ext}>B_{\rm sat}\). This is the
	canting relation used in Eq.~\eqref{eq:hard_axis_texture}.
	
	The parameters \(\mathcal J_{\rm AF}\), \(K_h\), and \(K_e\) can be
	estimated from measurements of the interlayer magnetic exchange, magnetic
	anisotropy, and saturation field in CrSBr
	\cite{Telford2020,Lee2021,Scheie2022,Yang2021,Ziebel2024}.
	
	\section{Low-energy units and parameter extraction}
	\label{app:parameterization}
	
	We obtain the low-energy lattice parameters by matching
	Eq.~\eqref{eq:crsbr_dispersion} to an anisotropic two-band
	\(\mathbf k\cdot\mathbf p\) Hamiltonian near the \(\Gamma\) point
	\cite{Klein2023,Semina2024,Smolenski2025}. In the orbital basis
	\(\{\rm A,B\}\), the canonical momentum convention is
	\begingroup
	\small
	\setlength{\arraycolsep}{2pt}
	\begin{equation}
		H_{\mathbf k\cdot\mathbf p}(\mathbf k)
		=
		\begin{pmatrix}
			\bareps^{\rm A}+\dfrac{\hbar^2k_x^2}{2m_{{\rm A},x}}
			+\dfrac{\hbar^2k_y^2}{2m_{{\rm A},y}}
			&\dfrac{\hbar k_y p_{cv}}{m_0}
			\\[8pt]
			\dfrac{\hbar k_y p_{cv}^{*}}{m_0}
			&\bareps^{\rm B}+\dfrac{\hbar^2k_x^2}{2m_{{\rm B},x}}
			+\dfrac{\hbar^2k_y^2}{2m_{{\rm B},y}}
		\end{pmatrix},
		\label{eq:app_kp_hamiltonian}
	\end{equation}
	\endgroup
	where \(p_{cv}=\langle \rm A|\hat p_y|\rm B\rangle\) is the canonical interband
	momentum matrix element and \(m_0\) is the free-electron mass.
	The quantities \(m_{{\rm A},\nu}\) and \(m_{{\rm B},\nu}\) are the signed
	curvature masses of the uncoupled A and B orbital sectors. They
	are not identical to the physical electron and hole masses because the
	interorbital \(k_yp_{cv}\) term also contributes to the band curvature
	along \(y\).
	
	We use the two-dimensional excitonic units of Ref.~\cite{Semina2024} at
	the fixed reference dielectric \(\epsilon_{\rm ref}=4.5\),
	\begin{align}
		a_B^{2D}
		&=\frac{4\pi\epsilon_0\epsilon_{\rm ref}\hbar^2}
		{2\mu_{\rm ref}e^2},
		\nonumber\\
		Ry^{2D}
		&=\frac{2\mu_{\rm ref}e^4}
		{(4\pi\epsilon_0\epsilon_{\rm ref})^2\hbar^2}
		=\frac{\hbar^2}{2\mu_{\rm ref}(a_B^{2D})^2},
		\qquad \mu_{\rm ref}\equiv\mu_y.
		\label{eq:app_effective_units}
	\end{align}

	Here, \(a_B^{2D}\) and \(Ry^{2D}\) are the length and energy units
	used to nondimensionalize the exciton calculation. The reference mass
	\(\mu_{\rm ref}=\mu_y\) is the electron-hole reduced mass along
	\(y\parallel b\), with \(\mu_y\) defined explicitly in
	Eq.~\eqref{eq:app_reduced_masses}. These units are fixed by the reference
	parameter set and are kept unchanged when a mass or dielectric parameter
	is varied below.
	
	We use \(\nu=x,y\) below for a spatial direction and reserve
	\(\mu_x\) and \(\mu_y\) for the electron-hole reduced masses.
	The dimensionless variables are
	\begin{align}
		\widetilde{\mathbf k}&=a_B^{2D}\mathbf k,
		&
		\widetilde H&=\frac{H}{Ry^{2D}},
		\nonumber\\
		\widetilde m_{\tau,\nu}&=\frac{m_{\tau,\nu}}{\mu_{\rm ref}},
		&
		\widetilde{\bareps}^{\,\tau}&=\frac{\bareps^\tau}{Ry^{2D}}.
		\label{eq:dimensionless_variables}
	\end{align}
	The resulting dimensionless Hamiltonian is
	\begingroup
	\small
	\setlength{\arraycolsep}{2pt}
	\begin{equation}
		\widetilde H_{\mathbf k\cdot\mathbf p}
		(\widetilde{\mathbf k})
		=
		\begin{pmatrix}
			\widetilde{\bareps}^{\,\rm A}
			+\dfrac{\widetilde k_x^2}{\widetilde m_{{\rm A},x}}
			+\dfrac{\widetilde k_y^2}{\widetilde m_{{\rm A},y}}
			&\widetilde k_y\widetilde p_{cv}
			\\[7pt]
			\widetilde k_y\widetilde p_{cv}^{*}
			&\widetilde{\bareps}^{\,\rm B}
			+\dfrac{\widetilde k_x^2}{\widetilde m_{{\rm B},x}}
			+\dfrac{\widetilde k_y^2}{\widetilde m_{{\rm B},y}}
		\end{pmatrix},
		\label{eq:app_dimensionless_kp}
	\end{equation}
	\endgroup
	where
	\begin{equation}
		\widetilde p_{cv}
		=\frac{\hbar p_{cv}}
		{m_0a_B^{2D}Ry^{2D}}
		=2\frac{\mu_{\rm ref}}{m_0}
		\frac{a_B^{2D}p_{cv}}{\hbar}.
		\label{eq:app_dimensionless_pcv}
	\end{equation}
	Equation~\eqref{eq:app_dimensionless_pcv} follows by matching
	\(\hbar k_yp_{cv}/m_0\) to
	\(Ry^{2D}\widetilde k_y\widetilde p_{cv}\).
	
	Define the two diagonal matrix elements of
	Eq.~\eqref{eq:app_dimensionless_kp} by
	\begin{align}
		h_{\rm A}(\widetilde{\mathbf k})
		&=\widetilde{\bareps}^{\,\rm A}
		+\frac{\widetilde k_x^2}{\widetilde m_{{\rm A},x}}
		+\frac{\widetilde k_y^2}{\widetilde m_{{\rm A},y}},
		\nonumber\\
		h_{\rm B}(\widetilde{\mathbf k})
		&=\widetilde{\bareps}^{\,\rm B}
		+\frac{\widetilde k_x^2}{\widetilde m_{{\rm B},x}}
		+\frac{\widetilde k_y^2}{\widetilde m_{{\rm B},y}},
	\end{align}
	and set
	\begin{equation}
		\bar\varepsilon_{\vk}
		=\frac{h_{\rm A}+h_{\rm B}}{2},
		\qquad
		\mathcal M_{\vk}
		=\frac{h_{\rm A}-h_{\rm B}}{2}.
		\label{eq:app_kp_scalar_mass_functions}
	\end{equation}
	The relative phase of the A and B orbitals may be chosen so that
	\(\widetilde p_{cv}\) is real. In this gauge,
	\begin{equation}
		\widetilde H_{\mathbf k\cdot\mathbf p}
		=
		\bar\varepsilon_{\vk}\tau_0
		+
		\mathcal M_{\vk}\tau_z
		+
		\widetilde k_y\widetilde p_{cv}\tau_x.
		\label{eq:app_kp_pauli}
	\end{equation}
	
	The phases of the two orbital basis states are arbitrary. If
	\(p_{cv}=|p_{cv}|e^{i\phi}\), replacing
	\(\lvert \rm B\rangle\) by \(e^{-i\phi}\lvert \rm B\rangle\), with
	\(\lvert \rm A\rangle\) unchanged, makes \(p_{cv}\) real. This change of basis
	does not alter the energy spectrum or the magnitudes of optical matrix
	elements. Every A-B matrix element acquires the same relative phase
	under this transformation, so the same orbital basis convention must be
	used for \(p_{cv}\) and \(g_y\).
	
	For a complex momentum matrix element, the final term is replaced by
	\[
	\widetilde k_y
	\left[
	\operatorname{Re}\widetilde p_{cv}\,\tau_x
	-
	\operatorname{Im}\widetilde p_{cv}\,\tau_y
	\right].
	\]
	
	The eigenvalues of Eq.~\eqref{eq:app_dimensionless_kp} are
	\begin{equation}
		\widetilde E_{\beta}(\widetilde{\mathbf k})
		=
		\bar\varepsilon_{\vk}
		+
		\beta
		\sqrt{
			\mathcal M_{\vk}^{2}
			+
			\widetilde k_y^2
			\left|
			\widetilde p_{cv}
			\right|^2
		},
		\qquad
		\beta=\pm1.
		\label{eq:app_kp_eigenvalues}
	\end{equation}
	Here, \(\beta=\pm1\) labels the upper and lower eigenvalue
	branches and is distinct from the spatial-direction index \(\nu=x,y\).
	
	Expanding about \(\Gamma\)-point gives
	\begin{align}
		\frac{1}{\widetilde m_{c,x}}
		&=
		\frac{1}{\widetilde m_{{\rm A},x}},
		&
		\frac{1}{\widetilde m_{c,y}}
		&=
		\frac{1}{\widetilde m_{{\rm A},y}}
		+
		\frac{
			|\widetilde p_{cv}|^2
		}{
			\widetilde\Delta_\Gamma^0
		},
		\nonumber\\
		\frac{1}{\widetilde m_{v,x}}
		&=
		\frac{1}{\widetilde m_{{\rm B},x}},
		&
		\frac{1}{\widetilde m_{v,y}}
		&=
		\frac{1}{\widetilde m_{{\rm B},y}}
		-
		\frac{
			|\widetilde p_{cv}|^2
		}{
			\widetilde\Delta_\Gamma^0
		},
		\label{eq:app_effective_masses}
	\end{align}
	where
	\begin{equation}
		\widetilde\Delta_\Gamma^0
		=
		\widetilde{\bareps}^{\,\rm A}
		-
		\widetilde{\bareps}^{\,\rm B}
		=
		\frac{\Delta_\Gamma^0}{Ry^{2D}}.
		\label{eq:app_dimensionless_gap}
	\end{equation}
	
	Let \(m_{e,\nu}>0\) and \(m_{h,\nu}>0\) denote the physical electron and
	hole masses. Since the signed valence-band curvature satisfies
	\(m_{v,\nu}=-m_{h,\nu}\), the orbital masses entering the bare
	two-orbital model are
	\begin{align}
		\widetilde m_{{\rm A},x}
		&=
		\widetilde m_{e,x},
		&
		\widetilde m_{{\rm A},y}
		&=
		\left[
		\widetilde m_{e,y}^{-1}
		-
		\frac{
			|\widetilde p_{cv}|^2
		}{
			\widetilde\Delta_\Gamma^0
		}
		\right]^{-1},
		\nonumber\\
		\widetilde m_{{\rm B},x}
		&=
		-\widetilde m_{h,x},
		&
		\widetilde m_{{\rm B},y}
		&=
		\left[
		-\widetilde m_{h,y}^{-1}
		+
		\frac{
			|\widetilde p_{cv}|^2
		}{
			\widetilde\Delta_\Gamma^0
		}
		\right]^{-1}.
		\label{eq:app_orbital_masses}
	\end{align}
	The negative sign of \(m_{{\rm B},\nu}\) describes the downward curvature
	of the valence-like B orbital. The physical hole masses
	\(m_{h,\nu}\) remain positive.
	
	The reduced masses governing the relative electron-hole motion are
	\begin{equation}
		\mu_x^{-1}
		=
		m_{e,x}^{-1}+m_{h,x}^{-1},
		\qquad
		\mu_y^{-1}
		=
		m_{e,y}^{-1}+m_{h,y}^{-1}.
		\label{eq:app_reduced_masses}
	\end{equation}
	
	The lattice dispersion in Eq.~\eqref{eq:crsbr_dispersion} satisfies
	\begin{align}
		2t_\nu^\tau
		\left[
		1-\cos(k_\nu a_\nu)
		\right]
		&=
		t_\nu^\tau a_\nu^2k_\nu^2
		+
		O(k_\nu^4).
		\label{eq:app_lattice_expansion}
	\end{align}
	Defining
	\[
	\widetilde a_\nu=\frac{a_\nu}{a_B^{2D}},
	\qquad
	\widetilde t_\nu^\tau=\frac{t_\nu^\tau}{Ry^{2D}},
	\]
	comparison with Eq.~\eqref{eq:app_dimensionless_kp} gives
	\begin{equation}
		\widetilde t_\nu^\tau
		=
		\frac{
			1
		}{
			\widetilde m_{\tau,\nu}
			\widetilde a_\nu^2
		}.
		\label{eq:app_lattice_hopping}
	\end{equation}
	Thus, a positive curvature mass gives a positive hopping in the
	convention of Eq.~\eqref{eq:crsbr_dispersion}, whereas the negative
	curvature of the valence-like \(B\) sector gives a negative hopping.
	
	The momentum-odd interorbital coupling in the main text is
	\[
	d_y(\mathbf k)
	=
	-g_y\sin(k_ya_y)
	\simeq
	-g_ya_yk_y.
	\]
	With
	\(\widetilde g_y=g_y/Ry^{2D}\),
	matching its term linear in \(k_y\) to the
	\(\widetilde k_y\widetilde p_{cv}\tau_x\) term in
	Eq.~\eqref{eq:app_kp_pauli} gives
	
	\begin{equation}
		\widetilde g_y
		=
		-
		\frac{
			\widetilde p_{cv}
		}{
			\widetilde a_y
		}.
		\label{eq:app_lattice_interband}
	\end{equation}
	
	The minus sign in Eq.~\eqref{eq:app_lattice_interband} follows from
	matching
	\(d_y(\mathbf k)\simeq-g_ya_yk_y\)
	to the \(+\widetilde k_y\widetilde p_{cv}\tau_x\) term in
	Eq.~\eqref{eq:app_kp_pauli}, using the same A-B orbital basis in
	both descriptions. Changing the relative phase of the two orbital basis
	states changes the phases of \(g_y\) and \(p_{cv}\) consistently and leaves
	Eq.~\eqref{eq:app_lattice_interband} invariant. After choosing both
	matrix elements to be real, the remaining sign change
	\(\lvert \rm B\rangle\rightarrow-\lvert \rm B\rangle\) reverses the signs of both
	\(g_y\) and \(p_{cv}\), so only their relative sign is fixed by the
	matching.
	
	Finally, the orbital-preserving lattice dispersion in
	Eq.~\eqref{eq:crsbr_dispersion} can equivalently be written as
	\begin{align*}
		\epsilon_\tau(\mathbf k)
		&=
		\epsilon^\tau
		-
		2t_x^\tau\cos(k_xa_x)
		\nonumber\\
		&\quad
		-
		2t_y^\tau\cos(k_ya_y).
	\end{align*}
	Therefore,
	\begin{equation}
		\epsilon^\tau
		=
		\bareps^\tau
		+
		2t_x^\tau
		+
		2t_y^\tau.
		\label{eq:app_lattice_onsite}
	\end{equation}
	
	The parameters quoted in physical units in the main text and in
	Appendix~\ref{app:crsbr_parameter_registry} are obtained by converting
	these dimensionless quantities back to eV, \(\text{\AA}\), and \(m_0\).
	
	Equations~\eqref{eq:app_effective_masses}-\eqref{eq:app_lattice_onsite}
	determine the orbital curvature masses, lattice hoppings, and onsite energies
	from the band-edge masses, interband momentum matrix element, and direct gap.
	The quantities \(m_{e,\nu}\), \(m_{h,\nu}\), \(p_{cv}\), and
	\(\Delta_\Gamma^0\) should be taken from the same band description because
	\(p_{cv}\) and \(\Delta_\Gamma^0\) also enter the \(y\)-direction curvature
	in Eq.~\eqref{eq:app_effective_masses}. The screening parameters separately
	set the electron-hole attraction used in the BSE and should correspond to
	the dielectric environment being modeled. Mixing band parameters or
	screening parameters obtained for different structures or dielectric
	environments can therefore change the calculated dispersion and exciton
	binding energy
	\cite{Watson2024,Klein2023,Semina2024,Smolenski2025}.
	
	\section{Proof of spin-resolved layer-parity conservation}
	\label{app:srlp_proof}
	
	To verify SRLP conservation, write
	\[
	H(\mathbf k)\equiv H_{\rm CrSBr}(\mathbf k)
	\]
	and decompose Eq.~\eqref{eq:Hcrsbr} as
	\begin{equation}
		H(\mathbf k)
		=
		H_{\rm orb}(\mathbf k)
		+
		H_z(\alpha)
		+
		H_y(\alpha)
		+
		H_\perp,
		\label{eq:app_hamiltonian_decomposition}
	\end{equation}
	with
	\begin{align*}
		H_{\rm orb}(\mathbf k)
		&=
		\eta_0\otimes
		\left[
		\left(
		\epsilon_{\rm A}(\mathbf k)\tau_{\rm A}
		+
		\epsilon_{\rm B}(\mathbf k)\tau_{\rm B}
		+
		d_y(\mathbf k)\tau_x
		\right)
		\otimes\sigma_0
		\right],
		\\
		H_z(\alpha)
		&=
		-\eta_0\otimes
		\left[
		J_{\rm orb}
		\otimes
		\sin\alpha\,\sigma_z
		\right],
		\\
		H_y(\alpha)
		&=
		-\eta_z\otimes
		\left[
		J_{\rm orb}
		\otimes
		\cos\alpha\,\sigma_y
		\right],
		\\
		H_\perp
		&=
		-\eta_x\otimes
		\left[
		\mathcal T_\perp
		\otimes\sigma_0
		\right],
	\end{align*}
	where
	\begin{equation}
		J_{\rm orb}
		=
		J^{\rm A}\tau_{\rm A}+J^{\rm B}\tau_{\rm B},
		\qquad
		\mathcal T_\perp
		=
		t_\perp^{\rm A}\tau_{\rm A}+t_\perp^{\rm B}\tau_{\rm B}.
		\label{eq:app_orbital_matrices}
	\end{equation}
	
	The SRLP operator is
	\begin{equation}
		\zeta
		=
		\eta_x\otimes\tau_0\otimes\sigma_z.
		\label{eq:app_srlp_operator}
	\end{equation}
	The orbital term commutes with \(\zeta\) because it is proportional to
	\(\eta_0\) and \(\sigma_0\), while \(\zeta\) is proportional to
	\(\tau_0\) in orbital space.
	The uniform exchange term \(H_z\) commutes with \(\zeta\) because its
	layer and spin factors are \(\eta_0\) and \(\sigma_z\). The interlayer
	hopping term \(H_\perp\) also commutes with \(\zeta\) because its layer
	factor is \(\eta_x\) and it is proportional to \(\sigma_0\) in spin space.
	
	The only nontrivial contribution is \(H_y\), which is proportional to
	\(\eta_z\otimes J_{\rm orb}\otimes\sigma_y\). Using
	\[
	\eta_x\eta_z=-\eta_z\eta_x,
	\qquad
	\sigma_z\sigma_y=-\sigma_y\sigma_z,
	\]
	we obtain
	\begin{align}
		\zeta
		\left(
		\eta_z\otimes
		J_{\rm orb}
		\otimes\sigma_y
		\right)
		&=
		(\eta_x\eta_z)
		\otimes
		J_{\rm orb}
		\otimes
		(\sigma_z\sigma_y)
		\nonumber\\
		&=
		(\eta_z\eta_x)
		\otimes
		J_{\rm orb}
		\otimes
		(\sigma_y\sigma_z)
		\nonumber\\
		&=
		\left(
		\eta_z\otimes
		J_{\rm orb}
		\otimes\sigma_y
		\right)
		\zeta.
		\label{eq:app_nontrivial_commutator}
	\end{align}
	The anticommutations in layer and spin space cancel. Therefore,
	\begin{equation}
		[\zeta,H(\mathbf k)]=0
		\label{eq:app_srlp_commutator}
	\end{equation}
	for every \(\mathbf k\) and every canting angle \(\alpha\).
	
	This result applies to the CrSBr Hamiltonian in
	Eq.~\eqref{eq:Hcrsbr}. If an additional term \(\delta H\) is included, the
	same SRLP quantum number remains conserved if and only if
	\[
	[\zeta,\delta H]=0,
	\]
	because the Hamiltonian already satisfies
	\([\zeta,H_{\rm CrSBr}]=0\). Spin-dependent hopping, spin-flip tunneling,
	and additional spin-orbit terms do not have a universal SRLP character.
	Their layer and spin matrix structure must be examined separately to
	determine whether they commute with \(\zeta\)
	\cite{Smiertka2026,Shao2025}.
	
	\section{CrSBr material parameters and numerical conventions}
	\label{app:crsbr_parameter_registry}
	
	Table~\ref{tab:crsbr_active_registry} lists the
	parameters used in the CrSBr calculations.
	The masses, dielectric constant, screening length, and reference layer
	separation are taken from Ref.~\cite{Semina2024}.
	The exchange fields, interlayer hoppings, odd interorbital coupling,
	and bare orbital separation belong to the reduced Hamiltonian introduced in
	the main text. Their values depend on the electronic structure used to
	construct that Hamiltonian. The parameters entering the band Hamiltonian and
	the screened electron-hole interaction must therefore correspond to the same
	material geometry and dielectric environment. The length and energy units are defined in
	Appendix~\ref{app:parameterization} and are kept fixed in all parameter
	simulations.
	
	\begin{table*}[t]
		\caption{Parameters used in the CrSBr calculations.
			The Rytova-Keldysh screening radius is
			\(r_*=\rho_0/\epsilon_{\rm env}\).
			In the fixed reference units its dimensionless value is
			\(\widetilde r_*=r_*/a_B^{2D}\).
			For \(\epsilon_{\rm ref}=4.5\), the values used here give
			\(\widetilde r_*=0.778\) and \(r_*=0.87~\mathrm{nm}\).
			The parameters entering the dielectric and mass simulations in
			Fig.~\ref{fig:binding_sensitivity} are varied from the values listed here.}
		\label{tab:crsbr_active_registry}
		
		\begin{ruledtabular}
			\begin{tabular}{llll}
				Quantity & Symbol & Value & Status/use \\
				\hline
				
				Lattice constants
				& \(a_x,a_y,a_z\)
				& \(3.50,4.75,7.94~\text{\AA}\)
				& crystallographic inputs \\
				
				Electron masses
				& \(m_x^e,m_y^e\)
				& \(7.31,0.14\,m_0\)
				& literature values \\
				
				Hole masses
				& \(m_x^h,m_y^h\)
				& \(2.84,0.45\,m_0\)
				& literature values \\
				
				Reduced masses
				& \(\mu_x,\mu_y\)
				& \(2.045,0.1068\,m_0\)
				& derived \\
				
				Reference units
				& \(a_B^{2D},Ry^{2D}\)
				& \(1.12~\mathrm{nm},286~\mathrm{meV}\)
				& defined in Appendix~\ref{app:parameterization} \\
				
				Reference dielectric
				& \(\epsilon_{\rm ref}\)
				& \(4.5\)
				& dielectric environment \\
				
				Effective RK radius
				& \(r_*\)
				& \(0.87~\mathrm{nm}\)
				\(\left(\widetilde r_*=0.778\right)\)
				& screening radius \\
				
				Reference layer separation
				& \(d\)
				& \(8.0~\text{\AA}\)
				& interlayer interaction \\
				
				Reference optical alignment
				& \(E_X^{\rm AFM}\)
				& \(1.344~\mathrm{eV}\)
				& chosen exciton energy \\
				
				Aligned AFM continuum gap
				& \(E_g^{\Gamma,{\rm AFM}}\)
				& \(1.56510~\mathrm{eV}\)
				& resulting continuum edge \\
				
				Conduction exchange
				& \(J^{\rm A}\)
				& \(+0.260~\mathrm{eV}\)
				& effective parameter \\
				
				Valence exchange
				& \(J^{\rm B}\)
				& \(-0.450~\mathrm{eV}\)
				& effective parameter \\
				
				Conduction interlayer hopping
				& \(t_\perp^{\rm A}\)
				& \(+0.029~\mathrm{eV}\)
				& effective parameter \\
				
				Valence interlayer hopping
				& \(t_\perp^{\rm B}\)
				& \(+0.064~\mathrm{eV}\)
				& effective parameter \\
				
				Odd \(b\)-axis coupling
				& \(g_y\)
				& \(0.350~\mathrm{eV}\)
				& optical and band-curvature parameter \\
				
			\end{tabular}
		\end{ruledtabular}
	\end{table*}

	At the reference dielectric constant, the dimensionless screened
	interaction is
	\begin{equation}
		\widetilde W_{\rm RK}(\widetilde q)
		=
		\frac{2\pi}
		{\widetilde q+0.778\,\widetilde q^2}.
		\label{eq:dimensionless_isotropic_kernel_registry}
	\end{equation}
	
	In the layer-resolved BSE, the intralayer part of the interaction is
	multiplied by \(\mathcal F^{\rm intra}\), while the interlayer part is multiplied by
	\(e^{-\widetilde d\widetilde q}\mathcal F^{\rm inter}\). Increasing \(d\) therefore
	weakens the attraction between an electron and a hole in different layers
	without changing the intralayer attraction in this calculation. This is the
	dependence on \(d\) used in the spectra and in
	Fig.~\ref{fig:binding_sensitivity}(a).
	
	For the exchange fields and interlayer hoppings listed in
	Table~\ref{tab:crsbr_active_registry},
	\begin{equation}
		R_{\rm A}=0.261612~\mathrm{eV},
		\qquad
		R_{\rm B}=0.454528~\mathrm{eV}.
	\end{equation}
	They follow from
	\begin{equation}
		R_\tau
		=
		\sqrt{(J^\tau)^2+(t_\perp^\tau)^2}
	\end{equation}
	at the AFM state.
	
	The energy-aligned bare orbital separation is
	\begin{equation}
		\Delta_\Gamma^0
		=
		E_g^{\Gamma,{\rm AFM}}+R_{\rm A}+R_{\rm B}
		=
		2.28124~\mathrm{eV}.
		\label{eq:AFM_gap_calibration_registry}
	\end{equation}
	It is chosen before the BSE solution so that the \(41\times41\),
	\(d=8~\text{\AA}\) AFM equal-SRLP exciton lies at
	\(1.344~\mathrm{eV}\). The associated threshold binding is
	\(221.14~\mathrm{meV}\).
	
	The value of \(\Delta_\Gamma^0\) also enters the relation between the
	physical \(y\)-direction masses and the bare orbital masses in
	Eq.~\eqref{eq:app_effective_masses}. Changing
	\(\Delta_\Gamma^0\) while keeping the physical electron and hole masses
	fixed therefore changes the \(A\)- and \(B\)-orbital curvatures. For this
	reason, the optical energy is aligned before the lattice Hamiltonian is
	constructed rather than shifted after solving the BSE.
	
	Including the contribution of \(g_y\) to the \(y\)-direction band
	curvature, the orbital hoppings are
	\begin{align}
		t_x^{\rm A}&=0.04255~\mathrm{eV},&
		t_y^{\rm A}&=1.15247~\mathrm{eV},
		\nonumber\\
		t_x^{\rm B}&=-0.10951~\mathrm{eV},&
		t_y^{\rm B}&=-0.32155~\mathrm{eV}.
	\end{align}
	Their signs follow the dispersion convention in
	Eq.~\eqref{eq:crsbr_dispersion}.
	The minus sign multiplying
	\(\eta_x\mathcal T_\perp\) in Eq.~\eqref{eq:Hcrsbr}
	fixes the sign convention for the interlayer hopping.
	
	The four electron-hole channels are ordered as
	\begin{equation}
		(c_+v_-,\,c_-v_-,\,c_+v_+,\,c_-v_+).
	\end{equation}
	The equal-SRLP block contains
	\((c_-v_-)\) and \((c_+v_+)\).
	The opposite-SRLP block contains
	\((c_+v_-)\) and \((c_-v_+)\).
	
	At the AFM state, the states with SRLP eigenvalues
	\(\lambda=+1\) and \(-1\) are exactly degenerate within each band pair.
	A numerical diagonalization can therefore return any orthonormal
	combination within the degenerate subspace.
	For the numerical simulations, we label the AFM states by following
	the SRLP-resolved eigenstates continuously as
	\(\alpha\rightarrow0^+\). This fixes the labels used in the numerical BSE
	without changing the exact degeneracy at \(\alpha=0\).
	The analytical results at \(\alpha=0\) refer to the exact AFM Hamiltonian
	and to eigenstates with definite SRLP.
	
	The physical layer probabilities are obtained from
	\begin{align}
		\mathcal P_{\rm intra}
		&=
		\mathbb P_1^{(e)}\mathbb P_1^{(h)}
		+
		\mathbb P_2^{(e)}\mathbb P_2^{(h)},
		\nonumber\\
		\mathcal P_{\rm inter}
		&=
		\mathbb P_1^{(e)}\mathbb P_2^{(h)}
		+
		\mathbb P_2^{(e)}\mathbb P_1^{(h)},
	\end{align}
	with
	\begin{equation}
		w_{\rm intra}+w_{\rm inter}=1.
	\end{equation}
	
	The expectation values of
	\(\mathcal P_{\rm intra}\) and \(\mathcal P_{\rm inter}\) give the
	probabilities that the electron and hole occupy the same layer or different
	layers. These probabilities are different from the equal- and opposite-SRLP
	labels used to separate the BSE.

	\section{SRLP-resolved \texorpdfstring{\(\Gamma\)}{Gamma}-point
		Hamiltonian and eigenstates}
	\label{app:gamma_solution}
	
	At the \(\Gamma\) point, the odd interorbital coupling vanishes,
	\(d_y(\Gamma)=0\), so the A and B orbital sectors decouple.
	For a fixed orbital \(\tau=\rm A,B\), Eq.~\eqref{eq:Hcrsbr} reduces to
	\begin{equation}
		H_\tau(\Gamma)
		=
		\bareps^\tau
		-
		J^\tau
		\left(
		\sin\alpha\,\sigma_z
		+
		\cos\alpha\,\eta_z\sigma_y
		\right)
		-
		t_\perp^\tau\eta_x .
		\label{eq:app_gamma_hamiltonian}
	\end{equation}
	
	\subsection{Fixed-SRLP block}
	
	The layer-parity states and the two SRLP sectors are given in
	Eq.~\eqref{eq:srlp_basis}. Keeping the state order specified there, the
	\(\Gamma\)-point Hamiltonian for orbital \(\tau=\rm A,B\) and SRLP eigenvalue
	\(\lambda=\pm1\) is
	\begin{equation}
		H_{\tau\lambda}(\Gamma)
		=
		\begin{pmatrix}
			\bareps^\tau-\lambda t_\perp^\tau-J^\tau\sin\alpha
			&
			iJ^\tau\cos\alpha
			\\
			-iJ^\tau\cos\alpha
			&
			\bareps^\tau+\lambda t_\perp^\tau+J^\tau\sin\alpha
		\end{pmatrix}.
		\label{eq:app_gamma_block}
	\end{equation}
	
	The interlayer hopping has opposite signs for the two layer-parity
	states. The uniform \(z\)-directed exchange field also has opposite signs
	for the two basis states because their \(\sigma_z\) eigenvalues are
	opposite. The staggered \(y\)-directed exchange term changes both layer
	parity and spin and therefore couples the two states without changing
	their SRLP eigenvalue.
	
	Diagonalizing Eq.~\eqref{eq:app_gamma_block} gives
	\begin{align}
		R_{\tau\lambda}(\alpha)
		&=
		\sqrt{
			\left(
			\lambda t_\perp^\tau
			+
			J^\tau\sin\alpha
			\right)^2
			+
			\left(
			J^\tau\cos\alpha
			\right)^2
		}\nonumber\\
		&=
		\sqrt{
			(J^\tau)^2
			+
			(t_\perp^\tau)^2
			+
			2\lambda J^\tau t_\perp^\tau\sin\alpha
		}.
		\label{eq:app_gamma_norm}
	\end{align}
	
	The first expression separates the two contributions to the level
	splitting. The term
	\(\lambda t_\perp^\tau+J^\tau\sin\alpha\) contains the interlayer hopping
	and the layer-uniform component of the exchange field. The term
	\(J^\tau\cos\alpha\) comes from the layer-staggered component of the
	exchange field.
	
	The eigenvalues are
	\begin{equation}
		E_{\tau\lambda}^{\beta}(\Gamma)
		=
		\bareps^\tau
		+
		\beta R_{\tau\lambda}(\alpha),
		\qquad
		\beta=\pm1.
		\label{eq:app_gamma_eigenvalues}
	\end{equation}
	The two states in a given orbital and SRLP sector are therefore
	separated by \(2R_{\tau\lambda}\).
	
	\subsection{Normalized eigenstates at arbitrary canting}
	For \(R_{\tau\lambda}\neq0\), the relative weights of the two
	states in each SRLP sector can be written in terms of
	\begin{equation}
		\theta_{\tau\lambda}
		=
		\operatorname{atan2}
		\left(
		J^\tau\cos\alpha,
		\lambda t_\perp^\tau+J^\tau\sin\alpha
		\right),
		\qquad
		-\pi<\theta_{\tau\lambda}\leq\pi .
		\label{eq:app_gamma_mixing_angle}
	\end{equation}
	This gives
	\begin{equation}
		\cos\theta_{\tau\lambda}
		=
		\frac{
			\lambda t_\perp^\tau+J^\tau\sin\alpha
		}{
			R_{\tau\lambda}
		},
		\qquad
		\sin\theta_{\tau\lambda}
		=
		\frac{
			J^\tau\cos\alpha
		}{
			R_{\tau\lambda}
		}.
		\label{eq:app_gamma_mixing_components}
	\end{equation}
	
	For \(\lambda=+1\), the normalized conduction-band states are
	\begin{equation}
		\ket{c_+}
		=
		\cos\left(\frac{\theta_{{\rm A}+}}{2}\right)
		\ket{p_+,{\rm A},\uparrow_z}
		+
		i\sin\left(\frac{\theta_{{\rm A}+}}{2}\right)
		\ket{p_-,{\rm A},\downarrow_z}.
		\label{eq:app_cplus_state}
	\end{equation}
	\begin{equation}
		\ket{c'_+}
		=
		i\sin\left(\frac{\theta_{{\rm A}+}}{2}\right)
		\ket{p_+,{\rm A},\uparrow_z}
		+
		\cos\left(\frac{\theta_{{\rm A}+}}{2}\right)
		\ket{p_-,{\rm A},\downarrow_z}.
		\label{eq:app_cpplus_state}
	\end{equation}
	
	The normalized valence-band states are
	\begin{equation}
		\ket{v'_+}
		=
		\cos\left(\frac{\theta_{{\rm B}+}}{2}\right)
		\ket{p_+,{\rm B},\uparrow_z}
		+
		i\sin\left(\frac{\theta_{{\rm B}+}}{2}\right)
		\ket{p_-,{\rm B},\downarrow_z}.
		\label{eq:app_vpplus_state}
	\end{equation}
	\begin{equation}
		\ket{v_+}
		=
		i\sin\left(\frac{\theta_{{\rm B}+}}{2}\right)
		\ket{p_+,{\rm B},\uparrow_z}
		+
		\cos\left(\frac{\theta_{{\rm B}+}}{2}\right)
		\ket{p_-,{\rm B},\downarrow_z}.
		\label{eq:app_vplus_state}
	\end{equation}
	
	For \(\lambda=-1\), the normalized conduction-band states are
	\begin{equation}
		\ket{c_-}
		=
		\cos\left(\frac{\theta_{{\rm A}-}}{2}\right)
		\ket{p_-,{\rm A},\uparrow_z}
		+
		i\sin\left(\frac{\theta_{{\rm A}-}}{2}\right)
		\ket{p_+,{\rm A},\downarrow_z}.
		\label{eq:app_cminus_state}
	\end{equation}
	\begin{equation}
		\ket{c'_-}
		=
		i\sin\left(\frac{\theta_{{\rm A}-}}{2}\right)
		\ket{p_-,{\rm A},\uparrow_z}
		+
		\cos\left(\frac{\theta_{{\rm A}-}}{2}\right)
		\ket{p_+,{\rm A},\downarrow_z}.
		\label{eq:app_cpminus_state}
	\end{equation}
	
	The normalized valence-band states are
	\begin{equation}
		\ket{v'_-}
		=
		\cos\left(\frac{\theta_{{\rm B}-}}{2}\right)
		\ket{p_-,{\rm B},\uparrow_z}
		+
		i\sin\left(\frac{\theta_{{\rm B}-}}{2}\right)
		\ket{p_+,{\rm B},\downarrow_z}.
		\label{eq:app_vpminus_state}
	\end{equation}
	\begin{equation}
		\ket{v_-}
		=
		i\sin\left(\frac{\theta_{{\rm B}-}}{2}\right)
		\ket{p_-,{\rm B},\uparrow_z}
		+
		\cos\left(\frac{\theta_{{\rm B}-}}{2}\right)
		\ket{p_+,{\rm B},\downarrow_z}.
		\label{eq:app_vminus_state}
	\end{equation}
	
	The coefficients
	\(\cos^2(\theta_{\tau\lambda}/2)\) and
	\(\sin^2(\theta_{\tau\lambda}/2)\) give the probabilities of the two
	spin and layer-parity basis states within the corresponding SRLP sector.
	Their variation with \(\alpha\) describes the change of the
	single-particle wave functions during magnetic canting.
	
	The two states with the same orbital character and SRLP eigenvalue
	become degenerate when \(R_{\tau\lambda}=0\). For \(J^\tau\neq0\) and
	\(0\leq\alpha\leq\pi/2\), Eq.~\eqref{eq:app_gamma_norm} gives
	\begin{equation}
		\alpha
		=
		\frac{\pi}{2},
		\qquad
		t_\perp^\tau
		=
		-\lambda J^\tau .
		\label{eq:app_gamma_degeneracy_condition}
	\end{equation}
	The CrSBr parameters used here do not satisfy this condition.
	At \(R_{\tau\lambda}=0\), the two states are degenerate and
	\(\theta_{\tau\lambda}\) is undefined.
	
	\subsection{Antiferromagnetic limit}
	
	At \(\alpha=0\),
	\begin{equation}
		R_{\tau\lambda}
		=
		R_\tau
		\equiv
		\sqrt{
			(J^\tau)^2+(t_\perp^\tau)^2
		}
		\label{eq:app_afm_norm}
	\end{equation}
	is independent of \(\lambda\). The two SRLP sectors are therefore exactly
	degenerate.
	
	This degeneracy is enforced by the additional unitary operator
	\begin{equation}
		\mathcal U_{\rm AFM}
		=
		\eta_x\otimes\tau_0\otimes\sigma_x,
		\label{eq:app_afm_partner_operator}
	\end{equation}
	which satisfies
	\begin{equation}
		\left[
		\mathcal U_{\rm AFM},
		H(\mathbf k,\alpha=0)
		\right]
		=
		0,
		\qquad
		\left\{
		\mathcal U_{\rm AFM},
		\zeta
		\right\}
		=
		0.
		\label{eq:app_afm_partner_relations}
	\end{equation}
	Consequently, \(\mathcal U_{\rm AFM}\) maps every eigenstate with SRLP
	eigenvalue \(\lambda\) to a degenerate eigenstate with SRLP eigenvalue
	\(-\lambda\).
	
	At \(\alpha=0\), \(\sigma_y\) is conserved. For a fixed eigenvalue
	\(s_y=\pm1\),
	\begin{equation}
		H_{\tau,s_y}^{\rm AFM}
		=
		\bareps^\tau
		-
		s_yJ^\tau\eta_z
		-
		t_\perp^\tau\eta_x.
		\label{eq:app_afm_fixed_spin}
	\end{equation}
	The two eigenvalues are
	\begin{equation}
		E_{\tau,\pm}^{\rm AFM}
		=
		\bareps^\tau
		\pm
		R_\tau.
		\label{eq:app_afm_energies}
	\end{equation}
	
	For
	\(\lvert t_\perp^\tau\rvert\ll\lvert J^\tau\rvert\),
	\begin{equation}
		E_{\tau,\pm}^{\rm AFM}
		=
		\bareps^\tau
		\pm
		\lvert J^\tau\rvert
		\pm
		\frac{
			(t_\perp^\tau)^2
		}{
			2\lvert J^\tau\rvert
		}
		+
		O\!\left[
		\frac{
			(t_\perp^\tau)^4
		}{
			\lvert J^\tau\rvert^3
		}
		\right].
		\label{eq:app_afm_energy_expansion}
	\end{equation}
	
	At \(t_\perp^\tau=0\), the layer states satisfy
	\begin{equation}
		E_{1,s_y}^{(0)}
		=
		\bareps^\tau-s_yJ^\tau,
		\qquad
		E_{2,s_y}^{(0)}
		=
		\bareps^\tau+s_yJ^\tau.
		\label{eq:app_afm_unperturbed_layer_energies}
	\end{equation}
	For a fixed spin \(s_y\), the states in layers \(1\) and \(2\) are
	therefore separated in energy by \(2|J^\tau|\) before interlayer hopping
	is included.
	
	Their leading wave-function corrections are
	\begin{align}
		\ket{1,s_y}
		&\longrightarrow
		\ket{1,s_y}
		+
		\frac{
			t_\perp^\tau
		}{
			2s_yJ^\tau
		}
		\ket{2,s_y},
		\nonumber\\
		\ket{2,s_y}
		&\longrightarrow
		\ket{2,s_y}
		-
		\frac{
			t_\perp^\tau
		}{
			2s_yJ^\tau
		}
		\ket{1,s_y}.
		\label{eq:app_afm_state_expansion}
	\end{align}
	
	Thus, a state localized in one layer at \(t_\perp^\tau=0\) acquires
	a component in the other layer already at first order in
	\(t_\perp^\tau/J^\tau\). Its energy changes only at second order in the
	interlayer hopping, with a correction of order
	\((t_\perp^\tau)^2/|J^\tau|\).
	
	\subsection{Ferromagnetic limit}
	
	At \(\alpha=\pi/2\),
	\begin{equation}
		H_\tau^{\rm FM}(\Gamma)
		=
		\bareps^\tau
		-
		J^\tau\sigma_z
		-
		t_\perp^\tau\eta_x .
		\label{eq:app_fm_hamiltonian}
	\end{equation}
	The operators \(\sigma_z\) and \(\eta_x\) commute separately with the
	Hamiltonian. The eigenstates can therefore be labeled by the spin
	eigenvalue \(s=\pm1\) and the layer-parity eigenvalue \(p=\pm1\),
	\begin{equation}
		\ket{\tau,s,p}
		=
		\ket{p}_{\eta}
		\otimes
		\ket{\tau}_{\tau}
		\otimes
		\ket{s}_{\sigma}.
		\label{eq:app_fm_eigenstates}
	\end{equation}
	The SRLP eigenvalue is
	\begin{equation}
		\lambda=sp.
		\label{eq:app_fm_lambda}
	\end{equation}
	
	The corresponding energies are
	\begin{equation}
		E_{\tau,s,p}^{\rm FM}
		=
		\bareps^\tau
		-
		sJ^\tau
		-
		p t_\perp^\tau .
		\label{eq:app_fm_energies}
	\end{equation}
	Equivalently, since \(p=\lambda s\),
	\begin{equation}
		E_{\tau,s,\lambda}^{\rm FM}
		=
		\bareps^\tau
		-
		sJ^\tau
		-
		\lambda s\,t_\perp^\tau .
		\label{eq:app_fm_srlp_energies}
	\end{equation}
	
	For a fixed spin \(s\), the exchange field shifts the states in layers
	\(1\) and \(2\) by the same energy \(-sJ^\tau\). The two layer states are
	therefore degenerate before interlayer hopping is included. Interlayer
	hopping mixes these degenerate states into the layer-parity combinations
	\(\ket{p_+}\) and \(\ket{p_-}\).
	
	The energies of the two layer-parity states differ by
	\begin{equation}
		\left|
		E_{\tau,s,p=+1}^{\rm FM}
		-
		E_{\tau,s,p=-1}^{\rm FM}
		\right|
		=
		2\left|t_\perp^\tau\right|.
		\label{eq:app_fm_bonding_splitting}
	\end{equation}
	
	Interlayer hopping therefore changes the FM energies already to
	first order in \(t_\perp^\tau\). In the AFM state, the same-spin states
	in layers \(1\) and \(2\) are separated by \(2|J^\tau|\) before hopping
	is included, and their leading energy shift is instead second order in
	\(t_\perp^\tau\).
	
	\subsection{Continuous labeling across the canting trajectory}
	
	For \(0<\alpha<\pi/2\), the SRLP eigenvalue \(\lambda\) remains exact,
	while the spin and layer composition
	changes continuously with the mixing angle in
	Eq.~\eqref{eq:app_gamma_mixing_angle}.
	The labels
	\(c_\lambda\), \(c'_\lambda\), \(v_\lambda\), and \(v'_\lambda\) are
	assigned using the energy-ordering convention in
	Eq.~\eqref{eq:band_ordering}.
	
	At the
	AFM state,
	the \(\lambda=\pm1\) sectors become degenerate but remain distinguishable
	as eigenstates of \(\zeta\). At the
	FM state,
	spin and layer parity are separately conserved and the SRLP eigenvalue
	becomes
	\(\lambda=sp\).
	The same SRLP eigenvalue therefore identifies each band-edge state
	throughout the canted regime even though spin and layer parity are not
	separately conserved for \(0<\alpha<\pi/2\).
	
	\subsection{Expansion in the interlayer mixing parameter}
	
	Writing $\mathcal B=\sin\alpha$, the exact SRLP-dependent splitting can be
	expressed as
	\begin{equation}
		R_{\tau\lambda}(\mathcal B)
		=
		R_\tau\sqrt{1+\lambda r_\tau \mathcal B},
		\qquad
		r_\tau
		=
		\frac{2J^\tau t_\perp^\tau}{R_\tau^2}.
		\label{eq:app_R_mixing_form}
	\end{equation}
	
	Since $|\mathcal B|\leq1$, the condition $|r_\tau|<1$ ensures
	$|r_\tau \mathcal B|<1$ throughout the canting range. The binomial expansion of
	Eq.~\eqref{eq:app_R_mixing_form} gives
	\begin{align}
		R_{\tau\lambda}(\mathcal B)
		&=
		R_\tau\left[
		1
		+\tfrac{1}{2}\lambda r_\tau \mathcal B
		-\tfrac{1}{8}r_\tau^2\mathcal B^2
		+O\!\left(|r_\tau \mathcal B|^3\right)
		\right]
		\nonumber\\
		&=
		R_\tau
		+\lambda\frac{J^\tau t_\perp^\tau}{R_\tau}\mathcal B
		-\frac{(J^\tau t_\perp^\tau)^2}{2R_\tau^3}\mathcal B^2
		+O\!\left(R_\tau|r_\tau \mathcal B|^3\right).
		\label{eq:app_R_mixing_expansion}
	\end{align}
	
	Substitution into Eq.~\eqref{eq:direct_gaps} gives
	Eq.~\eqref{eq:gap_weakfield} and the coefficients in
	Eq.~\eqref{eq:gap_coefficients}.
	For $\max_\tau|r_\tau|\ll1$, the expansion remains controlled
	throughout the canting range.

	\section{Analytical band-edge optical matrix elements}
	\label{app:selection_polarization}
	The momentum derivative of the Bloch Hamiltonian preserves SRLP symmetry
	because the SRLP operator is independent of crystal momentum. This fixes
	which interband matrix elements can remain finite before their
	polarization dependence is evaluated.
	
	\subsection{SRLP selection rule}
	
	Because the SRLP operator is momentum independent, the commutation
	$[\zeta,H_{\rm CrSBr}(\mathbf k)]=0$ implies
	\begin{equation}
		\left[
		\zeta,
		\partial_{k_\nu}H_{\rm CrSBr}(\mathbf k)
		\right]
		=
		\partial_{k_\nu}
		\left[
		\zeta,
		H_{\rm CrSBr}(\mathbf k)
		\right]
		=
		0.
		\label{eq:app_velocity_srlp_commutator}
	\end{equation}
	
	For conduction and valence states with SRLP eigenvalues
	$\lambda_c$ and $\lambda_v$,
	\begin{equation}
		\zeta\ket{c_{\lambda_c}}
		=
		\lambda_c\ket{c_{\lambda_c}},
		\qquad
		\zeta\ket{v_{\lambda_v}}
		=
		\lambda_v\ket{v_{\lambda_v}}.
		\label{eq:app_cv_srlp_eigenvalues}
	\end{equation}
	It follows that
	\begin{align}
		\lambda_c
		M_{c_{\lambda_c}v_{\lambda_v}}^\nu
		&=
		\mel{c_{\lambda_c}}
		{\zeta\,\partial_{k_\nu}H_{\rm CrSBr}}
		{v_{\lambda_v}}
		\nonumber\\
		&=
		\mel{c_{\lambda_c}}
		{\partial_{k_\nu}H_{\rm CrSBr}\,\zeta}
		{v_{\lambda_v}}
		\nonumber\\
		&=
		\lambda_v
		M_{c_{\lambda_c}v_{\lambda_v}}^\nu.
		\label{eq:app_srlp_matrix_element_derivation}
	\end{align}
	Therefore,
	\begin{equation}
		\left(
		\lambda_c-\lambda_v
		\right)
		M_{c_{\lambda_c}v_{\lambda_v}}^\nu(\mathbf k)
		=
		0.
		\label{eq:app_srlp_selection_rule}
	\end{equation}
	For opposite SRLP,
	\begin{equation}
		M_{c_{\lambda_c}v_{\lambda_v}}^\nu(\mathbf k)
		=
		0
		\qquad
		\mathrm{when}
		\qquad
		\lambda_c\neq\lambda_v.
		\label{eq:app_opposite_srlp_zero}
	\end{equation}
	
	Opposite-SRLP conduction and valence states therefore have zero
	interband matrix element for either optical polarization. Equal SRLP is
	necessary for a finite matrix element but does not by itself guarantee one.
	The polarization dependence is determined separately by the orbital
	structure of $\partial_{k_\nu}H_{\rm CrSBr}$.
	
	\subsection{Exact matrix elements at arbitrary canting}
	
	At the $\Gamma$ point, Eq.~\eqref{eq:gamma_velocity_operators} gives
	\begin{align}
		\partial_{k_y}H_{\rm CrSBr}(\Gamma)
		&=
		-a_yg_y\,
		\eta_0\otimes\tau_x\otimes\sigma_0,
		\nonumber\\
		\partial_{k_x}H_{\rm CrSBr}(\Gamma)
		&=
		0.
		\label{eq:app_gamma_optical_operator}
	\end{align}
	The $y$-polarized operator changes the orbital character
	$A\leftrightarrow B$ and is independent of layer and spin. Its matrix
	elements are therefore determined by the overlap of the layer and spin
	parts of the conduction and valence states.
	
	For states with the same SRLP eigenvalue, the relative mixing angle
	between the $B$- and $A$-orbital states is
	\begin{equation}
		\delta\theta_\lambda
		=
		\theta_{B\lambda}
		-
		\theta_{A\lambda}.
		\label{eq:app_delta_theta}
	\end{equation}
	
	For equal SRLP, direct evaluation gives
	\begin{align}
		M_{c_\lambda v_\lambda}^{y}(\Gamma)
		&=
		-\ii a_yg_y
		\sin\left(
		\frac{\delta\theta_\lambda}{2}
		\right),
		\nonumber\\
		M_{c'_\lambda v'_\lambda}^{y}(\Gamma)
		&=
		-\ii a_yg_y
		\sin\left(
		\frac{\delta\theta_\lambda}{2}
		\right),
		\nonumber\\
		M_{c'_\lambda v_\lambda}^{y}(\Gamma)
		&=
		-a_yg_y
		\cos\left(
		\frac{\delta\theta_\lambda}{2}
		\right),
		\nonumber\\
		M_{c_\lambda v'_\lambda}^{y}(\Gamma)
		&=
		-a_yg_y
		\cos\left(
		\frac{\delta\theta_\lambda}{2}
		\right).
		\label{eq:app_gamma_optical_matrix_elements}
	\end{align}
	The relative angle is fixed by the exchange fields, interlayer
	hoppings, and canting angle. Its cosine is
	\begin{equation}
		\cos\delta\theta_\lambda
		=
		\frac{
			J^A J^B
			+
			t_\perp^A t_\perp^B
			+
			\lambda\sin\alpha
			\left(
			J^A t_\perp^B
			+
			J^B t_\perp^A
			\right)
		}{
			R_{A\lambda}R_{B\lambda}
		}.
		\label{eq:app_delta_theta_cos}
	\end{equation}
	
	Its sine is
	\begin{equation}
		\sin\delta\theta_\lambda
		=
		\frac{
			\lambda\cos\alpha
			\left(
			J^B t_\perp^A
			-
			J^A t_\perp^B
			\right)
		}{
			R_{A\lambda}R_{B\lambda}
		}.
		\label{eq:app_delta_theta_sin}
	\end{equation}
	
	The phases of the matrix elements depend on the phase convention chosen for
	the individual Bloch states. Their absolute values are gauge independent,
	\begin{align}
		\left|
		M_{c_\lambda v_\lambda}^{y}
		\right|^2
		=
		\left|
		M_{c'_\lambda v'_\lambda}^{y}
		\right|^2
		&=
		a_y^2g_y^2
		\sin^2\left(
		\frac{\delta\theta_\lambda}{2}
		\right),
		\nonumber\\
		\left|
		M_{c'_\lambda v_\lambda}^{y}
		\right|^2
		=
		\left|
		M_{c_\lambda v'_\lambda}^{y}
		\right|^2
		&=
		a_y^2g_y^2
		\cos^2\left(
		\frac{\delta\theta_\lambda}{2}
		\right).
		\label{eq:app_gamma_optical_weights}
	\end{align}
	
	All opposite-SRLP matrix elements vanish by
	Eq.~\eqref{eq:app_opposite_srlp_zero}. All $x$-polarized band-edge matrix
	elements vanish because
	$\partial_{k_x}H_{\rm CrSBr}(\Gamma)=0$.
	
	The equal-SRLP matrix elements satisfy
	\begin{align}
		\left|
		M_{c_\lambda v_\lambda}^{y}
		\right|^2
		+
		\left|
		M_{c'_\lambda v_\lambda}^{y}
		\right|^2
		&=
		a_y^2g_y^2,
		\nonumber\\
		\left|
		M_{c_\lambda v'_\lambda}^{y}
		\right|^2
		+
		\left|
		M_{c'_\lambda v'_\lambda}^{y}
		\right|^2
		&=
		a_y^2g_y^2.
		\label{eq:app_optical_sum_rules}
	\end{align}
	
	Canting transfers $y$-polarized matrix-element weight between
	$c_\lambda$ and $c'_\lambda$ for a fixed valence state while their sum
	remains $a_y^2g_y^2$. Opposite-SRLP transitions remain forbidden.
	The oscillator strength also contains the transition-energy factor
	$1/(E_c-E_v)$ and therefore changes accordingly.
	
	\subsection{Antiferromagnetic and ferromagnetic limits}
	
	In the AFM state, $\alpha=0$, and the mixing angles in
	Eq.~\eqref{eq:app_gamma_mixing_components} satisfy
	\begin{align}
		\cos\theta_{\tau\lambda}
		&=
		\frac{\lambda t_\perp^\tau}{R_\tau},~
		\sin\theta_{\tau\lambda}
		=
		\frac{J^\tau}{R_\tau},~
		R_\tau
		=
		\sqrt{(J^\tau)^2+(t_\perp^\tau)^2}.
		\label{eq:app_afm_angle_parameters}
	\end{align}
	The relative angle between the A and B-orbital states then obeys
	\begin{equation}
		\cos\delta\theta_\lambda
		=
		\frac{
			t_\perp^{\rm A} t_\perp^{\rm B}+J^{\rm A} J^{\rm B}
		}{
			R_{\rm A} R_{\rm B}
		}.
		\label{eq:app_afm_delta_theta}
	\end{equation}
	This quantity is independent of the sign of $\lambda$. The squared
	$y$-polarized matrix elements are
	\begin{align}
		\left|
		M_{c_\lambda v_\lambda}^{y}
		\right|^2
		=
		\left|
		M_{c'_\lambda v'_\lambda}^{y}
		\right|^2
		&=
		\frac{a_y^2g_y^2}{2}
		\left[
		1-
		\frac{
			t_\perp^{\rm A} t_\perp^{\rm B}+J^{\rm A} J^{\rm B}
		}{
			R_{\rm A} R_{\rm B}
		}
		\right],
		\nonumber\\
		\left|
		M_{c'_\lambda v_\lambda}^{y}
		\right|^2
		=
		\left|
		M_{c_\lambda v'_\lambda}^{y}
		\right|^2
		&=
		\frac{a_y^2g_y^2}{2}
		\left[
		1+
		\frac{
			t_\perp^{\rm A} t_\perp^{\rm B}+J^{\rm A} J^{\rm B}
		}{
			R_{\rm A} R_{\rm B}
		}
		\right].
		\label{eq:app_afm_optical_weights}
	\end{align}
	For the CrSBr parameters, $J^{\rm A} J^{\rm B}<0$ and the exchange energies are
	larger than the interlayer hoppings. The transitions
	$c_\lambda\leftrightarrow v_\lambda$ and
	$c'_\lambda\leftrightarrow v'_\lambda$ consequently have larger
	matrix elements than
	$c'_\lambda\leftrightarrow v_\lambda$ and
	$c_\lambda\leftrightarrow v'_\lambda$.
	
	In the FM state, $\alpha=\pi/2$, and
	\begin{equation}
		R_{\tau\lambda}
		=
		\left|
		t_\perp^\tau+\lambda J^\tau
		\right|.
		\label{eq:app_fm_R}
	\end{equation}
	Equation~\eqref{eq:app_gamma_mixing_components} then gives
	\begin{equation}
		\cos\theta_{\tau\lambda}
		=
		\lambda\,
		\operatorname{sgn}
		\left(
		t_\perp^\tau+\lambda J^\tau
		\right),
		\qquad
		\sin\theta_{\tau\lambda}
		=
		0.
		\label{eq:app_fm_angle_parameters}
	\end{equation}
	The squared matrix elements become
	\begin{align}
		\left|
		M_{c_\lambda v_\lambda}^{y}
		\right|^2
		&=
		\left|
		M_{c'_\lambda v'_\lambda}^{y}
		\right|^2\\
		&=
		\frac{a_y^2g_y^2}{2}
		\left[
		1-
		\operatorname{sgn}
		\left(
		t_\perp^{\rm A}+\lambda J^{\rm A}
		\right)
		\operatorname{sgn}
		\left(
		t_\perp^{\rm B}+\lambda J^{\rm B}
		\right)
		\right].\nonumber
		\label{eq:app_fm_optical_weights_same}
	\end{align}
	The other two equal-SRLP transitions satisfy
	\begin{align}
		\left|
		M_{c'_\lambda v_\lambda}^{y}
		\right|^2
		&=
		\left|
		M_{c_\lambda v'_\lambda}^{y}
		\right|^2\\
		&=
		\frac{a_y^2g_y^2}{2}
		\left[
		1+
		\operatorname{sgn}
		\left(
		t_\perp^{\rm A}+\lambda J^{\rm A}
		\right)
		\operatorname{sgn}
		\left(
		t_\perp^{\rm B}+\lambda J^{\rm B}
		\right)
		\right].\nonumber
		\label{eq:app_fm_optical_weights_cross}
	\end{align}
	
	For the CrSBr parameters,
	\begin{equation}
		\operatorname{sgn}
		\left(
		t_\perp^{\rm A}+\lambda J^{\rm A}
		\right)
		\operatorname{sgn}
		\left(
		t_\perp^{\rm B}+\lambda J^{\rm B}
		\right)
		=
		-1
		\qquad
		(\lambda=\pm1).
		\label{eq:app_crsbr_fm_sign_condition}
	\end{equation}
	Hence
	\begin{align}
		\left|
		M_{c_\lambda v_\lambda}^{y}(\Gamma)
		\right|
		=
		\left|
		M_{c'_\lambda v'_\lambda}^{y}(\Gamma)
		\right|
		&=
		a_y|g_y|,
		\nonumber\\
		M_{c'_\lambda v_\lambda}^{y}(\Gamma)
		=
		M_{c_\lambda v'_\lambda}^{y}(\Gamma)
		&=
		0.
		\label{eq:app_crsbr_fm_optical_limit}
	\end{align}
	
	The vanishing of the two cross transitions in the FM state follows from
	the separate conservation of spin and layer parity. It is therefore
	stronger than the SRLP selection rule alone.
	
	In the AFM state, the matrix-element magnitudes are identical for
	$\lambda=+1$ and $\lambda=-1$ because
	$\cos\delta\theta_\lambda$ is independent of $\lambda$. For the CrSBr
	parameters, the same equality holds in the FM state because
	Eq.~\eqref{eq:app_crsbr_fm_sign_condition} has the same value for both
	SRLP sectors.

	\section{Optical response and numerical implementation details}
	\label{app:numerics}
	
	The calculation samples the full rectangular Brillouin zone of the effective
	lattice Hamiltonian on an odd centered grid,
	\begin{align}
		k_{x,i}&=\frac{2\pi i}{a_xN_x},
		& i&=-\frac{N_x-1}{2},\ldots,\frac{N_x-1}{2},
		\nonumber\\
		k_{y,j}&=\frac{2\pi j}{a_yN_y},
		& j&=-\frac{N_y-1}{2},\ldots,\frac{N_y-1}{2}.
		\label{eq:full_bz_grid}
	\end{align}
	
	Direct Euclidean differences
	\(|\mathbf k_m-\mathbf k_{m'}|\) are used in
	the effective continuum interaction.
	For the
	\(41\times41\) grid,
	the largest sampled magnitudes are
	\(k_{x,\max}=0.87571~\text{\AA}^{-1}\) and
	\(k_{y,\max}=0.64526~\text{\AA}^{-1}\).
	
	The use of a continuum interaction with direct Euclidean momentum differences
	is appropriate only when the exciton is concentrated near \(\Gamma\) and has
	negligible weight at the grid boundary.
	The calculation should therefore be interpreted as a
	\(\Gamma\)-centered effective-model discretization rather than a periodic
	lattice Coulomb kernel over the full zone.
	
	For a uniform grid,
	\begin{equation}
		w_m=\frac{\Delta k_x\Delta k_y}{(2\pi)^2},
		\qquad
		\widetilde{\Psi}_{n,ma}=\sqrt{w_m}\Psi_{n,a}(\mathbf k_m),
		\label{eq:app_grid_amplitude_definition}
	\end{equation}
	so \(\sum_{m,a}|\widetilde{\Psi}_{n,ma}|^2=1\).
	In this weighted basis, the Hermitian BSE matrix is
	\begin{align}
		\left[\mathcal H^{\rm BSE}\right]_{ma,m'b}
		&=
		\Delta_a(\mathbf k_m)\delta_{mm'}\delta_{ab}
		\nonumber\\
		&\quad
		-
		\sqrt{w_mw_{m'}}\,
		W_{\rm RK}(q_{mm'})
		\mathcal F_{ab}(d;\mathbf k_m,\mathbf k_{m'}),
		\label{eq:weighted_BSE_kernel}
	\end{align}
	where
	\(q_{mm'}=|\mathbf k_m-\mathbf k_{m'}|\).
	
	For the dimensionless interaction factor
	\(1/(\widetilde q+\widetilde r_*\widetilde q^2)\),
	its uniform-grid prefactor is
	\begin{equation}
		\frac{\Delta\widetilde k_x\Delta\widetilde k_y}{2\pi}.
		\label{eq:stored_kernel_prefactor}
	\end{equation}
	
	The singular element is evaluated by averaging the same RK interaction over
	the rectangular central momentum cell.
	\begin{equation}
		\overline W_0
		=
		\frac{1}{\Delta\widetilde k_x\Delta\widetilde k_y}
		\int_{\mathcal C_0}d^2\widetilde q\,
		\widetilde W_{\rm RK}(|\widetilde{\mathbf q}|),
		\label{eq:q0_cell_average_registry}
	\end{equation}
	where
	\(\mathcal C_0=[-\Delta\widetilde k_x/2,\Delta\widetilde k_x/2]
	\times[-\Delta\widetilde k_y/2,\Delta\widetilde k_y/2]\).
	
	Figure~\ref{fig:exciton_energy_opt_con} uses a \(51\times51\) grid,
	61 canting points, and 800 BSE eigenstates
	per SRLP sector.
	Figure~\ref{fig:two_channel_model_weights} uses
	\(41\times41\) sampling.
	The one-parameter simulations in
	Fig.~\ref{fig:binding_sensitivity} also use \(41\times41\) sampling, so the
	reported reference bindings correspond to the same grid.
	
	An input \(\alpha=0\) to the
	canting calculation
	is evaluated at the regularized value
	\(\alpha_{\rm reg}=10^{-7}\).
	Varying \(\alpha_{\rm reg}\) from \(10^{-4}\) to \(10^{-10}\) changes the
	AFM energies
	by less than \(10^{-10}~\mathrm{eV}\).
	The regularization only selects the
	states continuously connected to positive canting
	and does not act as a physical symmetry-breaking field.
	
	For the AFM equal-SRLP state at \(d=8~\text{\AA}\), the binding
	energy changes with the momentum-grid size as
	\begin{center}
		\begin{tabular}{c|ccccc}
			\(N_x=N_y\) & 21 & 25 & 31 & 35 & 41 \\
			\hline
			\(E_b\) (meV) & 218.431 & 218.495 & 219.548 & 220.253 & 221.140
		\end{tabular}
	\end{center}
	
	The \(35\rightarrow41\) change is \(0.89~\mathrm{meV}\).
	At the
	FM state
	the lowest equal-SRLP binding changes only from
	\(195.26~\mathrm{meV}\) at \(N=21\) to
	\(196.66~\mathrm{meV}\) at \(N=41\).
	The binding energies are therefore stable to within a few meV. The residual upward drift is limited by the \(k_y\) sampling of the
	quasi-one-dimensional exciton, while the sensitivity trends are robust.

	\subsection{Screening of the intralayer interaction by the second layer}
	\label{app:bilayer_kernel}
	
	In Sec.~\ref{sec:exciton},
	the intralayer attraction is the monolayer
	Rytova-Keldysh potential
	\(W_{\rm intra}(q)=W_{\rm RK}(q)\), and the
	interlayer attraction is
	\(W_{\rm inter}(q)=e^{-\widetilde d\widetilde q}
	W_{\rm RK}(q)\).
	With these interactions
	the interlayer separation \(d\) enters only the interlayer term,
	so increasing \(d\) weakens the attraction between electron and hole
	configurations in different layers while leaving the intralayer attraction
	unchanged
	[Eq.~\eqref{eq:form_factor}].
	
	The two-layer Rytova-Keldysh interaction of Semina
	\emph{et al.}~[their Eqs.~(9a,b)]~\cite{Semina2024}
	includes the electrostatic screening produced by both layers.
	With
	\(\xi=e^{-\widetilde d\widetilde q}\) and
	\(r_1=r_2=\widetilde r_*\),
	\begin{align}
		W_{11}(q)
		&=
		\frac{2\pi}{\widetilde q}\,
		\frac{1+\widetilde q\, r_2(1-\xi^2)}{D},
		\qquad
		W_{12}(q)
		=
		\frac{2\pi}{\widetilde q}\,
		\frac{\xi}{D},
		\nonumber\\
		D
		&=
		(1+\widetilde q\,r_1)
		(1+\widetilde q\,r_2)
		-
		\widetilde q^2 r_1 r_2\,\xi^2.
		\label{eq:bilayer_kernel}
	\end{align}
	
	Here,
	\(W_{\rm intra}=W_{11}\) and
	\(W_{\rm inter}=W_{12}\).
	The two
	screening treatments
	enter the BSE through the \emph{identical} layer-resolved form factors
	[Eqs.~\eqref{eq:F_intra} and \eqref{eq:F_inter}]. Their momentum-dependent interaction coefficients differ.
	
	At \(d\to0\), the
	two-layer interaction
	reduces to a single sheet with the combined screening length
	\(r_1+r_2\), with denominator
	\(1+(r_1+r_2)\widetilde q\).
	The interaction that neglects screening by the second layer
	retains the single-layer radius
	\(\widetilde r_*\), with denominator
	\(1+\widetilde r_*\widetilde q\).
	The two interactions therefore do \emph{not} coincide at small
	\(d\).
	
	They approach the same isolated-monolayer limit only as
	\(d\to\infty\),
	with
	\(W_{11}\to W_{\rm RK}(\widetilde r_*)\) and
	\(W_{12}\to0\).
	At intermediate \(d\),
	the \(\xi^2\) term in \(D\) accounts for the screening of an
	intralayer electron-hole pair by the second layer. Consequently,
	\(W_{11}\) changes with \(d\), whereas
	\(W_{\rm intra}=W_{\rm RK}\) remains independent of \(d\) when this
	additional screening is neglected.
	
	At the CrSBr reference
	(\(d=8~\text{\AA}\), \(41\times41\) grid), the two
	screening treatments give
	\begin{widetext}
		\begin{center}
			\begin{tabular*}{\textwidth}{@{\extracolsep{\fill}}lcc}
				\hline
				Quantity
				&
				fixed intralayer screening
				&
				two-layer screening
				\\
				\hline
				FM \(\Lambda_{eh}=+1\) binding
				& \(196.7\)
				& \(176.9~\mathrm{meV}\) \\
				AFM splitting \(2\Omega_{+}\)
				& \(102.6\)
				& \(99.1~\mathrm{meV}\) \\
				AFM\(\to\)FM redshift
				& \(4.37\)
				& \(3.27~\mathrm{meV}\) \\
				Interlayer weight, AFM\(\to\)FM
				& \(0.8\to7.1\%\)
				& \(0.8\to5.4\%\) \\
				\hline
			\end{tabular*}
		\end{center}
	\end{widetext}
	
	The avoided crossing, the quadratic AFM redshift, and the increase of
	interlayer character from the AFM to the FM state are all preserved. Including the screening produced by the second layer lowers the
	reference FM binding by approximately
	\(20~\mathrm{meV}\), or \(10\%\), while the AFM splitting changes by
	approximately \(3.6~\mathrm{meV}\), or \(3.5\%\).
	The spectra of
	Figs.~\ref{fig:exciton_energy_opt_con} and
	\ref{fig:two_channel_model_weights}, and the screening- and mass-dependence
	panels
	Fig.~\ref{fig:binding_sensitivity}(b) to (d), are therefore qualitatively
	unchanged.
	
	The one qualitative difference is in the dependence on the interlayer
	separation
	[Fig.~\ref{fig:binding_sensitivity}(a)].
	The single-particle energies, Bloch states, and equal-SRLP continuum
	threshold
	\(E_{\rm th}^{(+)}=1.53624~\mathrm{eV}\)
	are the same in both cases. The opposite dependence on \(d\) therefore comes
	from the screened electron-hole interaction.
	
	In the
	calculation that neglects the modification of intralayer screening by
	the second layer,
	increasing \(d\) modifies only the interlayer interaction through
	\(e^{-\widetilde d\widetilde q}\), while the intralayer interaction is held
	fixed. The FM binding therefore decreases monotonically
	from \(222\) to \(193~\mathrm{meV}\).
	The decrease follows from the weakening of the attraction between an
	electron and a hole in different layers while the same-layer attraction
	remains unchanged.
	
	In the
	two-layer screening calculation,
	increasing \(d\) simultaneously weakens the interlayer attraction and reduces
	the screening of the intralayer interaction by the second layer.
	For the lowest exciton, the latter effect dominates beyond an extremely
	small-\(d\)
	range.
	The
	two-layer
	binding is nearly flat and weakly nonmonotonic at very small \(d\),
	with a \(0.06~\mathrm{meV}\) dip between
	\(d=0\) and \(0.1~\text{\AA}\), and then increases toward the isolated-layer
	limit from \(146\) to \(189~\mathrm{meV}\).
	The opposite-SRLP exciton has the same broad dependence on \(d\).
	
	The two
	screening treatments
	agree only as \(d\to\infty\), not as \(d\to0\).
	Because the single-particle threshold is \(d\)-independent,
	the exciton energy
	\(E_X=E_{\rm th}-E_b\)
	changes in the opposite direction to the binding energy.
	The hopping amplitudes, exchange energies, effective masses, and
	Bloch states remain fixed as \(d\) changes. A physical change of the layer
	spacing can also modify these electronic quantities.
	
\begin{figure}[t]
		\centering
		\includegraphics[width=\columnwidth]{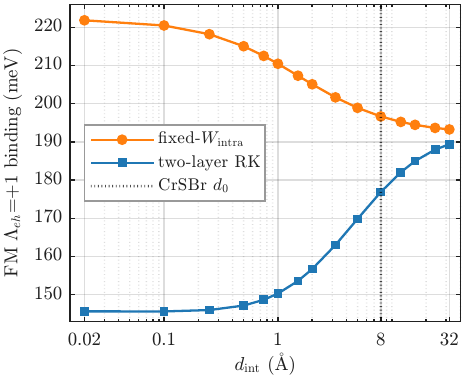}
		\caption{FM binding energy of the lowest equal-SRLP
			(\(\Lambda_{eh}=+1\)) exciton as a function of the electron-hole layer
			separation \(d\).
			The circular data points correspond to
			\(W_{\rm intra}=W_{\rm RK}\) and
			\(W_{\rm inter}=e^{-\widetilde d\widetilde q}W_{\rm RK}\), for which the
			screening of the intralayer interaction by the second layer is neglected.
			The square data points use the two-layer interaction in
			Eq.~\eqref{eq:bilayer_kernel}, which includes this additional screening.
			When the intralayer interaction is kept fixed, the binding decreases as the
			interlayer attraction weakens with increasing \(d\).
			With two-layer screening, the binding increases over most of the range
			because increasing \(d\) reduces the screening of the intralayer attraction
			by the second layer.
			Both interactions approach the isolated-layer limit at large \(d\).
			The dotted line marks the CrSBr value
			\(d_0=8~\text{\AA}\).}
		\label{fig:kernel_compare}
	\end{figure}

	Figure~\ref{fig:kernel_compare} compares the two binding curves directly.

	With the velocity convention of Sec.~\ref{sec:selectionrules},
	\begin{equation}
		r^\alpha_{cv}(\mathbf k)
		=
		-\ii
		\frac{
			\langle c,\mathbf k|
			\partial_{k_\alpha}H
			|v,\mathbf k\rangle
		}{
			E_c(\mathbf k)-E_v(\mathbf k)
		}.
	\end{equation}
	
	For Euclidean-normalized eigenvectors, the discrete optical amplitude on the momentum grid
	is
	\begin{equation}
		X^\alpha_{n0}
		=
		\sum_{m,a}
		\sqrt{w_m}\,
		\widetilde{\Psi}_{n,ma}^*
		r^\alpha_a(\mathbf k_m),
		\qquad
		X^\alpha_{0n}
		=
		(X^\alpha_{n0})^*.
		\label{eq:grid_optical_amplitude}
	\end{equation}
	
	The normalized response uses Eq.~\eqref{eq:sigma} with
	\(\gamma=0.3~\mathrm{meV}\).
	The spectra are normalized, so the overall optical-conductivity
	prefactor is omitted.
	For each SRLP sector, the exciton binding energy is measured from the
	lowest energy of an unbound electron-hole pair in that sector, as in
	Eqs.~\eqref{eq:binding_def} and \eqref{eq:binding_threshold}.

	\section{Two-particle symmetry and projection of the exciton Hamiltonian}
	\label{app:exciton_projection}
	For equivalent layers, the direct electron-hole interaction conserves
	the two-particle SRLP eigenvalue and separates the four near-gap
	electron-hole channels into two independent sectors. The two-channel exciton
	Hamiltonian in Sec.~\ref{sec:bse_four_channel_projection} follows after
	projection onto the two band-pair states within each sector.
	The symmetry decomposition is exact for the direct interaction retained in
	the manuscript. The analytical coupling and
	layer-probability
	formulas require the additional projection assumptions stated below.
	
	\subsection{Two-particle SRLP selection rule}
	
	The single-particle operator satisfies
	\begin{equation}
		\zeta^2=\mathbb{I},
		\qquad
		[\zeta,H_{\rm CrSBr}(\mathbf k)]=0.
	\end{equation}
	
	For a one-particle operator \(O\),
	the hole-space representation is
	\(O^{(h)}=O^*\), so that
	\begin{equation}
		\langle h_v|O^{(h)}|h_{v'}\rangle
		=
		\langle v'|O|v\rangle.
		\label{eq:app_hole_operator_convention}
	\end{equation}
	
	The corresponding two-particle SRLP operator is
	\begin{equation}
		\mathcal Z_{eh}
		=
		\zeta^{(e)}\otimes\zeta^{(h)},
		\qquad
		\zeta^{(h)}=\zeta^*.
		\label{eq:app_eh_srlp_operator}
	\end{equation}
	Because the SRLP eigenvalues are real,
	\begin{equation}
		\mathcal Z_{eh}
		|c_{\lambda_c}v_{\lambda_v}\rangle
		=
		\lambda_c\lambda_v
		|c_{\lambda_c}v_{\lambda_v}\rangle.
		\label{eq:app_eh_srlp_eigenvalue}
	\end{equation}
	Thus, the two-particle eigenvalue is
	\begin{equation}
		\Lambda_{eh}
		=
		\lambda_c\lambda_v.
		\label{eq:app_eh_srlp_product}
	\end{equation}
	
	Conjugation by \(\zeta=\eta_x\sigma_z\) exchanges the two
	one-particle layer projectors of
	Eq.~\eqref{eq:full_layer_projectors},
	\begin{equation}
		\zeta\mathbb P_1\zeta
		=
		\mathbb P_2,
		\qquad
		\zeta\mathbb P_2\zeta
		=
		\mathbb P_1.
		\label{eq:app_layer_projector_exchange}
	\end{equation}
	It follows that the physical
	intralayer and interlayer
	projectors obey
	\begin{align}
		\mathcal Z_{eh}\mathcal P_{\rm intra}\mathcal Z_{eh}^{-1}
		&=
		\mathcal P_{\rm intra},
		\nonumber\\
		\mathcal Z_{eh}\mathcal P_{\rm inter}\mathcal Z_{eh}^{-1}
		&=
		\mathcal P_{\rm inter}.
		\label{eq:app_pair_projector_invariance}
	\end{align}
	
	For a general layer-resolved direct interaction, this invariance requires
	\begin{equation}
		W_{11}(q)=W_{22}(q),
		\qquad
		W_{12}(q)=W_{21}(q).
		\label{eq:app_layer_interaction_symmetry}
	\end{equation}
	
	The quantity \(W_{\ell m}(q)\) is the screened direct interaction
	between an electron in layer \(\ell\) and a hole in layer \(m\).
	Before projection onto the Bloch states, the layer dependence of the
	attractive direct interaction can be written as
	\begin{equation}
		K^{\rm att}(q)
		=
		\sum_{\ell,m=1,2}
		W_{\ell m}(q)
		\mathbb P_\ell^{(e)}
		\mathbb P_m^{(h)} .
		\label{eq:app_Katt_definition}
	\end{equation}
	Its matrix elements give the layer-resolved direct-interaction term
	in Eq.~\eqref{eq:BSE}.
	
	The interaction forms in
	Sec.~\ref{sec:exciton} and Appendix~\ref{app:bilayer_kernel} both satisfy
	Eq.~\eqref{eq:app_layer_interaction_symmetry} for equivalent layers.
	The screened direct interaction does not act on spin.
	Therefore,
	\begin{equation}
		[\mathcal Z_{eh},K^{\rm att}]
		=
		0.
		\label{eq:app_kernel_eh_symmetry}
	\end{equation}
	This gives
	\begin{equation}
		K^{\rm att}_{\lambda_c\lambda_v,
			\lambda_c'\lambda_v'}
		=
		0
		\quad\text{unless}\quad
		\lambda_c\lambda_v
		=
		\lambda_c'\lambda_v'.
		\label{eq:app_kernel_selection_rule}
	\end{equation}
	
	The direct interaction therefore couples only electron-hole states
	with the same value of \(\Lambda_{eh}\), giving the block structure in
	Eq.~\eqref{eq:BSE_four_by_four_block}.
	If the two layers experience different intralayer screening, the pair of
	same-layer terms is no longer invariant under layer exchange and the two
	\(\Lambda_{eh}\) sectors can mix. Any additional interaction preserves the
	decomposition only if it commutes with \(\mathcal Z_{eh}\).
	
	\subsection{Layer-projector matrix elements in the SRLP basis}
	Because \(\eta_z\) reverses layer parity without acting on spin, it
	connects the \(\lambda=+1\) and \(\lambda=-1\) states with the same orbital
	character.
	For the conduction states,
	\begin{equation}
		Q_{\rm A}(\alpha)
		=
		\left|
		\langle c_+|\eta_z|c_-\rangle
		\right|.
		\label{eq:app_Q_definition_A}
	\end{equation}
	
	For the valence states,
	\begin{equation}
		Q_{\rm B}(\alpha)
		=
		\left|
		\langle v_+|\eta_z|v_-\rangle
		\right|.
		\label{eq:app_Q_definition_B}
	\end{equation}
	
	For either orbital \(\tau=A,B\), direct evaluation gives
	\begin{equation}
		Q_\tau^2(\alpha)
		=
		\frac{1}{2}
		\left[
		1+
		\frac{
			(J^\tau)^2-(t_\perp^\tau)^2
		}{
			R_{\tau,+}(\alpha)R_{\tau,-}(\alpha)
		}
		\right].
		\label{eq:app_Q_tau}
	\end{equation}
	
	At the
	AFM state,
	\begin{equation}
		Q_\tau(0)
		=
		\frac{|J^\tau|}
		{\sqrt{(J^\tau)^2+(t_\perp^\tau)^2}}.
		\label{eq:app_Q_tau_afm}
	\end{equation}
	For \(t_\perp^\tau=0\), \(Q_\tau(0)=1\). Finite interlayer hopping
	reduces \(Q_\tau(0)\) because the AFM eigenstates are no longer confined to
	a single layer.
	
	The relative phases of the two channel states can be chosen so that
	the matrix elements of \(\eta_z\) are real and positive. Their product is
	\begin{equation}
		Q(\alpha)
		=
		Q_{\rm A}(\alpha)Q_{\rm B}(\alpha).
		\label{eq:app_Q_product}
	\end{equation}
	
	In the ordered channel basis
	\begin{equation}
		\left\{
		|c_-v_-\rangle,
		|c_+v_+\rangle
		\right\},
	\end{equation}
	projection of the physical layer
	projectors gives
	\begin{align}
		\mathcal P_{\rm intra}^{\rm proj}
		&=
		\frac{1}{2}
		\begin{pmatrix}
			1 & Q\\
			Q & 1
		\end{pmatrix},
		\nonumber\\
		\mathcal P_{\rm inter}^{\rm proj}
		&=
		\frac{1}{2}
		\begin{pmatrix}
			1 & -Q\\
			-Q & 1
		\end{pmatrix}.
		\label{eq:app_projected_layer_operators}
	\end{align}
	
	Their sum is \(\mathbb I_2\). For \(Q<1\), the projected matrices are not
	idempotent because the two retained band-pair states do not span the full
	space of physical electron-hole layer configurations. Their expectation
	values nevertheless give the
	intralayer and interlayer
	probabilities within the retained two-channel subspace.
	
	\subsection{Attraction energies and coupling between equal-SRLP excitons}
	
	Let \(\varphi(\mathbf k;\alpha)\) be the common normalized relative-motion
	envelope
	of the two equal-SRLP excitons,
	\begin{equation}
		\int\frac{d^2\mathbf k}{(2\pi)^2}
		|\varphi(\mathbf k;\alpha)|^2=1.
		\label{eq:app_relative_envelope_normalization}
	\end{equation}
	
	The layer-even physical configurations are
	\begin{align}
		|X_{\rm intra}\rangle
		&=
		\frac{|1_e1_h\rangle+|2_e2_h\rangle}{\sqrt{2}},
		\nonumber\\
		|X_{\rm inter}\rangle
		&=
		\frac{|1_e2_h\rangle+|2_e1_h\rangle}{\sqrt{2}}.
	\end{align}
	
	The corresponding positive electron-hole attraction energies are
	\begin{align}
		\mathcal V_{\rm intra}(\alpha)
		&=
		\langle\varphi X_{\rm intra}|
		K^{\rm att}
		|\varphi X_{\rm intra}\rangle,
		\nonumber\\
		\mathcal V_{\rm inter}(\alpha)
		&=
		\langle\varphi X_{\rm inter}|
		K^{\rm att}
		|\varphi X_{\rm inter}\rangle.
		\label{eq:app_projected_attractions}
	\end{align}
	
	These quantities contain the momentum dependence of the screened
	interaction, the interlayer separation factor, and the common envelope.
	Their average and difference are
	\begin{align}
		\mathcal V(\alpha)
		&=
		\frac{
			\mathcal V_{\rm intra}(\alpha)+ \mathcal V_{\rm inter}(\alpha)
		}{2},
		\nonumber\\
		\Delta \mathcal V(\alpha)
		&=
		\frac{
			\mathcal V_{\rm intra}(\alpha)-\mathcal V_{\rm inter}(\alpha)
		}{2}.
		\label{eq:app_V_DeltaV}
	\end{align}
	
	Within the two-channel equal-SRLP subspace, the attractive
	electron-hole interaction is
	\begin{align}
		K_+^{\rm att}(\alpha)
		&=
		\mathcal V_{\rm intra}(\alpha)\mathcal P_{\rm intra}^{\rm proj}
		+
		\mathcal V_{\rm inter}(\alpha)\mathcal P_{\rm inter}^{\rm proj}
		\nonumber\\
		&=
		\begin{pmatrix}
			\mathcal V(\alpha) & Q(\alpha)\Delta \mathcal V(\alpha)\\
			Q(\alpha)\Delta \mathcal V(\alpha) & \mathcal V(\alpha)
		\end{pmatrix}.
		\label{eq:app_attraction_projection}
	\end{align}
	
	The average attraction \(\mathcal V(\alpha)\) shifts the two excitons by the
	same amount. The difference between the intralayer and interlayer
	attraction energies couples them.
	
	Because the attraction enters the BSE Hamiltonian with a minus sign, the
	coupling strength is
	\begin{equation}
		\Omega_+(\alpha)
		=
		Q(\alpha)\left|\Delta \mathcal V(\alpha)\right|.
		\label{eq:app_Omega_B_alpha}
	\end{equation}

	In the AFM state,
	\begin{equation}
		\Omega_+
		\equiv
		\Omega_+(0)
		=
		Q_0
		\frac{
			\left|\mathcal V_{\rm intra}(0)-\mathcal V_{\rm inter}(0)\right|
		}{2}.
		\label{eq:app_Omega_br}
	\end{equation}

	Equation~\eqref{eq:app_Omega_br} assumes the same normalized
	relative-motion envelope for the two equal-SRLP excitons and uses the layer
	matrix elements of their conduction and valence states at the
	\(\Gamma\) point. If their relative-motion envelopes differ, the coupling
	must be obtained from the off-diagonal BSE matrix element.
	
	\subsection{Expansion of the reduced two-channel exciton model}
	
	Along the physical canting trajectory, we use
	\begin{equation}
		\mathcal B=
		\sin\alpha=
		\frac{B_{\rm ext}}{B_{\rm sat}}\in[0,1].
		\label{eq:app_canting_coordinate}
	\end{equation}
	
	From Eq.~\eqref{eq:direct_gaps},
	\begin{equation}
		\Delta_{\lambda\lambda'}(\mathcal B)
		=
		\Delta_\Gamma^0
		-
		R_{A\lambda}(\mathcal B)
		-
		R_{B\lambda'}(\mathcal B).
		\label{eq:app_direct_gaps_x}
	\end{equation}
	The two
	equal-SRLP direct gaps
	satisfy
	\begin{align}
		\Delta_{-,-}(\mathcal B)
		&=
		\Delta_\Gamma'
		-
		\chi_+\mathcal B
		+
		O(\mathcal B^2),
		\nonumber\\
		\Delta_{+,+}(\mathcal B)
		&=
		\Delta_\Gamma'
		+
		\chi_+\mathcal B
		+
		O(\mathcal B^2),
		\label{eq:app_bright_gaps}
	\end{align}
	where \(\chi_+\) is given in Eq.~\eqref{eq:chi_br}.
	Thus, to leading order in the single-particle expansion, the two
	equal-SRLP direct gaps acquire opposite linear canting shifts
	$-\chi_+\mathcal B$ and $+\chi_+\mathcal B$.
	
	Choosing the relative phase of the two channel states so that the
	off-diagonal matrix element is real, the reduced two-channel Hamiltonian
	used in Eq.~\eqref{eq:Heffbright} is
	\begin{equation}
		H_+(\mathcal B)
		=
		\left(E_{0,X}+\xi_X\mathcal B^2\right)\mathbb I_2
		+
		\begin{pmatrix}
			-\chi_+\mathcal B & -\Omega_+\\
			-\Omega_+ & +\chi_+\mathcal B
		\end{pmatrix}.
		\label{eq:app_effective_bright_general}
	\end{equation}
	Here, \(E_{0,X}+\xi_X\mathcal B^2\) is the average energy of the two
	equal-SRLP excitons when their mutual off-diagonal coupling is omitted,
	\(2\chi_+\mathcal B\) is their energy difference, and \(\Omega_+\) is the magnitude
	of the off-diagonal coupling between them.
	
	The corresponding exciton energies are
	\begin{equation}
		E_{\pm,X}(\mathcal B)
		=
		E_{0,X}
		+
		\xi_X\mathcal B^2
		\pm
		\sqrt{
			\Omega_+^2
			+
			\chi_+^2\mathcal B^2
		}.
		\label{eq:app_general_eigenvalues}
	\end{equation}
	
	For the reduced model, the expansion parameter of the square-root
	term is \(|\chi_+\mathcal B/\Omega_+|\). Therefore, when
	\(\Omega_+>|\chi_+|\), the expansion converges throughout the physical
	canting range \(0\leq \mathcal B\leq1\). Retaining terms through second order gives
	\begin{equation}
		E_{-,X}(\mathcal B)
		=
		E_{0,X}
		-
		\Omega_+
		+
		\left(
		\xi_X
		-
		\frac{\chi_+^2}{2\Omega_+}
		\right)\mathcal B^2
		+
		O(\mathcal B^4).
		\label{eq:app_general_quadratic}
	\end{equation}
	The term
	\(-\chi_+^2/(2\Omega_+)\) is the additional quadratic contribution
	generated by the coupling between the two equal-SRLP excitons.
	
	\subsection{Opposite-SRLP two-channel exciton model}
	\label{app:dark_two_channel}
	
	The opposite-SRLP sector \(\Lambda_{eh}=-1\) contains the two
	electron-hole channels
	\(\{|c_+v_-\rangle,|c_-v_+\rangle\}\).
	Within the same common-envelope projection, the layer matrix elements
	connecting \(c_+\) with \(c_-\) and \(v_+\) with \(v_-\) have magnitudes
	\(Q_{\rm A}(0)\) and \(Q_{\rm B}(0)\), respectively. Their product is therefore the
	same band-edge factor
	\(Q_0=Q_{\rm A}(0)Q_{\rm B}(0)\)
	defined in Eq.~\eqref{eq:app_Q_product}.
	
	Using Eq.~\eqref{eq:gap_coefficients}, we define
	\begin{equation}
		\chi_-
		=
		\frac{J^{\rm A} t_\perp^{\rm A}}{R_{\rm A}}
		-
		\frac{J^{\rm B} t_\perp^{\rm B}}{R_{\rm B}}.
		\label{eq:app_chi_dark}
	\end{equation}
	
	The two opposite-SRLP direct gaps then have opposite leading
	canting shifts,
	\begin{align}
		\Delta_{+,-}(\mathcal B)
		&=
		\Delta_\Gamma'-\chi_-\mathcal B+O(\mathcal B^2),
		\nonumber\\
		\Delta_{-,+}(\mathcal B)
		&=
		\Delta_\Gamma'+\chi_-\mathcal B+O(\mathcal B^2).
		\label{eq:app_dark_gaps}
	\end{align}
	
	Let \(\mathcal V_{\rm intra}^{-}\) and \(\mathcal V_{\rm inter}^{-}\) denote the
	intralayer and interlayer electron-hole attraction energies obtained using
	the common relative-motion envelope for these two channels. Their difference
	couples the two opposite-SRLP excitons. The reduced two-channel Hamiltonian
	is
	\begin{align}
		H_{\rm eff}^{-}(\mathcal B)
		&=
		\left(E_{0,X}^{-}+\xi_X^{-}\mathcal B^2\right)\mathbb{I}_2
		+
		\begin{pmatrix}
			-\chi_-\mathcal B & -\Omega_-\\
			-\Omega_- & +\chi_-\mathcal B
		\end{pmatrix},
		\label{eq:app_dark_two_channel}\\
		\Omega_-
		&=
		Q_0\,
		\frac{
			\left|\mathcal V_{\rm intra}^{-}-\mathcal V_{\rm inter}^{-}\right|
		}{2}.
		\nonumber
	\end{align}
	
	Here, \(E_{0,X}^{-}\) is the common energy of the two opposite-SRLP
	excitons at the AFM state when their mutual coupling is omitted,
	\(\xi_X^{-}\mathcal B^2\) is the quadratic energy shift common to both excitons,
	and \(\Omega_-\) is the magnitude of their off-diagonal coupling.
	
	The corresponding exciton energies are
	\begin{equation}
		E_{\pm,X}^{-}(\mathcal B)
		=
		E_{0,X}^{-}
		+
		\xi_X^{-}\mathcal B^2
		\pm
		\sqrt{
			\Omega_-^2+\chi_-^2\mathcal B^2
		}.
		\label{eq:app_dark_eigenenergies}
	\end{equation}
	
	The expansion parameter of the square-root term is
	\(\left|\chi_-\mathcal B/\Omega_-\right|\). If
	\(\Omega_->|\chi_-|\), the expansion converges throughout the physical
	canting range \(0\leq \mathcal B\leq1\). Retaining terms through second order for
	the lower exciton gives
	\begin{equation}
		E_{-,X}^{-}(\mathcal B)
		=
		E_{0,X}^{-}
		-
		\Omega_-
		+
		\left(
		\xi_X^{-}
		-
		\frac{\chi_-^2}{2\Omega_-}
		\right)\mathcal B^2
		+
		O(\mathcal B^4).
		\label{eq:app_dark_quadratic}
	\end{equation}
	
	The term
	\(-\chi_-^2/(2\Omega_-)\) is the additional quadratic contribution
	generated by the coupling between the two opposite-SRLP excitons.
	
	At the AFM state \(\mathcal B=0\), the two opposite-SRLP excitons are
	degenerate when their mutual coupling is omitted. Including the coupling
	gives the energies
	\(E_{0,X}^{-}-\Omega_-\) and
	\(E_{0,X}^{-}+\Omega_-\), separated by \(2\Omega_-\).
	
	The opposite-SRLP curves in
	Fig.~\ref{fig:two_channel_model_weights}(a) are obtained from
	Eq.~\eqref{eq:app_dark_two_channel}. Within the minimal CrSBr Hamiltonian,
	these excitons remain optically dark because the conduction and valence
	states forming each channel have opposite SRLP eigenvalues.

	\subsection{
		Layer probabilities of the equal- and opposite-SRLP excitons}
	The physical intralayer and interlayer probabilities are the
	expectation values of the two-particle layer operators introduced in
	Eq.~\eqref{eq:layer_pair_projectors}. We evaluate these probabilities
	directly for the lower-energy eigenstate of each reduced two-channel
	Hamiltonian.
	
	For the equal-SRLP sector, we use the ordered basis
	$\{|c_-v_-\rangle,|c_+v_+\rangle\}$ and write the lower-energy exciton as
	\begin{equation}
		|X_{+,{\rm low}}(\mathcal B)\rangle
		=
		\mathfrak a_+(\mathcal B)|c_-v_-\rangle
		+
		\mathfrak b_+(\mathcal B)|c_+v_+\rangle.
		\label{eq:app_equal_lower_state}
	\end{equation}
	
	Diagonalizing Eq.~\eqref{eq:app_effective_bright_general} gives
	\begin{align}
		\mathfrak a_+^2(\mathcal B)
		&=
		\frac{1}{2}
		\left[
		1+
		\frac{\chi_+\mathcal B}
		{\sqrt{\Omega_+^2+\chi_+^2\mathcal B^2}}
		\right],
		\nonumber\\
		\mathfrak b_+^2(\mathcal B)
		&=
		\frac{1}{2}
		\left[
		1-
		\frac{\chi_+\mathcal B}
		{\sqrt{\Omega_+^2+\chi_+^2\mathcal B^2}}
		\right],
		\nonumber\\
		2\mathfrak a_+(\mathcal B) \mathfrak b_+(\mathcal B)
		&=
		\frac{\Omega_+}
		{\sqrt{\Omega_+^2+\chi_+^2\mathcal B^2}}.
		\label{eq:app_equal_channel_amplitudes}
	\end{align}
	
	Equation~\eqref{eq:app_projected_layer_operators} then gives the
	interlayer probability
	\begin{equation}
		w_{{\rm inter},+}^{\rm low}(\mathcal B)
		=
		\frac{1}{2}
		\left[
		1-
		Q_0
		\frac{\Omega_+}
		{\sqrt{\Omega_+^2+\chi_+^2\mathcal B^2}}
		\right].
		\label{eq:app_equal_interlayer_probability}
	\end{equation}
	
	The corresponding intralayer probability is
	\begin{equation}
		w_{{\rm intra},+}^{\rm low}(\mathcal B)
		=
		1-w_{{\rm inter},+}^{\rm low}(\mathcal B).
		\label{eq:app_equal_intralayer_probability}
	\end{equation}
	
	This is the expression used in
	Eq.~\eqref{eq:w_inter}. When $\Omega_+>|\chi_+|$, its expansion converges
	throughout the physical interval $0\leq \mathcal B\leq1$ and gives
	\begin{equation}
		w_{{\rm inter},+}^{\rm low}(\mathcal B)
		=
		\frac{1-Q_0}{2}
		+
		\frac{Q_0\chi_+^2}{4\Omega_+^2}\mathcal B^2
		+
		O(\mathcal B^4).
		\label{eq:app_equal_interlayer_expansion}
	\end{equation}
	
	For the opposite-SRLP sector, we use the ordered basis
	$\{|c_+v_-\rangle,|c_-v_+\rangle\}$ and write
	\begin{equation}
		|X_{-,{\rm low}}(\mathcal B)\rangle
		=
		\mathfrak a_-(\mathcal B)|c_+v_-\rangle
		+
		\mathfrak b_-(\mathcal B)|c_-v_+\rangle.
		\label{eq:app_dark_lower_state}
	\end{equation}
	
	Diagonalizing Eq.~\eqref{eq:app_dark_two_channel} gives
	\begin{align}
		\mathfrak a_-^2(\mathcal B)
		&=
		\frac{1}{2}
		\left[
		1+
		\frac{\chi_-\mathcal B}
		{\sqrt{\Omega_-^2+\chi_-^2\mathcal B^2}}
		\right],
		\nonumber\\
		\mathfrak b_-^2(\mathcal B)
		&=
		\frac{1}{2}
		\left[
		1-
		\frac{\chi_-\mathcal B}
		{\sqrt{\Omega_-^2+\chi_-^2\mathcal B^2}}
		\right],
		\nonumber\\
		2\mathfrak a_-(\mathcal B)\mathfrak b_-(\mathcal B)
		&=
		\frac{\Omega_-}
		{\sqrt{\Omega_-^2+\chi_-^2\mathcal B^2}}.
		\label{eq:app_dark_channel_amplitudes}
	\end{align}
	
	The off-diagonal layer matrix element in this basis contains the same
	product $Q_{\rm A}(0)Q_{\rm B}(0)=Q_0$ as in the equal-SRLP sector. The interlayer
	probability of the lower opposite-SRLP exciton is therefore
	\begin{equation}
		w_{{\rm inter},-}^{\rm low}(\mathcal B)
		=
		\frac{1}{2}
		\left[
		1-
		Q_0
		\frac{\Omega_-}
		{\sqrt{\Omega_-^2+\chi_-^2\mathcal B^2}}
		\right].
		\label{eq:app_dark_interlayer_probability}
	\end{equation}
	
	The corresponding intralayer probability is
	\begin{equation}
		w_{{\rm intra},-}^{\rm low}(\mathcal B)
		=
		1-w_{{\rm inter},-}^{\rm low}(\mathcal B).
		\label{eq:app_dark_intralayer_probability}
	\end{equation}
	
	Expanding about the AFM state gives
	\begin{equation}
		w_{{\rm inter},-}^{\rm low}(\mathcal B)
		=
		\frac{1-Q_0}{2}
		+
		\frac{Q_0\chi_-^2}{4\Omega_-^2}\mathcal B^2
		+
		O(\mathcal B^4).
		\label{eq:app_dark_interlayer_expansion}
	\end{equation}
	
	This expansion converges when
	$|\chi_-\mathcal B/\Omega_-|<1$. It therefore applies throughout
	$0\leq \mathcal B\leq1$ only when $\Omega_->|\chi_-|$. The exact expression in
	Eq.~\eqref{eq:app_dark_interlayer_probability} does not require this
	condition.
	
	Equations~\eqref{eq:app_equal_interlayer_probability} and
	\eqref{eq:app_dark_interlayer_probability} give the layer probabilities
	plotted for the lower equal- and opposite-SRLP excitons in
	Fig.~\ref{fig:two_channel_model_weights}(b).
	\subsection{
		Electron-hole continuum edge and exciton binding energy}
	
	The direct BSE Hamiltonian can be written as
	\begin{equation}
		H_{\rm BSE}
		=
		H_0^{eh}
		-
		K^{\rm att},
		\label{eq:app_bse_decomposition}
	\end{equation}
	where $H_0^{eh}$ contains the independent electron-hole energies and
	$K^{\rm att}$ denotes the attractive direct electron-hole interaction.
	
	For a normalized eigenstate,
	\begin{equation}
		E_{n,X}
		=
		\langle\Psi_n|H_0^{eh}|\Psi_n\rangle
		-
		\langle\Psi_n|K^{\rm att}|\Psi_n\rangle.
		\label{eq:app_bse_expectation_identity}
	\end{equation}
	Therefore,
	\begin{equation}
		\langle\Psi_n|H_0^{eh}|\Psi_n\rangle
		-
		E_{n,X}
		=
		\langle\Psi_n|K^{\rm att}|\Psi_n\rangle.
		\label{eq:app_attraction_expectation}
	\end{equation}
	
	The right-hand side is the mean attractive interaction energy of the
	exciton. It is not its binding energy, because the electron-hole wave
	function also has an independent-particle energy above the bottom of the
	continuum.
	
	For each SRLP sector $\Lambda_{eh}=\pm1$, the continuum edge is the
	lowest energy of an independent electron-hole pair in that sector,
	\begin{equation}
		E_{\rm th}^{(\Lambda_{eh})}(\mathcal B)
		=
		\min_{\mathbf k,\,(c,v)\in\mathcal C_{\Lambda_{eh}}}
		\left[
		E_c(\mathbf k,\mathcal B)-E_v(\mathbf k,\mathcal B)
		\right].
		\label{eq:app_sector_threshold}
	\end{equation}
	
	An exciton is bound when
	$E_{n,X}^{(\Lambda_{eh})}(\mathcal B)<E_{\rm th}^{(\Lambda_{eh})}(\mathcal B)$.
	Its binding energy is
	\begin{equation}
		E_{b,n}^{(\Lambda_{eh})}(\mathcal B)
		=
		E_{\rm th}^{(\Lambda_{eh})}(\mathcal B)
		-
		E_{n,X}^{(\Lambda_{eh})}(\mathcal B).
		\label{eq:app_binding_energy}
	\end{equation}
	
	Combining Eqs.~\eqref{eq:app_bse_expectation_identity} and
	\eqref{eq:app_binding_energy} gives
	\begin{equation}
		E_{b,n}^{(\Lambda_{eh})}
		=
		\langle\Psi_n|K^{\rm att}|\Psi_n\rangle
		-
		\left[
		\langle\Psi_n|H_0^{eh}|\Psi_n\rangle
		-
		E_{\rm th}^{(\Lambda_{eh})}
		\right].
		\label{eq:app_binding_vs_attraction}
	\end{equation}
	
	The binding energy is therefore the attractive interaction energy
	reduced by the independent electron-hole energy required to localize the
	relative motion above the continuum edge.
	\subsection{
		Validity of the two-channel exciton model}
	The separation of the BSE into equal- and opposite-SRLP sectors remains
	exact for the layer-symmetric direct electron-hole interaction. The reduced
	two-channel exciton models require three additional conditions.
	
	First, the layer composition of the conduction and valence states must
	vary weakly over the momenta that contribute appreciably to the exciton.

	For $\xi=c,v$, we write
	\begin{equation}
		U_c(\mathbf k)
		=
		\bigl(
		|c_-(\mathbf k)\rangle,
		|c_+(\mathbf k)\rangle
		\bigr).
	\end{equation}
	
	\begin{equation}
		U_v(\mathbf k)
		=
		\bigl(
		|v_-(\mathbf k)\rangle,
		|v_+(\mathbf k)\rangle
		\bigr).
	\end{equation}
	
	\begin{equation}
		\mathcal L_\ell^{(\xi)}(\mathbf k)
		=
		U_\xi^\dagger(\mathbf k)
		\mathbb P_\ell
		U_\xi(\mathbf k).
		\label{eq:app_layer_matrix_validity}
	\end{equation}
	
	The band-edge layer matrices are representative of the states sampled
	by the exciton when
	\begin{equation}
		\max_{\substack{
				\mathbf k\in{\rm supp}\,\varphi\\
				\ell=1,2;\,\xi=c,v
		}}
		\left\|
		\mathcal L_\ell^{(\xi)}(\mathbf k)
		-
		\mathcal L_\ell^{(\xi)}(\Gamma)
		\right\|
		\ll1.
		\label{eq:app_gamma_locality_condition}
	\end{equation}
	
	Under this condition, the layer matrix elements evaluated at
	$\Gamma$ can be used over the momentum range occupied by the exciton
	wave function.
	
	Second, electron-hole states involving $c'_\lambda$ or
	$v'_\lambda$ must remain sufficiently separated in energy from the four
	near-gap channels retained in the BSE. For the two reduced SRLP models, a
	sufficient condition is
	\begin{equation}
		\Delta_{\rm remote}
		\gg
		\max\left(
		E_b,
		\Omega_+,
		\Omega_-,
		|\chi_+\mathcal B|,
		|\chi_-\mathcal B|
		\right).
		\label{eq:app_spectral_isolation}
	\end{equation}
	
	For the CrSBr
	parameters, the smaller band-edge separation to a remote
	single-particle
	band
	is
	$2R_{\rm A}=0.523~\mathrm{eV}$, compared with an
	equal-SRLP exciton binding energy
	near $0.197~\mathrm{eV}$ and an equal-SRLP coupling
	$\Omega_+$ near $0.051~\mathrm{eV}$.
	
	The separation is therefore much larger than the equal-SRLP coupling
	but only a few times larger than the binding energy. The accuracy of the
	two-channel reduction is consequently assessed directly by comparison with
	the full BSE in Fig.~\ref{fig:two_channel_model_weights}.
	
	Third, the two excitons retained within each SRLP sector must have
	similar relative-motion wave functions. For the equal-SRLP sector, this
	requires
	\begin{equation}
		\left|
		\langle
		\varphi_{-,-}
		|
		\varphi_{+,+}
		\rangle
		\right|
		\simeq1.
		\label{eq:app_envelope_overlap}
	\end{equation}
	
	For the opposite-SRLP sector, the corresponding condition is
	\begin{equation}
		\left|
		\langle
		\varphi_{+,-}
		|
		\varphi_{-,+}
		\rangle
		\right|
		\simeq1.
		\label{eq:app_dark_envelope_overlap}
	\end{equation}
	
	These conditions allow the two excitons within each SRLP sector to
	share a common relative-motion envelope while retaining their different
	band and layer-spin compositions.
	
	If these conditions are not satisfied, the equal- and opposite-SRLP
	sectors remain separated by symmetry, but the two-channel Hamiltonians need
	not reproduce the exciton energies or layer probabilities quantitatively.
	The couplings and the intralayer and interlayer probabilities must then be
	obtained directly from the full momentum-dependent BSE eigenstates.

	\bibliography{references}
	
\end{document}